\documentclass[twocolumn,superscriptaddress,floatfix,prb,aps,showpacs]{revtex4}
\usepackage{graphicx}
\begin{document}
\bibliographystyle{apsrev}
\def\sa{\section} 
\def\sb{\subsection}
\def\sc{\subsubsection}
\def\ind{\ \ \ \ }
\def\be{\begin{equation}}
\def\ee{\end{equation}}
\def\bea{\begin{eqnarray}}
\def\eea{\end{eqnarray}}
\def\ba{\begin{array}}
\def\ea{\end{array}}
\def\nn{\nonumber}
\def\ben{\begin{enumerate}}
\def\een{\end{enumerate}}
\def\fn{\footnote}
\def\rd{\partial}
\def\rot{\nabla\times}
\def\r{\right}
\def\l{\left}
\def\gt{\rightarrow}
\def\cf{\leftarrow}
\def\bw{\leftrightarrow}
\def\ra{\rangle}
\def\la{\langle} 
\def\bla{\big\langle}
\def\bra{\big\rangle}
\def\Bla{\Big\langle}
\def\Bra{\Big\rangle}
\def\ddt{{d\over dt}}
\def\rdt{{\rd\over\rd t}}
\def\rdx{{\rd\over\rd x}}
\def\bb{\begin{thebibliography}{99}}
\def\eb{\end{thebibliography}}
\def\bit{\bibitem}
\def\bc{\begin{center}}
\def\ec{\end{center}}

\title{Disorder effect on 3-dimensional $Z_2$ quantum 
spin Hall systems}
\author{Ryuichi Shindou}
\affiliation{Furusaki Condensed Matter Theory Laboratory, RIKEN, 
2-1 Hirosawa, Wako, Saitama 351-0198, Japan}
\author{Shuichi Murakami} 
\affiliation{Department of Physics, Tokyo Institute of Technology, 
2-12-1 Ookayama, Meguro-ku, Tokyo 152-8551, Japan} 
\affiliation{PRESTO, Japan Science and Technology Agency (JST),
Kawaguchi, Saitama 332-0012, Japan} 
\begin{abstract}
{In this paper, we address ourselves to the nonmagnetic 
disorder effects onto the quantum critical point which 
intervenes the 3-dimensional $Z_2$ quantum spin Hall insulator
(topological insulator) and an ordinary insulator. 
The minimal model describing this type of the quantum critical point 
is the single-copy of the $3+1$ Dirac fermion, whose topological mass $m$ 
induces the phase transition between the topological 
insulator and an ordinary one. 
We first derive the phase diagram spanned by the 
mass-term $m$, chemical potential $\mu$ and 
strength of the disorder within the self-consistent Born approximation.  
By way of this, we find a finite density of state appears 
even at the zero-energy and
at the phase transition point, i.e. $m=\mu=0$, if the strength of the 
disorder potential exceeds some critical value. To infer the structure 
of the low-energy effective theory around these zero-energy states, 
we further calculated the weak localization (WL) correction 
to the conductivity. 
To be more specific, we have found that the diffuson is 
dominated by the charge diffusion mode and parity diffusion mode. 
While the charge diffusion mode always carries the diffusion pole,  
the parity diffusion mode becomes massless only 
at $m=0$, but suffers from the infrared 
cutoff for non-zero $m$. Corresponding to 
this feature of the diffuson, the 
Cooperon is also composed of two 
quasi-degenerate contributions. We found that these 
two give rise to the same magnitude 
of the anti-weak-localization (AWL) correction 
with each other at $m=0$.   
As a result, when the topological mass $m$ is fine-tuned 
to be zero (but for generic $\mu$), the AWL 
correction becomes doubled (quantum correction doubling). 
Based on this observation, we will discuss the possible 
microscopic picture of the ``levitation and pair annihilation'' 
phenomena, recently discovered by Onoda {\it et al}~\cite{OAN}.} 
\end{abstract}
\maketitle

\section{Introduction}
Physics of spin transport
has been a matter of intensive 
research in recent years. One of the 
topics of current interest is the 
spin Hall effect. This has been originally 
proposed theoretically 
\cite{Murakami03,Sinova04} and later followed by 
various experimental results \cite{Kato04,Wunderlich05}.
The research on the spin Hall effect opens a new field 
of Hall effects in time-reversal invariant systems.
This has also led us to a new concept of the quantum 
spin Hall effect, which is the natural ``spin"
extension of the quantum Hall effect \cite{BZ,KM,KM2}. 
In the quantum spin Hall
effect in two dimensions, the 
bulk is gapped while there are gapless edge states carrying 
spin current. In this case, the external magnetic field is
zero, while the spin-orbit coupling acts as a ``spin-dependent
magnetic field", giving rise to the effect analogous to the 
quantum Hall effect. Such insulators showing the 
quantum spin Hall effect 
are characterized by the $Z_2$ topological number \cite{KM}. 
We shall call them  as $Z_2$ quantum 
spin Hall insulators (QSHI). 

The simplest system for the 
two-dimensional (2-d) $Z_2$ QSHI is 
realized as a superposition of the 
wavefunctions 
of two quantum Hall subsystems \cite{BZ,KM,KM2}, 
one with spin up and the other with 
spin down, having 
opposite Chern numbers.
The system respects not only the time-reversal $({\cal T})$ 
invariance but also the spin-conservation. Such an 
insulator supports the same numbers of right-moving up-spin edge states and 
left-moving down-spin edges.

This Kramers pair of chiral edge 
states is often called as the helical 
edge state. By its construction, the number of 
this Kramers pairs of edge states 
correspond to the Chern integer  
associated with its bulk wavefunction~\cite{Hatsugai93}. 
The ${\cal T}$ symmetry guarantees the 
double degeneracy between right-moving 
up-spin and left moving down-spin states. 
Thus, the stability of {\it each} Kramers 
pair is supported by this ${\cal T}$-symmetry.  

Spin non-conserving 
(but ${\cal T}$ invariant) perturbations, however, 
introduce level repulsions 
between two {\it different} Kramers pairs. 
Namely, they usually let two pairs 
annihilate with each other, and 
open a gap.
Accordingly, in the 
presence of generic 
spin-non-conserving perturbations,  those 
wavefunctions having {\it even} numbers of Kramers pairs, reduce to 
{\it trivial} insulators, which have no 
gapless edge states~\cite{Wu06,Xu06a,Fu06a,Fukui07}. 
Meanwhile, wavefunctions having {\it odd} numbers of pairs still can 
have one active helical edge mode. The latter 
is dubbed as the $Z_2$ quantum 
spin Hall (topological) insulator. Thus, stability of 
such a gapless edge state is protected only by the ${\cal T}$-symmetry, 
while does not require spin-conservations 
anymore~\cite{Wu06,Xu06a,Fu06a,Fukui07,Bernevig07a,Koenig07}.  
 
The three-dimensional (3-d) version \cite{Roy,FKM,Moore,Murakami06a,Hsieh,Schnyder}
of the $Z_2$ QSHI  
carries same characters as that of 2-d does.  
The 3-d $Z_2$ QSHI also allows any 
spin-nonconservering perturbations, while  
always requires the ${\cal T}$ symmetry. 
Simultaneously, however, it is not   
a mere extension of the 2-d $Z_2$ QSHI, in a sense  
that 3-d $Z_2$ QSHI has {\it no $U(1)$-analogue} of QSHI. Namely, 
they support a $2+1$ massless Dirac fermion as its surface 
state ~\cite{FKM,Hsieh},   
instead of a helical edge state. In the 2-d  
surface Brillouin zone, say $k_x$-$k_y$ plane, this massless Dirac fermion 
has a spin which depends 
on the (surface) crystal momentum. It is clear that such an insulator cannot 
be adiabatically connected into a composite of two 
spinless wavefunctions. In  such 
$Z_2$ QSHI, the ${\cal T}$ symmetry therefore guarantees the 
massless nature of each $2+1$ surface Dirac fermion. 
  
\begin{figure}
\begin{center}
\includegraphics[width=0.45\textwidth]{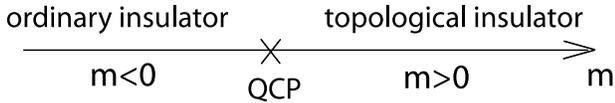}
\end{center}
\caption{A schematic phase diagram for the quantum critical point 
intervening the $Z_2$ QSHI and an ordinary insulator.
$m$ is a system parameter driving the phase transition.
When $m=0$, the system is in a critical phase. 
In the Fu-Kane-Mele model, it corresponds to the relative strength of one 
out of the four NN transfer integrals emitting from a single site.}
\label{t1}
\end{figure}

$Z_2$ QSHI 
always has a quantum critical point at its phase boundary to 
any ordinary insulators, in both 2-d and 3-d. 
For example, from 
the 3-d tight-binding model proposed by Fu, Kane and Mele \cite{FKM},
we can explicitly see this; when a certain ${\cal T}$ 
symmetric parameter is varied in their model, 
3d $Z_2$ QSHI is driven into an 
ordinary insulator, latter of which does not support any 
surface states (see Fig.~\ref{t1}).  
Observing this, a following question naturally arises;    
during this tuning, the ${\cal T}$  symmetry is {\it always} preserved, 
so that the massless nature of 
the surface Dirac fermion is supposed to be 
protected by this. At the same time, 
however, this $2+1$ surface fermion should have 
become ``massive'', when a system enter an ordinary insulator phase.  
Thus, one might ask how this single surface Dirac fermion 
could acquire a finite mass, with keeping the ${\cal T}$ 
symmetry?   

The answer is simple;  we have
{\it two} sample boundaries, say $z=+L$ and $z=-L$. Each boundary 
supports one $2+1$ surface massless Dirac fermion respectively.  
They are localized at each boundary, when the bulk gap  
is sufficiently large. In such a situation,  
a mixing between these two $2+1$ surface massless 
Dirac fermions is tiny, i.e. ${\cal O}(e^{-L/\xi})$ 
with $\xi$ being the localization length.  
However, when a system becomes close
to the quantum critical point, a mixing between these two surface states 
becomes substantial, with increasing $\xi$. When a bulk eventually reaches 
the quantum critical point, two surface massless Dirac fermions 
readily communicate via extended bulk states.  
Thus, they generally annihilate in pairs, just as
in those insulators having even number of $2+1$ surface 
Dirac fermions at one boundary.

\begin{figure}
\begin{center}
\includegraphics[width=0.45\textwidth]{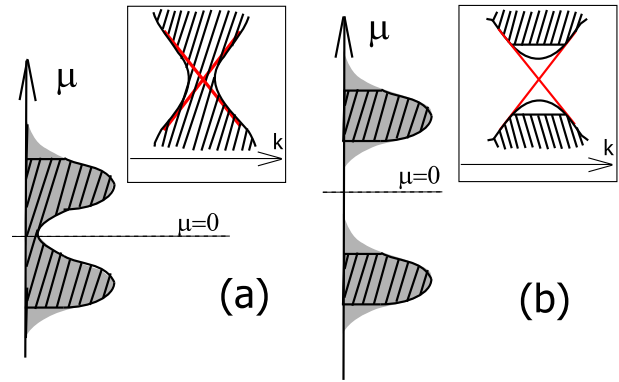}
\end{center}
\caption{A schematic picture of the density of state and mobility edges, where 
hatched region corresponds to the extended state.  Inset represents the 
energy dispersion as a function of surface crystal momentum, where the red 
line corresponds to the surface state at $z=\pm L$. (a) $m=0$; at quantum critical 
point. (b) $m>0$; in the topological insulator phase.}
\label{t2}
\end{figure}

This simple picture in the clean limit raises the following  
non-trivial speculations about the disorder effects  
on the $Z_2$ QSHI.
Suppose that ${\cal T}$-symmetric random potentials are introduced 
in the topological insulator phase. When the 
corresponding bulk gap is sufficiently large, 
we could begin with two separate bands. 
The scaling argument in 3-d~\cite{fourgang} 
tells us that each band should always have 
two mobility edges, respectively (see Fig.~\ref{t2}(b)). Namely,   
there is no delocalized bulk-wavefunction near the zero energy. 
Starting from this phase,  let us change some ${\cal T}$-invariant 
model-parameters, so that a bulk transits from this topological 
insulator to an ordinary one. 
From the argument in the clean limit,  one can then expect 
that {\it a delocalized bulk-wavefunction should emerge  
at the zero-energy region at the quantum critical 
point}, i.e. $\mu=m=0$ (see Fig.~\ref{t2}(a)). If it were not, 
the two surface states localized at the two sample  
boundaries could not communicate at all 
and they could not annihilate with 
each other. As a result, the system 
was unable to smoothly enter an ordinary  
band insulator, since the latter one 
does not support any surface state at all. 

To put this reversely, the existence of 
the quantum critical point (QCP) 
having extended bulk 
wavefunctions is always required, whenever this critical 
point separates an ordinary insulator and the topological insulator.  
This is because these two insulating phases support 
different numbers of Kramers pairs of surface states.
Moreover, provided that these surface states are stable 
by itself, this QCP should be 
also stable, {\it however small} 
the density of state (DOS) at  
the zero energy is and {\it however strong}  
the disorder strength is. Otherwise, the topological 
insulator could be adiabatically connected into an  
ordinary band insulator, which contradicts
the different $Z_2$ topological numbers for the two phases.

In this paper, we will uncover several novel features 
associated with the non-magnetic disorder effects onto  
this topological quantum critical point.   
The organization of this paper is summarized as follows.  
In the next section, we will briefly review the   
effective continuum model for the quantum critical point 
intervening the $Z_2$ topological insulator and an ordinary 
insulator. The effective model is known to be described by 
the $3+1$ Dirac fermion, whose mass term brings 
about the topological quantum phase transition. Namely, when 
the mass term changed from positive to negative, a system 
transits from the topological insulator to an ordinary insulator. 
As such, we call this mass term especially as the topological 
mass term. Based on this effective model, we will 
next introduce various types of the on-site random potentials 
respecting the  ${\cal T}$-symmetry. Note that, in this paper,  
we restrict ourselves to ${\cal T}$-symmetric cases and 
exclude magnetic impurities, because, in the absence of 
the ${\cal T}$  symmetry, the two phases are no longer 
topologically distinct. 

Based on the self-consistent Born approximation,  
we first work over the single-particle Green 
function in the section~III. 
The phase diagram spanned by the (bare) chemical potential 
$\mu$, (bare) mass term $m$ 
and strength of the disorder $\alpha$ is derived. In particular, 
at the critical point, i.e. $m=0$, 
we found some critical value of the 
disorder strength, $\alpha_{c}$, above which 
the zero-energy state, i.e. $\mu=0$, acquires a finite life-time $\tau$; 

\begin{eqnarray}
\frac{1}{\tau} {\rm ArcTan} \big[\tau\big] = 1-\frac{\alpha_{c}}{\alpha} \label{e1}.  
\end{eqnarray} 
Since the density of state in our model is always proportional 
to the inverse of the life-time (see below), non zero $\tau^{-1}$ 
simply means that a system is in a compressible phase.

\begin{figure}
\begin{center}
\includegraphics[width=0.45\textwidth]{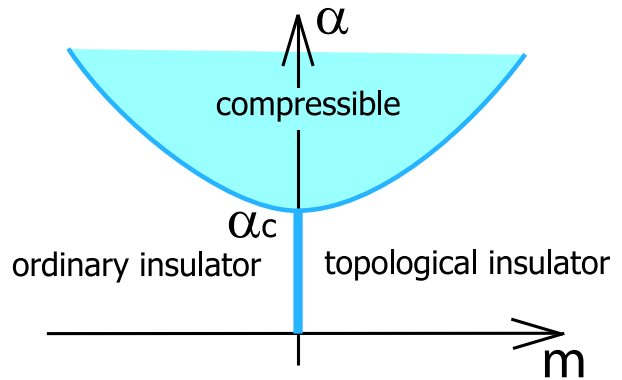}
\end{center}
\caption{A schematic phase diagram for $\mu=0$. A blue shaded 
region corresponds to a compressible phase, which separates 
two gapped phases, i.e. an ordinary insulator and the topological insulator. 
This phase boundary for $\alpha>\alpha_c$ is given by 
$1-\frac{\alpha_c}{\alpha}\equiv\frac{m}{2}{\rm ArcTan}\big[\frac{2}{m}\big]$.}
\label{t3}
\end{figure}

When a finite but small topological mass $m$ is introduced for 
$\alpha>\alpha_c$, the life-time $\tau$ and the renormalized 
mass  $\overline{m}$ becomes as follows; 
\begin{eqnarray}
\left(\frac{1}{\tau},\overline{m}\right) = 
\left(\sqrt{\frac{1}{\tau^2_0}-\frac{m^2}{4}},\frac{m}{2}\right), \label{e2} 
\end{eqnarray}   
where $\tau_0$ is given as a function only of 
$\alpha$ via Eq.~(\ref{e1}). Thus, when the bare  
topological mass exceeds the critical value 
$m_c\equiv2\tau^{-1}_0$, 
the density of state vanishes, so that 
a system enters an incompressible phase. 
This gapped phase 
can be adiabatically 
connected into band insulator phases in the clean limit.  
Accordingly, we will reach the phase diagram for the zero-energy state 
as depicted in Fig.~\ref{t3}. 
In the section~III, we also describe the behavior of the one-particle 
Green function for a finite $\mu$ (see below), in which we  
observe that the compressible phase (not necessarily metallic phase) 
always intervenes the topological insulator 
phase and an ordinary insulator phase as in Fig.~\ref{t3}. 
    
Focusing on this intervening compressible phase, 
especially for $\alpha < \alpha_c$, 
we will derive in the section.~IV  
the diffuson, Cooperon and the  
weak localization correction to 
the electric conductivity.  We will first observe that 
the diffuson is composed of two quasi-degenerate  
low-energy modes;
\begin{eqnarray}
\hat{\Gamma}^d(q,\omega) \propto \frac{1}{\omega + iDq^2}\hat{\Gamma}^d_1 
+ \frac{1}{\omega + iDq^2 + i\tau^{-1}_{\rm topo}} \hat{\Gamma}^d_2 + \cdots,  
\label{punch1-0} 
\end{eqnarray} 
(see Fig.~(\ref{t12-1-0}) or eq.~(\ref{identity}) for the 
definition of $\hat{\Gamma}^d(q,\omega)$). 
The first term participates in usual charge diffuson mode, 
and therefore always has the diffusion pole, 
i.e. $[\omega + i Dq^2]^{-1}$ ($\omega$ and $q$ 
stand for the frequency and momentum of the density fluctuation 
respectively). The other low-energy mode, however, 
becomes massless only in the  
absence of the topological mass $m$.  Namely, its  
low-energy and long wavelength behavior is generally 
truncated by the infrared cutoff $\tau^{-1}_{\rm topo}$,  
while this infrared cutoff reduces to zero at $m=0$ 
(but generic $\mu$), i.e. 
$\tau^{-1}_{\rm topo}\propto m^2$.        

Physically speaking, this second mode describes 
the diffusion of the parity density degree of 
freedom, which becomes 
a conserved quantity of our effective hamiltonian 
at 
$m=0$.  
Namely, the parity-density correlation function 
exhibits the diffusion pole structure at the 
critical point ($m=0$), while it becomes massive 
in the presence of the finite topological mass. 
Consequently, the diffuson acquires one additional 
low-energy, i.e. the 2nd term of eq.~(\ref{punch1-0}), 
into which the information of 
this parity-density correlation function is separately 
encoded.    

When the hole-line of the diffuson time-reversed, these 
two-mode features are transcribed into the Cooperon:  
the Cooperon thus obtained is also composed of 
two quasi-degenerate dominant contributions;
\begin{eqnarray}
\hat{U}^{\rm coop}(k+k',\omega) &\propto& \frac{1}{\omega + iD(k+k')^2}
\hat{U}^{\rm c}_1 \nn \\
&&\hspace{-2cm} + \ \frac{1}{\omega + iD(k+k')^2 
+ i\tau^{-1}_{\rm topo}} \hat{U}^{\rm c}_2 + \cdots, \label{punch2-0}
\end{eqnarray}
(see Fig.~\ref{t12-1-0} and its caption for the definition of  
$\hat{U}^{\rm coop}(k+k',\omega)$).  
In section.~IV, we will see that, 
at the critical point ($m=0$), 
these two contributions give rise to 
the same amplitude of the  
anti-weak-localization (AWL) correction 
to the electric conductivity. 
When the finite topological  
mass is introduced, however, the Cooperon associated  
with the parity mode channel becomes ineffective, 
since its backward scattering behaviour 
is truncated by the cutoff 
$\tau^{-1}_{\rm topo}$.   
Meanwhile, the Cooperon obtained 
from the charge mode channel 
remains effective, even in the presence of 
finite $m$.  
As a result, the AWL correction at the critical 
point becomes precisely halved,  
on introducing the finite topological mass 
(quantum correction doubling).  

In terms of this novel behaviour of 
the parity diffusion mode and that of the 
corresponding AWL correction, we will argue  
in the section.~V the possible microscopic 
mechanism of how the delocalized bulk-wavefunction 
emerges at the critical point, i.e. $m=\mu=0$. 
To be more specific, we expect  
that 
the parity diffusion mode mentioned above generally  
becomes massless, when a system transits from 
the topological insulator side to the ordinary 
insulator side. Assuming that this is the case, 
we will attribute the emergence of the  
extended bulk wavefunction 
to the AWL correction obtained from 
this parity mode channel, i.e. the second 
term of eq.~(\ref{punch2-0}).   
For the systematic understanding, 
however, one generally needs to go beyond 
the theoretical approach employed in this 
paper. Several open issues will be 
also discussed in the section~V.  

A number of appendices describe other 
topics useful in understanding the main text in more detail.  
For clarity of the explanation, we have presented 
the results only in the case of 
chemical-potential type disorder in the text. 
The study in the presence of general ${\cal T}$ symmetric 
disorders becomes more cumbersome. But the basic feature  
such as the phase diagram is expected to 
be same. In the appendix~A, we 
will describe how the one-particle 
Green function at the zero-energy state 
behaves in the presence of these general 
${\cal T}$-invariant random potentials. 

Our weak-localization calculation described in the 
section.~IV is the controlled analysis, 
when it comes to the weakly disordered region, 
$\alpha<\alpha_c$. Namely, for this parameter region,  
one can confirm self-consistently the coupling constant 
$1/k_F l \equiv 1/\mu\tau$ to be sufficiently small 
around $\mu=0$ (see eq.~(\ref{control})).  
For $\alpha>\alpha_c$, however, this 
coupling constant generally diverges toward $\mu=0$, 
only to make the weak localization calculation 
(and scB calculation) an uncontrolled analysis. 
Thus, as the complementary analysis for this strongly 
disordered region, $\alpha>\alpha_c$, 
we employed the mode-mode coupling analysis in 
the appendices.~B-D. By taking into account the quantum 
interference effect due to the Cooperon terms,   
this theoretical framework gives us the 
gap equation for the diffusion constant. Main results  
in section.~IV such as the quantum correction doubling 
are also supported by the analysis in the appendices.~B-D. 

\section{Effective continuum model and disorder} 
\subsection{Effective continuum model}
We consider a system with both ${\cal T}$-symmetry and 
the spatial inversion (${\cal I}$)-symmetry.
Under this symmetry requirement, Murakami {\it et al.} recently derived 
the minimal model for an arbitrary 
quantum critical point intervening the topological insulator 
and an ordinary insulator on a quite general 
ground~\cite{Murakami07,Murakami08}.  
It turns out to be always described by   
the $3+1$ Dirac fermion given as follows; 
\begin{eqnarray}
{\cal H}_0 &\equiv& \int d^3r 
\psi^{\dagger}(r)
\Big\{\sum_{\mu=1}^3 \hat{\gamma}_{\mu}
\big(-i\partial_{\mu}\big) 
- m \hat{\gamma}_5\Big\}
\psi(r),  \label{e5-t} 
\end{eqnarray}
where $m$ corresponds to the topological mass term.  
Without loss of generality, one can regard the topological insulator 
phase to be $m>0$ and an ordinary insulator phase to be 
$m<0$ (see Fig.~\ref{t1}).  
To see that ``$m$'' actually endows 
this Dirac fermion with a mass, we note that following  
five $4\times 4$ $\gamma$-matrices are 
anticommuting with one another; 
\begin{eqnarray}
&&\hat{\gamma}_1 \equiv \hat{\sigma}_y \otimes 1, \hat{\gamma}_2 
\equiv \hat{\sigma}_z \otimes \hat{s}_x, \hat{\gamma}_3 \equiv \hat{\sigma}_z \otimes \hat{s}_y, \nn \\ 
&& \hat{\gamma}_4 \equiv \hat{\sigma}_z\otimes \hat{s}_z, 
\hat{\gamma}_5 \equiv  \hat{\sigma}_x \otimes 1. \nn 
\end{eqnarray}
The matrices
$\hat{\sigma}_{\mu}$ and $\hat{s}_{\mu}$ are
Pauli matrices, representing the (generalized) sublattice index, 
and the spin index, respectively.
 In terms 
of these Pauli matrices,  
we will take the ${\cal T}$ operator 
as $i\hat{s}_y K$ with $K$ being the  
complex conjugation. Meanwhile, the 
${\cal I}$ operator will be taken 
as $\hat{\sigma}_x$. 
It follows from these conventions 
that $\hat{\gamma}_{1,2,3,4}$ are ${\cal T}$ odd and ${\cal I}$ odd, 
while $\hat{\gamma}_5$ is ${\cal T}$ even and ${\cal I}$ even (see table.~I). 
Together with the property that $-i\partial_{\mu}$ is ${\cal T}$ odd 
and ${\cal I}$ odd, we can easily see that our Hamiltonian in the clean case 
is indeed ${\cal T}$ even and  ${\cal I}$ even. 
This guarantees the Kramers degeneracy 
at each $k$-point, irrespectively of the topological 
mass $m$. 
We also note that eq.~(\ref{e5-t}) is indeed  
the low-energy effective continuum Hamiltonian 
for various lattice model recently discussed in 
literatures~\cite{FKM,Teo,Hsieh}. 
\begin{table}[htbp]
\begin{center} 
\begin{tabular}{|c|c|c|}
\hline
Dirac matrices  & ${\cal T}$ &  ${\cal I}$ \\ \hline \hline 
$\hat{\gamma}_0 \equiv 1 \otimes 1$ & $+$ & $+$ \\ \hline
$\hat{\gamma}_1 \equiv \hat{\sigma}_y \otimes 1$ & $-$ & $-$ \\
$\hat{\gamma}_2 \equiv \hat{\sigma}_z \otimes \hat{s}_x$ & $-$ & $-$ \\  
$\hat{\gamma}_3 \equiv \hat{\sigma}_z \otimes \hat{s}_y$ & $-$ & $-$ \\
$\hat{\gamma}_4 \equiv \hat{\sigma}_z \otimes \hat{s}_z$ & $-$ & $-$ \\ \hline 
$\hat{\gamma}_5 \equiv \hat{\sigma}_x \otimes 1$ & $+$ & $+$ \\ \hline 
\end{tabular} 
\end{center}
\caption{Dirac operators and their symmetries. 
}
\end{table}

Generally speaking, we can 
enumerate all Hermite matrices possible 
in this spin-sublattice space. Namely,  
using the commutator between these five Dirac matrices, 
we have other $10\equiv {}_5C_2$ associated Dirac matrices;   
\begin{eqnarray}
\hat{\gamma}_{ij} \equiv \frac{1}{2i}\big[\hat{\gamma}_i,\hat{\gamma}_j\big] = -i \hat{\gamma}_i \hat{\gamma}_j. 
\end{eqnarray}  
We can further classify these 
$10$ matrices into two classes;   
one is ${\cal T}$  invariant (even) matrices and the other is 
${\cal T}$  odd. Since the 
five Dirac matrices are always even under ${\cal I}\cdot{\cal T}$,   
these 10 associated Dirac matrices are by construction   
odd under ${\cal I}\cdot{\cal T}$. 
Thus the 
symmetries of these 10 matrices can be  
summarized as in Table~II. 
 
\begin{table}[htbp]
\begin{center} 
\begin{tabular}{|c|c|c|}
\hline
{\bf 10} matrices  & ${\cal T}$ &  ${\cal I}$ \\ \hline \hline
$\hat{\gamma}_{15} \equiv - \hat{\sigma}_z \otimes 1$ & $+$ & $-$ \\  
$\hat{\gamma}_{25} \equiv \hat{\sigma}_y \otimes \hat{s}_x$ & $+$ & $-$ \\
$\hat{\gamma}_{35} \equiv \hat{\sigma}_y \otimes \hat{s}_y$ & $+$ & $-$ \\ 
$\hat{\gamma}_{45} \equiv \hat{\sigma}_y \otimes \hat{s}_z$ & $+$ & $-$ \\ \hline
$\hat{\gamma}_{12} \equiv \hat{\sigma}_x \otimes \hat{s}_x$ & $-$ & $+$ \\  
$\hat{\gamma}_{13} \equiv \hat{\sigma}_x \otimes \hat{s}_y$ & $-$ & $+$ \\
$\hat{\gamma}_{14} \equiv \hat{\sigma}_x \otimes \hat{s}_z$ & $-$ & $+$ \\ 
$\hat{\gamma}_{23} \equiv 1 \otimes \hat{s}_x$ & $-$ & $+$ \\
$\hat{\gamma}_{34} \equiv 1 \otimes \hat{s}_y$ & $-$ & $+$ \\  
$\hat{\gamma}_{42} \equiv 1 \otimes \hat{s}_z$ & $-$ & $+$ \\ \hline 
\end{tabular} 
\end{center}
\caption{Dirac associated operators and their symmetries}
\end{table} 

let us introduce ${\cal T}$-symmetric ``on-site type'' 
random potentials as generally as possible;  
\begin{eqnarray}
{\cal H}_{\rm imp} \equiv \int dr
\psi^{\dagger}(r)\Big\{v_0 \hat{\gamma}_0 + v_{5} \hat{\gamma}_5 + 
\sum_{j=1}^{4} v_{j5} \hat{\gamma}_{j5} \Big\}\psi(r), \label{imp}
\end{eqnarray}
where all the 6 components of the vector $\vec{v}(r)$ 
are real-valued functions of $r$. Then,  
each single-particle eigenstate of 
${\cal H}_0 + {\cal H}_{\rm imp}$ always 
has a Kramers pair state;
\begin{eqnarray}
\langle \tilde{\phi}(r)| \equiv 
\hat{1}\otimes (-i)\hat{s}_y 
|\phi(r)\rangle .  
\end{eqnarray}
Namely, $|\tilde{\phi}(r)\rangle$ and $|\phi(r)\rangle $ 
are degenerate and orthogonal to 
each other.
Noting this, one 
can see that the 
retarded (advanced) Green function observes 
the following relation in each ensemble; 
\begin{eqnarray}
\hat{G}^{R(A)}(r,r';\mu) &\equiv& 
{\sum_n} \frac{|\phi_n(r) 
\rangle \langle \phi_{n}(r')|}
{\mu - \epsilon_n \pm i\delta}  \nn \\
&&\hspace{-3.0cm} = 
\hat{1}\otimes 
\hat{s}_y
\cdot  
\big\{
\hat{G}^{R(A)}(r',r;\mu)
\big\}^t 
\cdot 
\hat{1}\otimes  
\hat{s}_y. 
\label{e7} 
\end{eqnarray}

\subsection{Disorder averages, spatial inversion symmetry, 
rotational symmetry and engineering dimension} 
As usual, we will take 
the quenched-average of these ${\cal T}$ invariant 
impurities at the gaussian level; 
\begin{eqnarray}
\overline{\cdots}  &\equiv&  
\frac{1}{\cal N}\int {\cal D}[v] e^{P[v]}  \cdots, \label{e7-1} \\  
P[v]&\equiv&  \sum_{j,m\in \{0,5,15,\cdots,45\}}
\int \int d^3r d^3r' \nn \\ 
&&\hspace{0.6cm} 
[\hat{\Delta}^{-1}]_{(r,j|r',m)} v_j(r) v_m (r'), \label{e7-1-1}  
\end{eqnarray} 
with a proper normalization factor ${\cal N}$ and real-valued symmetric 
matrix $\hat{\Delta}$. 
For simplicity, an ``on-site type'' correlation will be assumed; 
\begin{eqnarray}
\Delta(r,j|r',m)\equiv \Delta_{jm}\delta^3(r-r').  \label{e8-0}
\end{eqnarray}
We also suppose that the translation symmetry 
and the spatial inversion symmetry 
are recovered {\it after} these quenched averages;
\begin{eqnarray}
\hat{G}^{R(A)}(r,r';\mu) 
&\equiv& 
\hat{G}^{R(A)}(r+b,r'+b;\mu),  
\label{e8} \\ 
\hat{G}^{R(A)}(r,r';\mu) 
&\equiv&  \hat{\sigma}_x\otimes\hat{1} \cdot 
\hat{G}^{R(A)}(-r,-r';\mu)
\cdot 
\hat{\sigma}_x\otimes\hat{1}. \label{e9}
\end{eqnarray}      

Then, the latter symmetry, i.e. 
eq.~(\ref{e9}), prohibits   
any matrix elements between 
$\gamma_{0,5}$ and $\gamma_{j5} (j=1,\cdots,4)$  
in the right hand side of eq.~(\ref{e8-0}). 
Namely, the $6\times 6$ matrix $\hat{\Delta}$ in its 
right hand side takes the following form;  
\begin{eqnarray}
\hat{\Delta} \equiv \left[\begin{array}{ccc}
\Delta_{00}&\Delta_{05} & {\bf 0}\\
\Delta_{50}&\Delta_{55} & {\bf 0}\\
{\bf  0} & {\bf 0} & \hat{\Delta}_{a} \\ 
\end{array} 
\right], \label{e9-1}
\end{eqnarray}
with 
\begin{eqnarray}
\hat{\Delta}_{a} \equiv \left[\begin{array}{ccc}
\Delta_{1515} & \cdots & \Delta_{1545} \\    
\vdots & \ddots & \vdots \\
\Delta_{4515} & \cdots & \Delta_{4545} \\
\end{array}\right].   \label{e9-1-1} 
\end{eqnarray}    
This is because $\hat{\gamma}_{0,5}$ are even under 
${\cal I}$, while $\hat{\gamma}_{j5}\ (j=1,2,3,4)$ are odd. 
In order that the gaussian 
integral in eq.~(\ref{e7-1}) converges, 
all the eigenvalues of $\hat{\Delta}$ 
have to be positive. Accordingly, 
the matrix elements described in eqs.~(\ref{e9-1}-\ref{e9-1-1}) 
must obey the following inequalities;   
\begin{eqnarray}
&& \Delta_{00}\Delta_{55} > \Delta_{05}\Delta_{50} = \Delta^2_{05}, \nn \\  
&& \Delta_{00} + \Delta_{55} > 0,  \ \  
{\rm Tr} \hat{\Delta}_{a} > 0, \ \  \cdots \label{e9-2}
\end{eqnarray}

We can study the 
effects of these 
general ${\cal T}$-invariant ``on-site type'' 
random potentials, without  
any further assumptions. As will be partly shown in the 
Appendix~A, however, 
such an analysis 
becomes very cumbersome and lengthy. Thus, 
we henceforth consider only the chemical potential 
type disorder $\Delta_{00}$, because it is
expected to be dominant among various types of disorder.
Those who are interested in 
the effects of other components 
such as $\Delta_{05},\Delta_{55}$ and 
$\hat{\Delta}_a$ may consult the appendix~A.
In Appendix~A we have studied 
the effect of the 
${\cal T}$-reversal invariant 
``on-site type'' disorder on 
a general ground, focusing on the 
zero-energy wavefunction at the
critical point.  

Being translationally invariant as in eq.~(\ref{e8}), 
the averaged Green functions
can be readily fourier-transformed by the use of  
the crystal momentum $k$. The resulting Green functions can be 
expanded in terms of Dirac matrices and its associates; 
\begin{eqnarray}
\hat{G}^R(k,\mu)\equiv\sum_{j\in \{0,1,\cdot,5,15,\cdots,42\}} 
\overline{\sf F}_{j}(k,\mu)\hat{\gamma}_{j}. 
\end{eqnarray} 
$\overline{\sf F}_{i}(k,\mu)$ stands for 
some complex-valued function of $k$ and $\mu$. 
In this momentum representation,  
${\cal T}$ and ${\cal I}$ invariance, 
i.e. eq.~(\ref{e7}) and eq.(\ref{e9}), read as follows;  
\begin{eqnarray}
\hat{\sigma}_x\otimes \hat{1} 
\cdot \hat{G}^{R(A)}(k,\mu)\cdot\hat{\sigma}_x \otimes \hat{1}
&=& \hat{G}^{R(A)}(-k,\mu), \label{e9-3} \\  
\hat{1}\otimes 
\hat{s}_y\cdot\hat{G}^{R(A)}(k,\mu)\cdot 
\hat{1}\otimes\hat{s}_y 
&=& \big\{\hat{G}^{R(A)}\big\}^{t}(-k,\mu). \label{e9-4} 
\end{eqnarray}
These two symmetries require that 
$\overline{\sf F}_{i=1,\cdots,4}(k,\mu)$ are odd functions of $k$, 
$\overline{\sf F}_{0,5}$ are even functions 
of $k$, and also that  
$\overline{\sf F}_{ij}\equiv 0$ 
for $i\ne j$ and $i,j=1,\cdots,5$. Namely, the 
retarded and advanced Green 
functions are given only in terms of the anti-commuting 
Dirac matrices;   
\begin{eqnarray}
\hat{G}^{R}(k,\mu) &\equiv& \overline{\sf F}_0(k,\mu)\ \hat{1} 
+ \sum_{\mu=1}^5\overline{\sf F}_{\mu}(k,\mu)\ \hat{\gamma}_{\mu},  \label{e9-5-r} \\ 
\hat{G}^{A}(k,\mu) &\equiv& \overline{\sf F}^{*}_0(k,\mu)\ \hat{1} 
+ \sum_{\mu=1}^5\overline{\sf F}^{*}_{\mu}
(k,\mu)\ \hat{\gamma}_{\mu}.  \label{e9-5-a} 
\end{eqnarray}

In addition to the ${\cal T}$-symmetry and ${\cal I}$-symmetry, 
the {\it pseudo-spin rotational symmetry} is also 
recovered {\it after the quenched average}. 
This is because only 
the chemical-potential type disorder 
$\Delta_{00}$ is considered now. 
Specifically, the 1-point Green function after 
the quenched average respects the 
simultaneous rotations of the spatial coordinate 
{\it and} the pseudo-spin coordinate;
\begin{eqnarray}
\hspace{0.2cm} \hat{U}_{n,\phi}\cdot 
\hat{G}^{R(A)}(k,\mu)\cdot \hat{U}^{\dagger}_{n,\phi} 
&\equiv& \hat{G}^{R(A)}(R_{n,\phi}k,\mu), \label{a80} \\ 
\hat{U}_{n,\phi}&\equiv& e^{\frac{\phi}{4}
\epsilon_{\mu\nu\rho}n_{\mu}\hat{\gamma}_{\nu}\hat{\gamma}_{\rho}}. 
\nn
\end{eqnarray}
$\mu$, $\nu$ and $\rho$ above run over $1,2$ and $3$. $R_{n,\phi}$ in 
the right hand side stands for the spatial rotation around the 
vector $n$ by the angle $\phi$.  
When combined with eqs.~(\ref{e9-3},\ref{e9-4}), this 
rotational symmetry further restricts the form
of the Green functions. For example, 
the coefficient of $\hat{\gamma}_4$ should be an odd 
function of $k$ due to eq.~(\ref{e9-3}), while 
it should be an even function of $k$ because of 
eq.~(\ref{a80}). As such, Green functions cannot contain 
$\hat{\gamma}_4$-component, under these 
two symmetry requirements. 
Moreover, eq.~(\ref{a80}) by itself compels  
$\overline{\sf F}_{1,2,3}(k,\mu)$ to be transformed as a vector under 
the rotation in the $k$-space;
\begin{eqnarray}
\overline{\sf F}_{\mu}(k,\mu) \equiv c_1 k_{\mu} + c_3 k^2 k_{\mu} + \cdots, \label{vector} 
\end{eqnarray} 
with $\mu=1,2,3$. 

So far, we have imposed several generic symmetries such as 
${\cal T}$-symmetry and ${\cal I}$-symmetry on the Green 
function after the quenched averaged. As a result of this,  
the Green function is 
given only in terms of the Dirac matrices. 
Since these 5 Dirac matrices are all anticommuting with one another, 
the inverse of the Green function 
can be easily calculated, 
\begin{eqnarray}
&&\hspace{-0.1cm}
\hat{G}^{{R},-1}(k,\mu) \equiv 
{\sf F}_0(k,\mu)\ \hat{1} + \sum^{5}_{\nu=1} 
{\sf F}_{\nu}(k,\mu)\ \hat{\gamma}_{\nu},  \label{e9-6} \\
&&\hspace{-0.4cm}\overline{\sf F}_0 = 
\frac{{\sf F}_0}{{\sf F}^2_0  -  
\sum_{\mu=1}^{5} {\sf F}^2_{\mu}}, \ \  
\overline{\sf F}_{\nu} = - \frac{{\sf F}_\nu}{{\sf F}^2_0 - 
\sum_{\mu=1}^5{\sf F}^2_{\mu}}.  \label{e9-7}  
\end{eqnarray}
Correspondingly, the inverse of the 
bare Green function is given as follows; 
\begin{eqnarray}
\hat{G}^{{R},-1}_{0}(k,\mu) &=& 
(\mu + i\delta)\ \hat{1} - \sum_{\lambda = 1,2,3} k_{\lambda} 
\ \hat{\gamma}_{\lambda} + m \hat{\gamma}_5  \nn \\
&\equiv& \sum_{\lambda=0,\cdots,5} 
{\sf f}_{\lambda}\ \hat{\gamma}_{\lambda}. \label{e9-8} 
\end{eqnarray}

Based on these simplifications, 
we will derive in the next 
two sections the electronic property 
of the disordered single copy of $3+1$ 
Dirac fermion described by eq.~(\ref{e5-t}). 
Before finalizing this section,  
however, it would be appropriate to 
summarize the engineering dimension of the various quantities 
introduced in this section. Comparing the impurity hamiltonian 
with the pure hamiltonian, one can first see that 
\begin{eqnarray}
m, k_{\mu}, \mu, {\sf f}_{\mu}, {\sf F}_{\mu}, v_i   
\sim [L^{-1}], 
 \ \ \overline{\sf F}_{\mu}  
\sim [L], 
\label{e10}
\end{eqnarray}  
where $L$ denotes the dimension of a length.
Out of this, we can further figure out the engineering 
dimension of $\Delta_{jm}$;   
\begin{eqnarray}
\Delta_{jm} \sim  [L],  \label{e11}
\end{eqnarray}
by requiring $P[v]$ in eq.~(\ref{e7-1-1}) to be dimensionless.


\section{Self-consistent Born approximation}

The self-consistent Born (scB) approximation 
simply equates the right hand sides of the 
following two; 
\begin{eqnarray}
&&\hspace{-0.3cm}\hat{\Sigma}^{R}(k,\mu) \equiv \hat{G}^{R,-1}_0 - 
\hat{G}^{R,-1}, \nonumber \\
&&\hspace{0.1cm} \equiv ({\sf f}_0 -{\sf F}_0)\ \hat{1} + \sum_{\nu=1}^{5}
({\sf f}_{\nu}-{\sf F}_{\nu})\ 
\hat{\gamma}_{\nu} \label{e12}, \\ 
&&\hspace{-0.3cm}
\hat{\Sigma}^{R}(k,\mu) = 
\Delta_{00}  
\int d^3 k' 
\hat{G}^{R}(k',\mu) \nn \\
&&\hspace{0.1cm} 
=  \Delta_{00} 
\int_{0<|k|<\Lambda} d^3k' 
 \left\{
\overline{\sf F}_{0}(k',\mu) \hat{\gamma}_0 
+\overline{\sf F}_{5}(k',\mu)    
\hat{\gamma}_5\right\}.\ \label{e14}
\end{eqnarray} 
We have already omitted those terms proportional 
to $\overline{\sf F}_{1,2,3,4}(k',\mu)$ 
in the integrand of eq.~(\ref{e14}), since 
they are odd functions of $k'$.  
Comparing the coefficients of each $\gamma$ matrix  
in eq.~(\ref{e12}) and eq.~(\ref{e14}), we can   
make the closed coupled equation for ${\sf F}_0$  
and ${\sf F}_5$;  
\begin{eqnarray}
&&\Delta_{00} \int_{0<|k|<\Lambda} d^3k\ \frac{ {\sf F}_0 }
{{\sf F}^2_0 - {\sf F}^2_5 - k^2} 
= {\sf f}_0 -{\sf F}_0, \label{e15} \\
&& - \Delta_{00} \int_{0<|k|<\Lambda} d^3k\ 
\frac{  {\sf F}_5}{{\sf F}^2_0 - {\sf F}^2_{5}- k^2}
={\sf f}_5 - {\sf F}_5, \label{e16} 
\end{eqnarray}
by the use of eq.~(\ref{e9-7}). 
We have already used the following relations also;  
\begin{eqnarray} 
&& {\sf F}_{1,2,3} \equiv {\sf f}_{1,2,3}= -k_{1,2,3}, \ \ 
{\sf F}_4 \equiv {\sf f}_4 \equiv 0. \label{e17}
\end{eqnarray}  

These integral equations in eqs.~(\ref{e15}-\ref{e16})
clearly depend on the ultraviolet cut-off $\Lambda$. Thus, 
rescaling the momentum by this cut-off $\Lambda$, let us 
introduce the {\it dimensionless quantities}, instead of ${\sf F}_{\mu}$, 
${\sf f}_{\mu}$, and $\Delta_{00}$.  
Eqs.~(\ref{e10}-\ref{e11}) indicate that they should be rescaled 
in the following way;    
\begin{eqnarray}
&&{\sf F}_{0,5} \rightarrow F_{0,5}\equiv 
{\sf F}_{0,5}{\Lambda}^{-1}, \label{e19-1} \\ 
&& {\sf f}_{0,5} \rightarrow f_{0,5} \equiv  
{\sf f}_{0,5}{\Lambda}^{-1}, \label{e20}  \\
&& {\sf \Delta}_{00} 
\rightarrow \alpha \equiv 2\pi{\Delta}_{00}\Lambda. \
\label{e19} 
\end{eqnarray} 
The factor $2\pi$ in the definition of $\alpha$ is just for
later convenience. 
In terms of these dimensionless quantities, the above coupled 
non-linear equations become;  
\begin{eqnarray}
(1+ \alpha G)F_0 &=& f_0\equiv\mu\pm i\delta,  \label{e26} \\
(1- \alpha G)F_5 &=& f_5\equiv m, \label{e27}   
\end{eqnarray}  
where $\mu$ and $m$ in the right hand side are supposed be also normalized by 
$\Lambda^{-1}$. $G$ used in the left hand side was   
also made dimensionless; 
\begin{eqnarray} 
&&\hspace{-0.4cm} 
G \equiv 
2\int_{0<k<1}\frac{1}{(a+ib)^{2} - k^2}k^2dk, \label{e22-1} \\ 
&& \hspace{-0.4cm} 
(a+ib)^2\equiv F^2_0-F^2_5. \label{e28}  
\end{eqnarray}
Eqs.~(\ref{e26}-\ref{e28}) thus 
determine $F_0$ and $F_5$ as a function of their bare values; $f_0$ 
and $f_5$. $F_{\mu}$ thus 
obtained should be by definition 
much smaller than the ``ultraviolet cut-off'' $1$;
\begin{eqnarray}
F_{\mu} \ll 1. \label{small}
\end{eqnarray}  
This also leads to $a,b \ll 1$.  
In the followings, we will frequently take full advantage 
of their smallness, which is always self-consistently 
verified later (see below).    

In the next subsection, we will present the solution 
of this coupled equation for general $\mu$ and $m$. 
Before doing this, however, it would be appropriate to 
express the imaginary part 
and real part of $G$ in terms of $a$ and $b$, 
so that one can roughly estimate these 
two quantities in small $a$ and $b$. The real part 
and the imaginary part  of $G$ read as follows;  
\begin{widetext}
\begin{eqnarray}
{\rm Re} G &\equiv& -2 -\frac{a}{2}\log\bigg[\frac{(1-a)^2+b^2}{(1+a)^2+b^2}
\bigg] + b\bigg({\rm ArcTan}\Big[\frac{1-a}{b}\Big] + {\rm ArcTan}\Big[\frac{1+a}{b}\Big]\bigg) 
,\label{e24} \\
{\rm Im} G &\equiv& -\frac{b}{2}\log\bigg[\frac{(1-a)^2+b^2}{(1+a)^2+b^2}\bigg]
-a\bigg({\rm ArcTan}\Big[\frac{1-a}{b}\Big] + {\rm ArcTan}\Big[\frac{1+a}{b}\Big]\bigg). 
\label{e25} 
\end{eqnarray}
\end{widetext}
Observing these two, please notice that 
the final two terms in eq.~(\ref{e24}-\ref{e25}),  
which are proportional to ${\rm ArcTan}$, are 
nothing but the pole contribution. Namely,     
the limit $b\rightarrow 0$ reduces them a 
finite constant with its sign identical to 
that of $b$, e.g. 
\begin{eqnarray}
{\rm ArcTan}\Big(\frac{1-a}{b}\Big) + 
{\rm ArcTan}\Big(\frac{1+a}{b}\Big)  \rightarrow \pi 
{\rm sgn}(b), \nonumber
\end{eqnarray}
where one should also note that $a, b \ll 1$. 
Bearing these in mind, one can then evaluate the 
leading order of ${\rm Re}G$ and ${\rm Im}G$ with 
respect to small $a$ and $b$;     
\begin{eqnarray}
{\rm Re}G &=& -2 + \pi |b| + {\cal O}(a^2), \label{e25-1} \\ 
{\rm Im}G &=& - {\rm sgn}(b) \pi a + {\cal O}(ab). \label{e25-2} 
\end{eqnarray} 
Namely, the second member of eq.~(\ref{e25-1}) and the first member 
of eq.~(\ref{e25-2}) are nothing but the pole contributions 
mentioned above.

\subsection{Solution for scB equations}
\subsubsection{$m=\mu=0$ case}
For the warming-up, consider first the case with $\mu=m=0$, i.e. 
the zero-energy state at the critical point. Equations (\ref{e26}-\ref{e28}) 
have three types of solutions;
\begin{eqnarray}
&&\hspace{-0.5cm}{\rm (i)}: \ F_0=F_5=0,  \label{ty1} \\ 
&&\hspace{-0.5cm}{\rm (ii)}: \ 1 + \alpha G = 0  \  \cap \  F_5 = 0, \label{ty2} \\
&&\hspace{-0.5cm}{\rm (iii)}: \ 1-\alpha G =0 \ \cap \ F_0 =0. \label{ty3} 
\end{eqnarray}
Observing the estimates given in eqs.~(\ref{e25-1}-\ref{e25-2}), 
please notice that type-(iii) solution cannot be satisfied for  
$\alpha>0$ and $a,b \ll 1$. Thus, we will ignore this henceforth. 

The type-(i) 
solution is always trivially satisfied. 
This solution indicates that the zero-energy state 
is not renormalized at all by the 
disorder, $F_0 = f_0 =0, F_5=f_5=0$. 
Thus, it describes the 
{\it diffusionless} zero-energy state.   

\begin{figure}
\begin{center}
\includegraphics[width=0.45\textwidth]{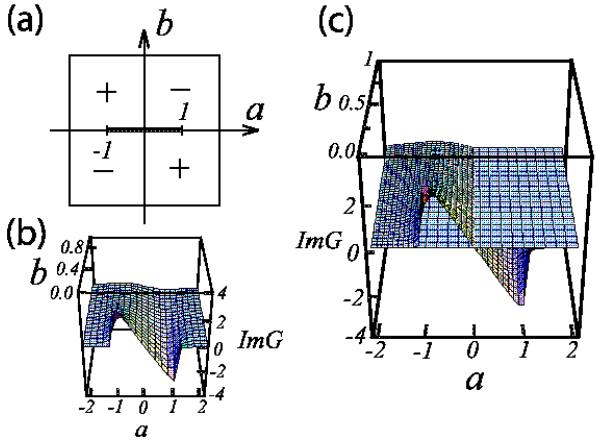}
\end{center}
\caption{${\rm Im}G$ as a function of $a$ and $b$. (a)  
The sign of ${\rm Im}G$, which is an odd 
function both in $a$ and in $b$. ``$+(-)$'' stands for  
the sign of ${\rm Im}G$ at the 4 regions, i.e. 
$a,b>0$, $a>0>b$, $b>0>a$ and $0>a,b$. 
The bold line which runs from $(-1,0)$ 
to $(1,0)$ denotes a sort of the branch cut. 
Namely, ${\rm Im}G$ jumps 
from $-2\pi^2 a$ to $+2\pi^2 a$ 
(from $b=+0$ to $b=-0$). (b) 
A side view plot of ${\rm Im}G$ 
only for $b>0$. (c) ${\rm Im}G \equiv 0$ is satisfied   
either when $a=0$, or when $b=0$ and $|a|>1$.}
\label{t5}
\end{figure}

The type-(ii) solution is a non-trivial 
solution, which turns out to describe 
the {\it diffusive} zero-energy state. To see this, 
let us begin with the first 
condition of eq.~(\ref{ty2}), i.e. $1+\alpha G = 0$. 
The imaginary part of this gives ${\rm Im}G=0$, 
which is satisfied either when $a=0$ or 
when $b=0$ and $|a|>1$ (see Fig.~\ref{t5}). 
Since 
$a,b\ll 1$ as noted earlier,
the only physical solution  
satisfying ${\rm Im}G=0$ is thus $a=0$. The remaining condition, 
$1+\alpha {\rm Re}G=0$, becomes then simple;
\begin{eqnarray}
b{\rm ArcTan}\big[b^{-1}\big] = 1- 
\frac{1}{2\alpha}.  \label{e29} 
\end{eqnarray}       
Since $F_5=0$ gives $a+ib=F_0$, $a$ and $b$ thus 
obtained stand for  the renormalized chemical 
potential $\overline{\mu}$   
and the inverse of the lifetime $\tau^{-1}$, respectively.   
Accordingly, the type-(ii) solution simply denotes that 
the zero-energy state acquires a finite lifetime $\tau$,  
while its chemical potential is free from renormalizations; 
\begin{eqnarray}
\overline{\mu} = 0, \ \ 
\tau^{-1} {\rm ArcTan}\big[\tau\big] = 1- \frac{\alpha_c}{\alpha}. 
\label{e30}
\end{eqnarray}
Namely, for $\alpha > \alpha_c\equiv \frac{1}{2}$, 
$\tau^{-1}$ can take a finite value. 

For a weak disorder region ($\alpha<\alpha_c$),  
Eq.~(\ref{e30}) 
cannot be satisfied for any $\tau$.  
Thus, the only solution therein is the type-(i)  
trivial solution.  
On the other hand, 
both the type-(i) 
solution 
and type-(ii) 
solution  
become possible, above this critical 
disorder strength ($\alpha>\alpha_c$). 
In the next three paragraphs, we will 
determine which solution is {\it physically sensible} 
for $\alpha>\alpha_c$.

To do this, 
we will extend these two solutions into a 
small but finite $\mu$ region.  Namely, by seeing 
how this chemical potential will be 
renormalized for each case, we will judge which solution 
is the physical one for $\alpha>\alpha_c$.  
Recall first that $F_5=0$ in either case.  
Thus, $a$ and $b$ correspond to 
$\overline{\mu}$ and $\tau^{-1}$ respectively,  
so that $a$ and $b$ should be an odd and an even 
function of the bare chemical potential 
$\mu$ respectively. 

Bearing these in mind, let us 
extend the type-(i) solution into a 
small 
$\mu$-region first. 
Namely, keeping the leading order 
in small $\mu$, we can evaluate 
the real part of eq.~(\ref{e26}); 
\begin{eqnarray}
(1-2\alpha) a + {\cal O}(\mu^3) = \mu, 
\label{e31} 
\end{eqnarray}  
where we used 
$a \propto {\cal O}(\mu)$ and
 $b \propto {\cal O}(\mu^2)$.  
Thus, the renormalized chemical 
potential is estimated up to ${\cal O}(\mu)$ as follows;   
\begin{eqnarray}
\overline{\mu} \equiv a = \frac{\mu}{1-2\alpha} + {\cal O}(\mu^3), \label{e31-1}
\end{eqnarray}
while $b$ will be determined up to $(\cal O)(\mu^2)$ from the imaginary 
part of eq.~(\ref{e26});
\begin{eqnarray}
\tau^{-1} \equiv 
b = \frac{\alpha\pi}{(1-2\alpha)^3}\mu^2 + {\cal O}(\mu^4). \label{e31-1-i}
\end{eqnarray}

This solution indicates that the negative 
eigen-energy state and the positive eigen-energy 
state are {\it inverted} energetically 
for $\alpha>\alpha_c$; 
${\rm sign} \overline{\mu} = - {\rm sign} \mu$.    
This is, however, clearly unphysical at least for small $\mu$.  

When the type-(ii) solution is 
extended into a small $\mu$-region, 
a similar algebra gives us the following 
expression for the real part of 
eq.~(\ref{e26}) up to ${\cal O}(\mu)$; 
\begin{eqnarray}
&&\hspace{-0.5cm} (1-2\alpha) a + 4\alpha  \tau^{-1} {\rm ArcTan}\big[\tau\big] 
a = \mu.  \label{e31-2}
\end{eqnarray}  
In this equation, we have already 
made implicit those contributions 
proportional to ${\cal O}(\mu^3)$ and ${\cal O}(\tau^{-2} \mu)$ 
while keeping those proportional to 
${\cal O}(\tau^{-1} \mu)$ explicit.  
Please also note that we have used 
$a \propto {\cal O}(\mu)$ and $b = \tau^{-1} + {\cal O}(\mu^2)$. 
Namely,  contrary to the type-(i) solution,  
$b$ converges to a {\it non-zero} $\tau^{-1}$ at the leading order 
in small $\mu$. As a result of this, 
eq.~(\ref{e31-2}) has acquired an additional ${\cal O}(\mu)$-term, 
i.e. $4\alpha \tau^{-1} 
{\rm ArcTan}^{-1}[\tau]\cdot a$, which was absent in   
eq.~(\ref{e31}). This additional term makes the sign of 
$\overline{\mu}$ to be same as that of $\mu$. Namely, 
by the use of $1-2\alpha = - 2\alpha \tau^{-1} 
{\rm ArcTan}^{-1}\big[\tau\big] + {\cal O}(\mu^2)$, 
eq.~(\ref{e31-2}) leads us to;  
\begin{eqnarray}
(2\alpha-1) a + {\cal O}(\mu^3) = \mu.
\end{eqnarray}   
Out of this, one can evaluate the renormalized chemical 
potential up to ${\cal O}(\mu)$ as follows; 
\begin{eqnarray}
\overline{\mu} \equiv a = \frac{\mu}{2\alpha-1} + {\cal O}(\mu^3), \label{e31-3}
\end{eqnarray}
whose sign is clearly same as that of the bare one for 
$\alpha>\alpha_c$;  ${\rm sgn}\overline{\mu}={\rm sgn}\mu$. 
Observing these two distinct behaviors for 
the finite $\mu$ region, i.e.  eq.~(\ref{e31-1}) and 
eq.~(\ref{e31-3}), we conclude that, for $\alpha>\alpha_c$, 
the type-(ii) solution is the physically sensible 
solution, while the type-(i) solution is an unphysical one.

To summarize so far, the physical solutions 
obtained at $m=\mu=0$ 
are two-fold, depending on the disorder strength 
$\alpha$. 
When $\alpha<\alpha_c=1/2$, the type-(i) trivial  
solution is the only possible solution;  
\begin{equation}
 {\rm (i)}:\  
 F_0 = F_5 = 0 \ \ \ {\rm for} \ \ \  \alpha < \alpha_c. \label{e31-4} 
\end{equation}
This means that the electronic 
state at the zero-energy is free from the disorder up 
to a certain critical disorder strength.

On the other hand, when its strength exceeds 
this critical value, i.e. $\alpha > \alpha_c$, 
the type-(ii) solution should be adopted; 
\begin{equation}
 {\rm (ii)}: \  
 F_0 = i\tau^{-1}, F_5=0  \ \ \ {\rm for} \ 
\ \  \alpha > \alpha_c. \label{e31-5} 
\end{equation}  
This solution means that the electronic state at 
the zero energy acquires a finite lifetime 
$\tau$ defined by eq.~(\ref{e30}). 

\subsubsection{$\mu=0$ and finite $m$ case}
\begin{figure}
\begin{center}
\includegraphics[width=0.42\textwidth]{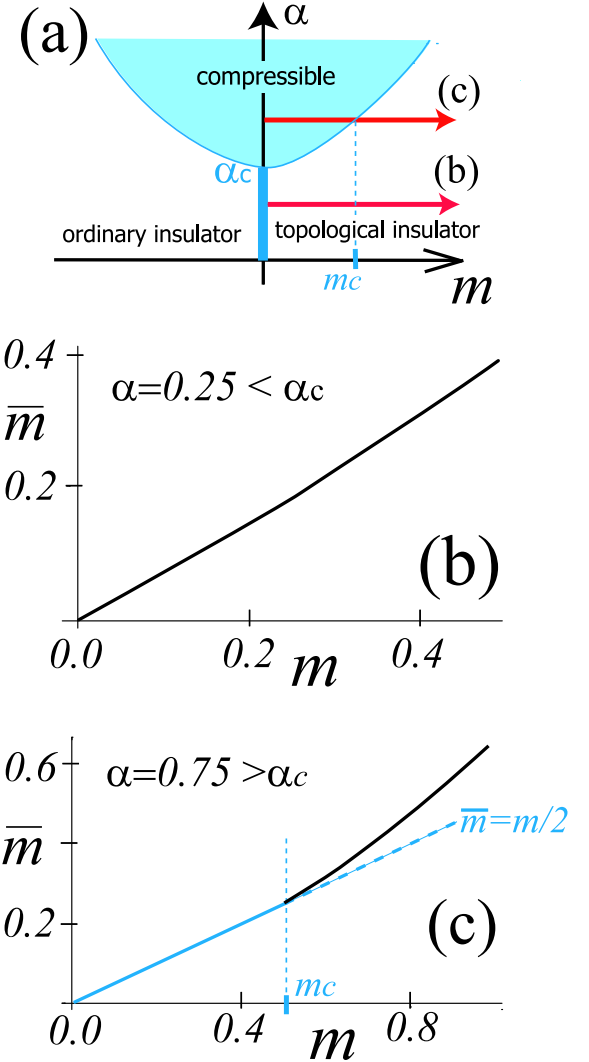}
\end{center}
\caption{(a); A schematic phase diagram at $\mu=0$. 
The white region corresponds to 
the incompressible phase, where no finite DOS exists at $\mu=0$. 
The blue region corresponds to the compressible 
states, where a finite DOS exists at $\mu=0$, i.e. eq.~(\ref{e33}). 
(b) The renormalized mass $\overline{m}$ as a 
function of the bare mass $m$ for $\alpha=0.25<\alpha_c$. 
(c) $\overline{m}$ as a function of $m$ for $\alpha=0.75>\alpha_c$. 
There exists the critical value of the bare mass $m$, below which 
$\overline{m}=m/2$, and above which $\overline{m}$ is determined by 
eq.~(\ref{e32}). These two values coincide with each other precisely 
at $m=m_{c}$.}
\label{t6}
\end{figure}

Let us introduce a finite topological mass $m$ into 
eq.~(\ref{e31-4}) and eq.~(\ref{e31-5}) respectively, with  
the chemical potential $\mu$ being still zero.  
We will first argue that 
the solution of eqs.~(\ref{e26}-\ref{e28}) 
in the presence of the finite mass  
is uniquely determined 
for $\alpha<\alpha_c$. Such a solution reads;  
\begin{eqnarray}
F_0 = 0, \ \ F_5 = \overline{m}, \label{e32+1}
\end{eqnarray} 
where $\overline{m}$ is given as a function of the bare mass;
\begin{eqnarray}
\overline{m} \big\{ 1+2 \alpha -2 \alpha \overline{m}
{\rm ArcTan}\big[\overline{m}^{-1}\big]\big\} = m. \label{e32} 
\end{eqnarray}
A typical behavior of $\overline{m}$ as a function 
of the bare mass is depicted in Fig.~\ref{t6}(b). 

To see that eqs.~(\ref{e32+1}-\ref{e32}) is the only possible 
solution for $\alpha<\alpha_c$, let us begin with the real 
part of $1+\alpha G$ appearing  
in eq.~(\ref{e26}). In the case of $\alpha<\alpha_c$, 
it is always positive definite for any $a<1$. 
As such, we must take $F_0$ to be zero, to satisfy 
eq.~(\ref{e26}). This leads to $F_5=b-ia$. Using this, 
consider next the imaginary part of eq.~(\ref{e27});
\begin{eqnarray}
\big(1-\alpha {\rm Re}G\big)\cdot a + \alpha {\rm Im} G \cdot b = 0. \label{h1} 
\end{eqnarray} 
Observing the leading-order estimates of ${\rm Re}G$ and ${\rm Im}G$, i.e. 
eqs.~(\ref{e25-1}-\ref{e25-2}), one can further see that eq.~(\ref{h1}) uniquely 
leads to $a=0$. The remaining condition, 
i.e. the real part of eq.~(\ref{e27}), then becomes simple; 
\begin{eqnarray}
(1-\alpha {\rm Re}G)\cdot b = 
\big(1+2\alpha - 2\alpha b{\rm ArcTan}\big[b^{-1}\big]\big)
\cdot b = m. \nn 
\end{eqnarray}   
Now that $(F_0,F_5)\equiv (0,b)$, this equation is nothing but 
eq.~(\ref{e32}), when  $b$ replaced by $\overline{m}$.    

Let us next consider the case of $\alpha>\alpha_c$. 
The solution of eqs.~(\ref{e26}-\ref{e28}) 
in this case is two fold; we have a certain 
critical mass value $m_{c}$, which 
is given as a function of $\alpha$; 
\begin{eqnarray}
m_{c} \equiv 2\tau^{-1}, \ \ \tau^{-1} {\rm ArcTan}\big[\tau\big] 
\equiv 1 -\frac{\alpha_c}{\alpha}. \label{e33+1} 
\end{eqnarray}
When the topological mass is less than 
this critical value ($m<m_c$), 
the solution of eqs.~(\ref{e26}-\ref{e28}) 
becomes;
\begin{eqnarray}
F_0 = + i\sqrt{\tau^{-2} - {\overline{m}}^2}, 
\ F_5 = \overline{m} \equiv \frac{m}{2}, \label{e33}
\end{eqnarray} 
where $\tau$ was already defined in eq.~(\ref{e33+1}). 
On the other hand, when the topological 
mass exceeds this critical value ($m>m_c$), the 
solution becomes eqs.~(\ref{e32+1}-\ref{e32}) again. 

To see that eq.~(\ref{e33}) is the solution of eqs.~(\ref{e26}-\ref{e28}) 
for $\alpha>\alpha_c$ and $m<m_c$, 
take $1+\alpha G\equiv 0$ first, so that eq.~(\ref{e26}) is satisfied. 
By the use of the same arguments described above eq.~(\ref{e29}), this  
immediately gives us $(a,b) \equiv (0,\tau^{-1})$, with 
$\tau$ being defined by 
eq.~(\ref{e30}). Since $1-\alpha G\equiv 2$, 
eq.~(\ref{e27}) leads to $F_5\equiv m/2$. Thus, using these two 
things, we obtain $F_0$ out of eq.~(\ref{e28}), which is 
nothing but eq.~(\ref{e33}). 
When $m$ exceeds $m_{c} \equiv 2\tau^{-1}$, 
eq.~(\ref{e33}) becomes an unphysical solution 
in a similar way as the type-(i) solution 
in the previous subsection did for 
$\alpha>\alpha_c$;   
\begin{eqnarray}
F_0 = \pm \sqrt{\overline{m}^2-\tau^{-2}}, 
F_5 = \overline{m} \equiv \frac{m}{2}. \nn  
\end{eqnarray}
Instead of this, it turns out that 
we should adopt the other solution
for $m>m_c$, i.e. eqs.~(\ref{e32+1}-\ref{e32}). 
 
A typical behavior of $\overline{m}$ in the case of $\alpha>\alpha_c$
is depicted in Fig.~\ref{t6}(c), where these two 
solutions, i.e. eq.~(\ref{e33}) and eqs.~(\ref{e32+1}-\ref{e32}),  
are indeed connected continuously at $m=m_c$.   
Since eq.~(\ref{e33}) always 
supports a finite density of state (DOS), we can regard 
that the compressible phase extends over $\alpha>\alpha_c$ and 
$m<m_{c}$. On the other hand,  
eqs.~(\ref{e32+1}-\ref{e32}) do not 
support any finite DOS. As such, we can regard 
that an incompressible phase extends over 
$\alpha<\alpha_c$ or  $m>m_{c}$  
(Fig.~\ref{t6}(a)).

\subsubsection{General $\mu$ and $m$ case} 
For finite $\mu$ and $m$, both $F_0$ and $F_5$ 
are in general nonzero and we cannot solve 
eqs.~(\ref{e26}-\ref{e28}) analytically. 
Accordingly, we have numerically 
solved the coupled equations with 
respect to $a$ and $b$, so that 
$F_0$ and $F_5$ are derived in terms of 
$\mu$ and $m$. 

\begin{figure}
\begin{center}
\includegraphics[width=0.4\textwidth]{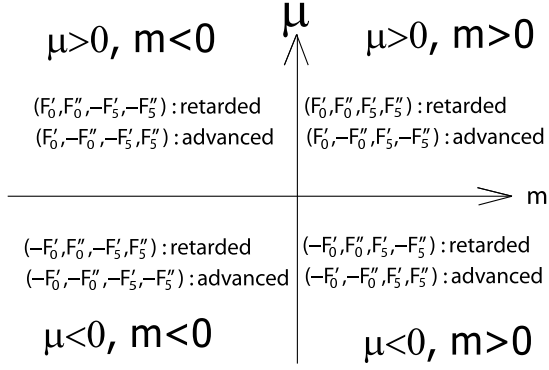}
\end{center}
\caption{$F_0$ and $F_5$ as a function of $m$ and $\mu$. $F^{\prime}_j$ and 
$F^{\prime\prime}_j$ are the real and imaginary part of a function $F_j$. 
At any parameter point, we generally have at least two solutions, 
which correspond to the retarded Green function and advanced one.}
\label{t7}
\end{figure}

Before describing the 
numerical solution, let us first argue 
about the generic features of such solutions.  
Notice first that ${\rm Re}G$ is an even function of 
both $a$ and $b$, while ${\rm Im}G$ is an odd function of  
both $a$ and $b$. 
Thus the following two should be degenerate  
at any given $\mu$ and $m$ as the 
solutions of eqs.~(\ref{e26}-\ref{e28});  
\begin{eqnarray}
(F^{\prime}_0,F^{\prime\prime}_0,F^{\prime}_5,F^{\prime\prime}_5), \ \ 
(F^{\prime}_0,-F^{\prime\prime}_0,F^{\prime}_5,-F^{\prime\prime}_5), 
\end{eqnarray}
where $F^{\prime}_{j}$ and $F^{\prime\prime}_j$ are the real and 
imaginary part of $F_j$. Namely, 
these two solutions correspond to 
the retarded Green function and advanced one respectively.

The above two solutions 
at given $m$ and $\mu$ can be further extended into the other 3 quadrants, 
i.e. $(-m,\mu)$, $(m,-\mu)$ and $(-m,-\mu)$;  
\begin{eqnarray}
&&\hspace{-1cm}({F_0}',\pm{F_0}'',{F_5}',\pm{F_5}'')_{|m,\mu} \nn \\
&=& ({F_0}',\pm{F_0}'',-{F_5}',\mp{F_5}'')_{|-m,\mu} \nn \\ 
&=& (-{F_0}',\pm{F_0}'',{F_5}',\mp{F_5}'')_{|m,-\mu} \nn \\
&=&  (-{F_0}',\pm{F_0}'',-{F_5}',\pm{F_5}'')_{|-m,-\mu}, \nn 
\end{eqnarray}   
where the upper sign corresponds to the retarded function 
for any of these four regions by construction (Fig.~\ref{t7}).
Observing this, 
please notice that both $F^{\prime}_5$ and 
$F^{\prime\prime}_5$ vanish 
at $m=0$, which is indeed the case with 
Sec.III-A1.  
Similarly, one can also see that  
$F^{\prime}_0$ and $F^{\prime\prime}_5$ should 
vanish at $\mu=0$ for any $m$. 
Both  eqs.~(\ref{e32+1}-\ref{e32}) and  
eq.~(\ref{e33}) actually observe this.

\begin{figure}
\begin{center}
\includegraphics[width=0.42\textwidth]{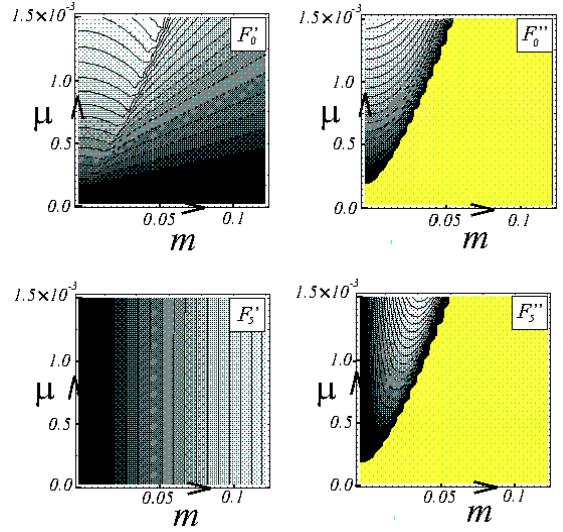}
\end{center}
\caption{$\alpha = 0.48$ ;   
(a): The contour plot of $F^{\prime}_0$ as a 
function of $\mu>0$ and $m>0$. The value of $F^{\prime}_0$ 
decreases toward the dark region, and becomes zero at $\mu=0$. 
The contour interval is $1.2 \times 10^{-3}$.    
(b): The contour plot of $F^{\prime\prime}_0$. The value of 
$F^{\prime\prime}_0$ decreases toward the darker region, and becomes
 zero at the yellow region. 
The contour interval is  $0.6 \times 10^{-3}$. 
(c): The contour plot  of $F^{\prime}_5$. 
$F^{\prime}_5$ decreases toward the darker region, and
becomes zero at $m=0$. The contour interval is $4.0\times 10^{-3}$.   
(d): The contour plot of $F^{\prime\prime}_5$. 
$F^{\prime\prime}_5$ increases toward the darker region 
and becomes zero at $\mu=0$, $m=0$ and the yellow region. 
The contour interval is  $-1.2\times 10^{-5}$.}
\label{t10}
\end{figure}
	
\begin{figure}
\begin{center}
\includegraphics[width=0.42\textwidth]{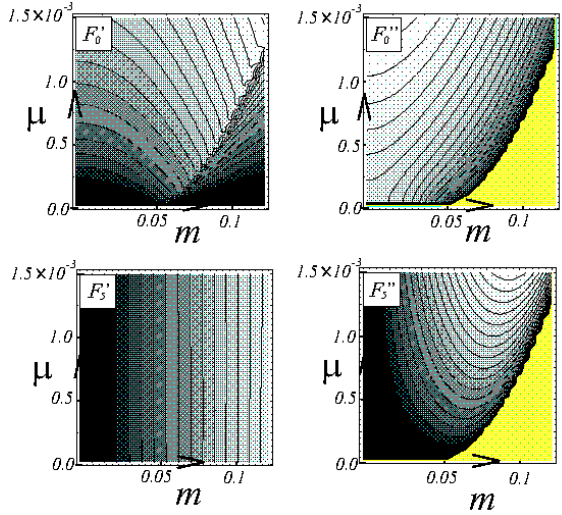}
\end{center}
\caption{$\alpha = 0.52$ ;  
(a): The contour plot of $F^{\prime}_0$ as a 
function of $\mu>0$ and $m>0$. The value of $F^{\prime}_0$ 
decreases toward the dark region, and becomes zero at $\mu=0$. 
The contour interval is  $1.8 \times 10^{-3}$.    
(b): The contour plot of $F^{\prime\prime}_0$. The value of 
$F^{\prime\prime}_0$ decreases toward the darker region, and becomes
 zero at the yellow region. 
The contour interval is  $1.8 \times 10^{-3}$. 
(c): The contour plot  of $F^{\prime}_5$. 
$F^{\prime}_5$ decreases toward the darker region, and becomes zero at $m=0$. The contour interval is $4.0\times 10^{-3}$.   
(d): The contour plot of $F^{\prime\prime}_5$. 
$F^{\prime\prime}_5$ increases toward the darker region 
and becomes zero at $\mu=0$, $m=0$ and the yellow region. 
The contour interval is  $-3.0\times 10^{-5}$.}
\label{t11}
\end{figure}

These considerations are also consistent with 
the numerical solution. 
In Fig.~\ref{t10} and \ref{t11}, we demonstrated numerically 
how $F_0$ and $F_5$ behave as a function 
of $\mu$ and $m$ (only for the first quadrant, $m>0$ and $\mu>0$), 
at specific values of $\alpha$. Fig.~\ref{t10} 
is for $\alpha<\alpha_c$, while  
Fig.~\ref{t11} is for $\alpha > \alpha_c$.  
In the limit of $\mu\rightarrow 0$, 
$F_0$ and $F_5$ in these two figures 
continuously connects with 
the two analytic solutions previously derived, i.e. eq.~(\ref{e32+1}-\ref{e32}) 
and eq.~(\ref{e33}) respectively.  

We have also checked that, whenever 
$F^{\prime\prime}_0=F^{\prime\prime}_5\equiv 0$, 
$F^{\prime}_5$ is always greater than 
$F^{\prime}_0$, i.e. $a^2-b^2<0$.     
As such, the spectral function is identically zero, 
provided that both 
$F^{\prime\prime}_0$ and $F^{\prime\prime}_5$ 
vanish. Such a phase should be regarded as 
an incompressible phase having no bound states.
 On the one hand, when either 
$F^{\prime\prime}_0$ or $F^{\prime\prime}_5$ 
is finite, the spectral weight 
is finite and such a phase is compressible.      

\begin{figure}
\begin{center}
\includegraphics[width=0.4\textwidth]{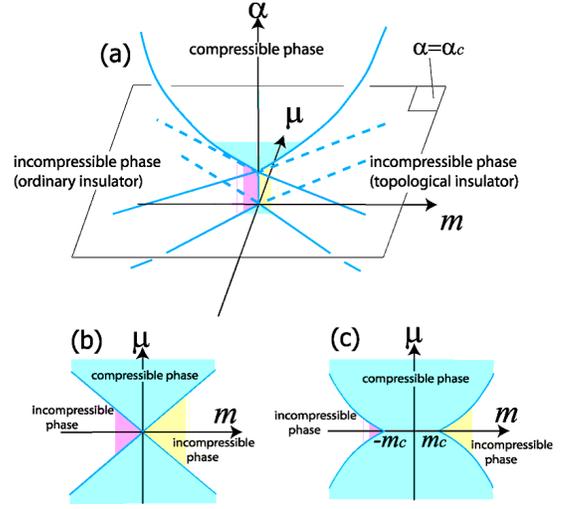}
\end{center}
\caption{(a)
A schematic phase diagram in the $\mu$-$m$-$\alpha$ space. 
Either $F^{\prime\prime}_0$ or $F^{\prime\prime}_5$ always remains 
finite in the compressible phase (blue), 
while both of them become zero at the remaining parameter region 
(incompressible phase), which is further divided into 
an ordinary insulator (red) and the topological insulator (yellow). 
(b) A schematic phase diagram in $\mu$-$m$ 
plane for $\alpha<\alpha_c$, and (c) that for $\alpha>\alpha_c$. 
These correspond to the numerical results shown 
in Fig.~\ref{t9}.}
\label{t8}
\end{figure}

\begin{figure}
\begin{center}
\includegraphics[width=0.45\textwidth]{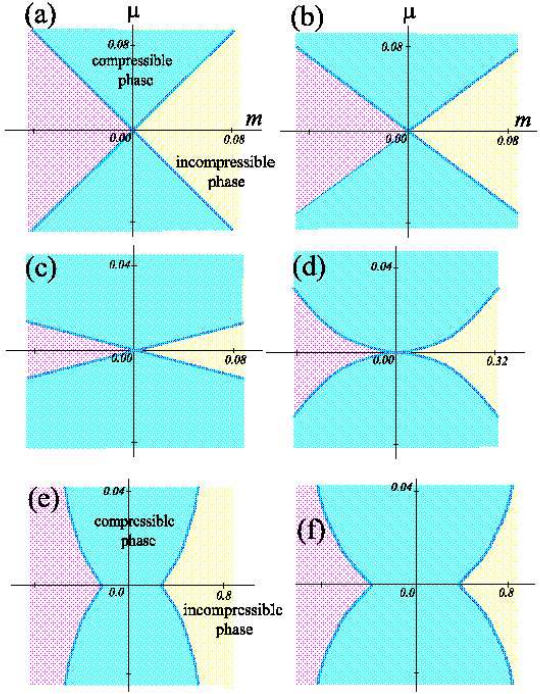}
\end{center}
\caption{Phase boundaries   
between the compressible phase and 
the incompressible (gapped) phase, in the $\mu-m$ plane, 
at several values of $\alpha$. 
(a) $\alpha =0.0$,(b) $\alpha=0.1$, (c) $\alpha =0.4$ 
(d) $\alpha = \alpha_c = 0.5$, (e) $\alpha=0.6$ 
(f) $\alpha = 0.7$.}
\label{t9}
\end{figure}

By seeing whether $F^{\prime\prime}_0$ and 
$F^{\prime\prime}_5$ totally vanishes or not, 
we have determined the phase diagram in the 
$\mu$-$m$-$\alpha$ space.  
The phase boundaries between the 
compressible phase and the incompressible phase 
thus obtained are schematically drawn 
in Fig~\ref{t8}, while accurately   
specified in Fig.~\ref{t9}. 
For $\alpha>\alpha_c$, we have a finite critical mass 
value, i.e. $m_c$, below which a compressible phase extends 
even at $\mu=0$ (Fig. \ref{t8}(c) and Fig. \ref{t9}(e,f)). 
This critical value goes to zero, when $\alpha$ goes to $\alpha_c$ 
from above (Fig. \ref{t9}(d)). For $\alpha < \alpha_c$, we  
have a compressible region not in 
the $\mu=0$ region anymore, but still 
in the nonzero $\mu$ region 
(Fig.~\ref{t8}(b) and Fig.~\ref{t9}(a-c)). 
The slope of the phase boundary in 
$\alpha<\alpha_c$ given as follows;
\begin{eqnarray}
{\frac{d\mu_c}{dm_c}}_{|\mu_c=m_c=0} 
\equiv \frac{1-2\alpha}{1+2\alpha}, \nn  
\end{eqnarray}  
increases when the disorder strength 
decreases (Fig~\ref{t9}(a-c)).

\section{Diffuson and Quantum Conductivity Correction} 
In the previous section, we have derived the 1-point 
Green function within the self-consistent Born approximation.  
In the 3-$d$ parameter space spanned by $\mu$, $m$ and $\alpha$, 
we have observed that 
the topological insulator and an ordinary insulator  
are always intervened by the compressible 
phase (see the blue region in Fig.~\ref{t8}).  
The topological insulator supports a 
single $2+1$ surface massless Dirac fermion  
on each boundary, while an ordinary insulator  
does not. As such, we expect that this intervening  
phase is composed by those wavefunctions which {\it extend} over 
an entire bulk (see section~I for its reason).   

\begin{figure}
\begin{center}
\includegraphics[width=0.45\textwidth]{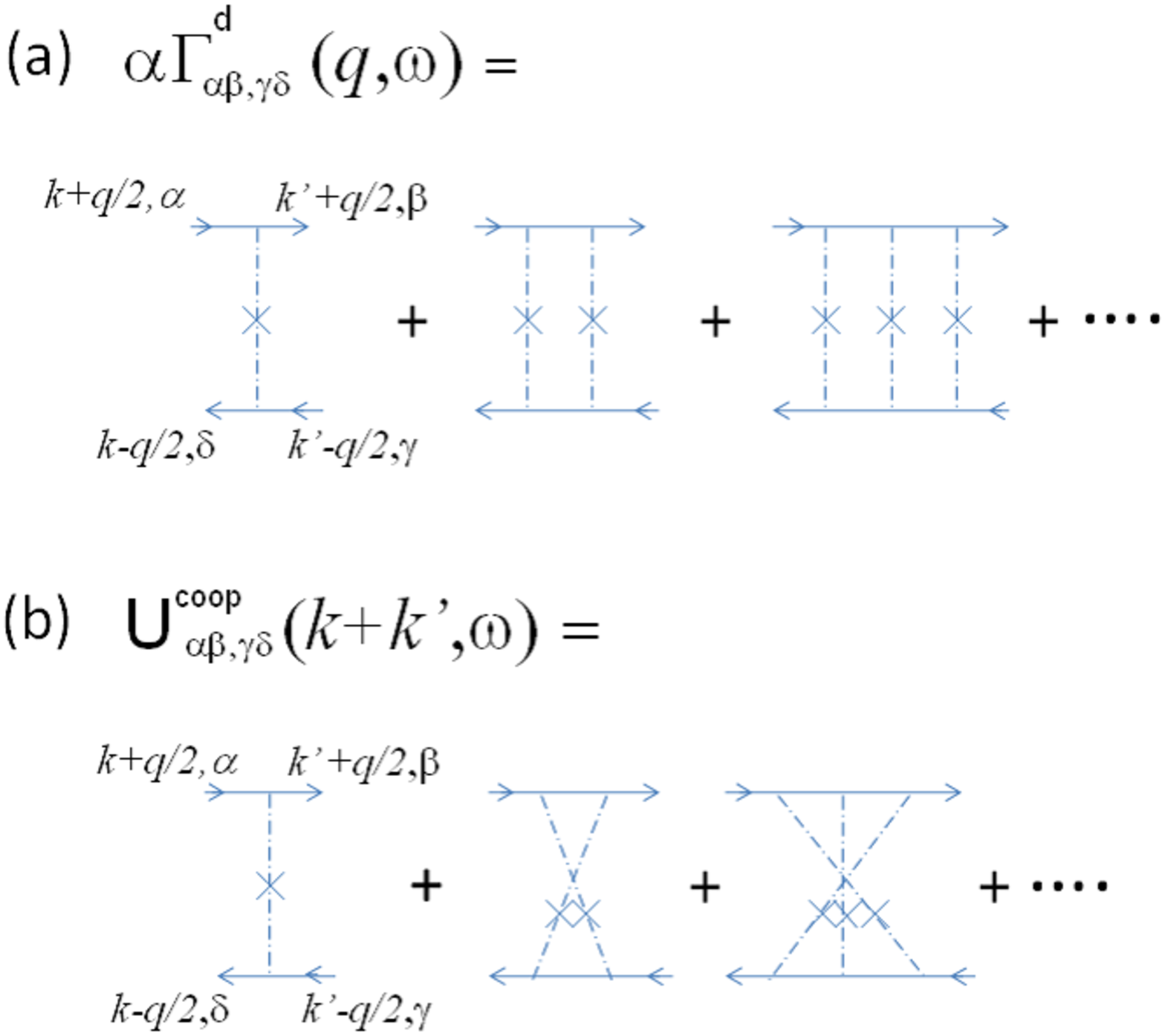}
\end{center}
\caption{(a) A series sum of the ladder-type diagrams 
$\Gamma^d_{\alpha\beta,\gamma\delta}(q,\omega)$, 
(b) A series sum of the ``fan''-type diagrams  
$\hat{U}^{\rm coop}_{\alpha\beta,\gamma\delta}(k+k',\omega) \equiv   
\alpha \sum_{\delta_1,\gamma_1} 
\{\hat{1}\otimes \hat{s}_y\}_{\gamma\gamma_1} 
\hat{\Gamma}^d_{\alpha\beta,\delta_1\gamma_1}(k+k',\omega)
\{\hat{1}\otimes \hat{s}_y\}_{\delta_1\delta}.$}
\label{t12-1-0}
\end{figure}

As the first step to understand the nature of this 
compressible phase especially 
for $\alpha<\alpha_c$, 
we will calculate  
the series sum of the ladder-type diagram 
$\hat{\Gamma}^d(q,\omega)$ (see Fig.~\ref{t12-1-0}(a)) 
, only to see that the diffuson 
thus obtained consists of two quasi-degenerate 
low-energy modes; 
\begin{eqnarray}
\hat{\Gamma}^d(q,\omega) \propto \frac{1}{\omega + iDq^2}\hat{\Gamma}^d_1 
+ \frac{1}{\omega + iDq^2 + i\tau^{-1}_{\rm topo}} \hat{\Gamma}^d_2 + \cdots 
\label{punch1}
\end{eqnarray}
with $\tau^{-1}_{\rm topo} \propto m^2$. 
The information of the charge diffusion is solely encoded into 
the first term, which thus always  
has the diffusion pole structure. On the other hand, 
the second term becomes massless 
only at $m=0$ (but generic $\mu$), while it suffers from 
the finite infrared cutoff $\tau^{-1}_{\rm topo}$ 
for the finite $m$ case. 
We will explicitly see that the second term    
is actually associated with the parity degree of freedom, 
which, at $m=0$, 
becomes a conserved quantity of 
our effective continuum model,  
i.e. eq.~(\ref{e5-t}).  

When the hole line of $\hat{\Gamma}^d(q,\omega)$ 
is time-reversed, these two-mode features are translated  
into the backward scattering channel associated with 
the ``fan''-type diagrams 
$\hat{U}^{\rm coop}(k+k',\omega)$ (see Fig.~\ref{t12-1-0}(b)). 
Namely, for small $\omega$ and $k+k'$, it  
is also dominated by two quasi-degenerate dominant 
contributions;  
\begin{eqnarray}
\hat{U}^{\rm coop}(k+k',\omega) &\propto& \frac{1}{\omega + iD(k+k')^2}
\hat{U}^{\rm c}_1 \nn \\
&&\hspace{-2cm} + \ \frac{1}{\omega + iD(k+k')^2 
+ i\tau^{-1}_{\rm topo}} \hat{U}^{\rm c}_2 + \cdots. \label{punch2}
\end{eqnarray}
One is obtained from the charge mode channel, i.e. $\Gamma^d_1$,  
with its hole-line time-reversed, 
while the other is from the parity 
mode channel, $\hat{\Gamma}^d_2$. In this section,  
we will further see that both of these two give rise to 
the same amplitude of the anti-weak-localization (AWL)   
correction to the electric conductivity at $m=0$. 
In the presence of the finite topological mass 
$m$, however, the second term in eq.~(\ref{punch2}) 
becomes less dominant, because of the finite infrared cut-off 
$\tau^{-1}_{\rm topo}$. Namely, half of the AWL  
correction becomes ineffective on increasing $m$      
(``quantum correction doubling'').   

Using these knowledges obtained in this section, 
we will propose in the next section  
the possible microscopic mechanism of 
how the bulk-critical region  
emerges between the topological insulator and 
an ordinary insulator.      
 
This section is organized as follows. In Sec.~IV-A, 
we will sum up the entire ladder type diagram in the particle-hole channel, 
using the 1-point Green function obtained in the self-consistent 
Born calculation;
\begin{eqnarray}
\hat{G}^{R,-1}(k;\mu,m) = F_0 \hat{1} 
- k_{\nu}\hat{\gamma}_{\nu} + F_5 \hat{\gamma}_5, \label{r} \\
\hat{G}^{A,-1}(k;\mu,m) = F^{\ast}_0 \hat{1} 
- k_{\nu}\hat{\gamma}_{\nu} + F^{\ast}_5 \hat{\gamma}_5, \label{a}
\end{eqnarray}  
with $F_0 \equiv \bar{\mu} + i\tau^{-1}$ and $F_5 \propto m$.   
Such a summand should contain those contributions which diverge 
at $\omega=0$ and $q=0$. We will identify this diverging 
contribution in the 
section~IV-B, only to see that 
$\hat{\Gamma}^d(q,\omega)$   
contain two {\it quasi}-degenerate dominant contributions, 
as in eq.~(\ref{punch1}).  Explicit expressions 
for $\hat{\Gamma}^{d}_1$,  $\hat{\Gamma}^d_2$ and 
$\tau^{-1}_{\rm topo}$ will be therefore given here. 
By calculating the parity-density correlation function, 
we will show in the section.~IV-C   
that $\hat{\Gamma}^{d}_2$ solely participates in    
the parity diffusion mode. Finally, the quantum 
conductivity corrections arising from these two terms 
are calculated in section.~IV-D, based on the Kubo formula. 

Comparing eqs.~(\ref{e31-1}-\ref{e31-1-i}) with 
eqs.~(\ref{e31-3},\ref{e31-5}), notice also that 
the weak-localization (WL) calculation in 
this section becomes a controlled analysis 
only for the weak disorder region, $\alpha<\alpha_c$.  
Namely, the renormalized chemical potential 
$\bar{\mu}$ and the life-time $\tau^{-1}$ determined in 
section.~III guarantee a sufficiently small $1/\bar{\mu}\tau$ 
around $\mu\simeq 0$ {\it only for this weak disorder region}  
;
\begin{eqnarray}
\bar{\mu}\tau =  
\frac{1}{\alpha\pi}
\frac{(\alpha_c-\alpha)^2} 
{\alpha^2_c}\frac{1}{\mu} + {\cal O}\big(\mu\big) \ \ 
\ {\rm for} \ \ \alpha < \alpha_c. \label{control} 
\end{eqnarray}
For the strong disorder region, however,  
$1/\bar{\mu}\tau$ readily diverges  
around the zero energy region;  
\begin{eqnarray}
\bar{\mu}\tau = \frac{\pi \alpha_c}{2 \alpha} \mu 
+ {\cal O}\big(\mu^3\big) \hspace{1.7cm} 
\ \ \ \ {\rm for} \ \ 
\alpha > \alpha_c.  
\label{uncontrol} 
\end{eqnarray} 
Thus, the NCA approximation employed in section.~III 
and the corresponding WL calculation described below acquire 
the small coupling constant $\frac{1}{\bar{\mu}\tau}$, 
only for $\alpha<\alpha_c$, but, for $\alpha>\alpha_c$, 
they generally don't.  
Bearing in mind especially this strong disorder 
region, we will demonstrate in the appendix.~B   
the mode-mode coupling calculation, which 
is complementary 
to the weak-localization studies described 
in this section. 
Without resorting to the 
Kubo formula, this theoretical framework 
gives us the gap equation for the  
diffusion constant, taking into account 
the interference effects due to the Cooperon term.  
The basic feature which we will observe  
in this section, such as the quantum 
correction doubling, are also confirmed 
by this analysis (see Appendix.~B).      
  
\subsection{Ladder-type diagrams in the particle-hole channel}   
For clarity of the following description, 
let us first define a tensor 
composed by two $4\times 4$ matrices;
\begin{eqnarray}
\hat{A} \equiv \hat{A}_{r}\times \hat{A}_{a}. \nn 
\end{eqnarray} 
The former $4\times 4$ matrix $\hat{A}_r$ is for the particle 
(retarded) line, while the other, i.e. $\hat{A}_a$, is for the 
hole (advance) line. Throughout this section, we 
distinguish this ``$\times$''-mark from 
the ``$\otimes$''-mark, latter of which 
separates the spin space and 
sublattice space. 
The product of two tensors is defined as follows; 
\begin{eqnarray}
\hat{A}\cdot \hat{B} \equiv 
\hat{A}_{r}\hat{B}_{r} \times \hat{B}_{a}\hat{A}_a.  \label{dc2-0}
\end{eqnarray} 
Note here that the order of the product in the 
hole line is reversed, compared with 
that of the left hand side.
Under this algebra, the series-sum  
of the ladder diagram in the particle-hole 
channel, i.e.~Fig.~\ref{t12-1-0}(a), is just the inverse of the following tensor,  
\begin{eqnarray}
&&\hspace{-0.3cm}
\big\{\hat{1} - \alpha 
\hat{\Pi}(q,\omega)\big\}_{\alpha\delta,\gamma\beta} \nn \\
&&\hspace{0.1cm} 
\equiv \big\{\hat{1}\times \hat{1} - \alpha 
\sum_{k}\hat{G}^{R}(0_{+})\times 
\hat{G}^{A}(0_{-})\big\}_{\alpha\delta,\gamma\beta} \nn \\ 
&&\hspace{0.1cm} 
\equiv \delta_{\alpha\delta}\delta_{\gamma\beta} - \alpha 
\sum_{k} \hat{G}^{R}_{\alpha\delta}(0_{+}) 
\hat{G}^{A}_{\gamma\beta}(0_{-}), \label{dc4}
\end{eqnarray} 
with $(0_{\pm})\equiv (k\pm\frac{q}{2},\mu\pm \frac{\omega}{2})$. 
Namely, the following identity can be readily checked;
\begin{eqnarray}
&&\hspace{-0.6cm} 
\big\{\hat{1} - \alpha 
\hat{\Pi}(q,\omega)\big\}_{\alpha\alpha_1,\beta_1\beta} 
\Gamma^{d}_{\alpha_1\delta,\gamma\beta_1}(q,\omega)
\equiv \delta_{\alpha\delta}\delta_{\gamma\beta}. \label{identity}
\end{eqnarray}
Notice also that we have already 
normalized the momentum by the ultraviolet 
cutoff $\Lambda$ as in eqs.~(\ref{e19-1}-\ref{e19}),  
so that $\alpha \equiv 2\pi\Delta_{00} \Lambda$ and 
$\sum_{k} \equiv 2\int^{1}_{0} k^2 dk$.  
This notation also holds true for 
eqs.~(\ref{dc6-1},\ref{dc10-1},\ref{dc10-2}).    

For simplicity of the explanation, 
we will calculate the inverse of eq.~(\ref{dc4}), 
with $q$ taken to be zero from the beginning (see also the 
description around eq.~(\ref{baredif})). 
Such an inverse diverges if $\omega=0$, while 
it does not for general $\omega$. When $q$ 
taken to be zero, the polarization part becomes simple; 
\begin{eqnarray}
&&\hspace{-0.6cm}\big\{\hat{1}-\alpha \hat{\Pi}(q\equiv 0,\omega)\big\}
_{\alpha\delta,\gamma\beta} \nn \\
&& \hspace{0.3cm} 
\equiv a_0\hat{1} + a_1 \hat{T}_1 + a_2 \hat{T}_2 
+ a_3 \hat{T}_3 + a_4 \hat{T}_4 \label{dc5}, \\ 
&&\hspace{-0.15cm} T_{1} \equiv \sum_{\mu=1}^3 \hat{\gamma}_{\mu}\times \hat{\gamma}_{\mu}, \ \ 
\hat{T}_2 \equiv \hat{1} \times \hat{\gamma}_{5}, \label{dc6+1} \\ 
&&\hspace{-0.15cm} \hat{T}_3 \equiv \hat{\gamma}_5 \times \hat{1}, \ \ 
\hat{T}_4 \equiv \hat{\gamma}_5 \times \hat{\gamma}_5. \label{dc6-0} 
\end{eqnarray}
Namely, it is just a linear combination of the five  
tensors with their coefficients  
defined as follows; 
\begin{eqnarray}
&&\hspace{-0.9cm}\big\{a_{0},a_1,a_2,a_3,a_4\big\} \equiv \nn \\
&& \hspace{-0.7cm} 
\bigg\{ 1 - \alpha \sum_{k} 
\frac{F_{0+}F^{*}_{0-}}
{(k^2 - F^2_{0+} + F^2_{5+})(k^2 - (F^*_{0-})^2 + (F^*_{5-})^2)}, \nn \\ 
&&\hspace{-0.6cm} 
-\alpha \sum_{k} 
\frac{k^2_x}{(k^2 - F^2_{0+} + F^2_{5+})(k^2 - (F^*_{0-})^2 + (F^*_{5-})^2)}, \nn \\
&&\hspace{-0.6cm} 
\alpha \sum_{k} \frac{F_{0+}F^{*}_{5-}} 
{(k^2 - F^2_{0+} + F^2_{5+})(k^2 - (F^*_{0-})^2 + (F^*_{5-})^2)}, \nn \\
&&\hspace{-0.6cm}
\alpha \sum_{k} \frac{F_{5+}F^{*}_{0-}}
{(k^2 - F^2_{0+} + F^2_{5+})(k^2 - (F^*_{0-})^2 + (F^*_{5-})^2)}
, \nn \\
&&\hspace{-0.6cm} 
- \alpha \sum_{k} \frac{F_{5+}F^{*}_{5-}}
{(k^2 - F^2_{0+} + F^2_{5+})(k^2 - (F^*_{0-})^2 + (F^*_{5-})^2)}\bigg\}.  
\label{dc6-1} 
\end{eqnarray}
The $\pm$ subscripts on $F_{i}$ above mean that the argument of 
$F_{i}(\mu)$ is shifted by $\pm \frac{\omega}{2}$;
\begin{eqnarray}    
F_{i\pm} \equiv F_{i}(\mu\pm\frac{\omega}{2}). \nn   
\end{eqnarray} 
As such, $a_0$, $a_1$ and $a_4$ become 
real-valued at $\omega=0$, while $a_2$ and $a_3$ become  
complex conjugate with each other;
\begin{eqnarray}
{a^{*}_{0,1,4}}_{|\omega=0} 
= {a_{0,1,4}}_{|\omega=0}, \ \  
{a^{*}_{2}}_{|\omega=0} ={a_3}_{|\omega=0}.  \label{rv}
\end{eqnarray}
One can evaluate the signs of the 
former three real-valued quantities,  
by noting that $F_{\mu}$ is much smaller 
than the ultraviolet cutoff. They read,   
\begin{eqnarray}
{a_{0}}_{|\omega=0} > 0, \ \ {a_{1}}_{|\omega=0} < 0, \ \ 
{a_4}_{|\omega=0} \le 0. \label{p100}
\end{eqnarray}
Notice also that all 
the integrands for $a_{2}$, $a_{3}$ and $a_{4}$  
contain $F_5$, which is proportional to the topological 
mass (see sec.~III). Thus these three quantities 
vanish when the topological 
mass $m$ is zero;
\begin{eqnarray}
a_{4} \propto {\cal O}(m^2) , \ \ a_{2,3} \propto {\cal O}(m). 
\label{p101}
\end{eqnarray}  

The inverse of eq.~(\ref{dc5}) 
also becomes a linear combination of a finite number of  
tensors composed of $\gamma$ 
matrices, because of the cyclic nature;  
$\gamma_1\gamma_2\gamma_3\gamma_4\gamma_5\equiv-\gamma_0$; 
\begin{eqnarray}
&&\hspace{-0.5cm} 
\hat{\Gamma}^d(q = 0,\omega) \nn \\
&&\hspace{0.1cm} 
 \equiv \beta_0 \hat{1} + \beta_1 \hat{T}_1 + \beta_2 \hat{T}_2  
+ \beta_3 \hat{T}_3 + \beta_4 \hat{T}_4 
 \nn \\ 
&&\hspace{0.4cm} 
+ \beta_{5} \hat{S}_1 + \beta_{6} \hat{S}_2  
+ \  \beta_{7} \hat{T}_4\cdot \hat{T}_1 
+ \beta_{8} \hat{T}_2\cdot \hat{S}_1 \nn \\
&&\hspace{0.4cm} 
 + \beta_{9}\hat{T}_3\cdot \hat{S}_1 
+ \beta_{10} \hat{T}_4\cdot \hat{S}_1 
+ \beta_{11} \hat{T}_4\cdot \hat{S}_2,  \label{dc7} 
\end{eqnarray}
where two additional tensors are introduced in the following way;  
\begin{eqnarray}
\hat{S}_1 &\equiv& \hat{\gamma}_{1}\hat{\gamma}_2 \times \hat{\gamma}_2\hat{\gamma}_1 
+ \hat{\gamma}_2\hat{\gamma}_3 \times \hat{\gamma}_3 \hat{\gamma}_2 + \hat{\gamma}_3\hat{\gamma}_1 
\times \hat{\gamma}_3\hat{\gamma}_1, \nn \\
\hat{S}_2 &\equiv& \hat{\gamma}_1 \hat{\gamma}_2\hat{\gamma}_3 \times 
\hat{\gamma}_3 \hat{\gamma}_2 \hat{\gamma}_1.  \label{dc7-0}
\end{eqnarray}
After lengthy algebra,
one can express its 12 coefficients $\beta_j$ 
in terms of those of eq.~(\ref{dc5}) as follows; 
\begin{eqnarray}
\left[
\begin{array}{c}
\beta_0 \\ 
\beta_1 \\
\beta_2 \\
\beta_3 \\
\beta_4 \\
\beta_{5} \\
\beta_{6} \\
\beta_{7} \\
\beta_{8} \\
\beta_{9} \\
\beta_{10} \\
\end{array}
\right]
\equiv \frac{1}{8} 
\left[
\begin{array}{cccc}
-3\delta a_{04} &-\delta a_{04}&-3a_{04} &-a_{04} \\ 
a_1&3a_1&a_1&3a_1 \\ 
3\delta a_{23}& \delta a_{23} & 3a_{23}& a_{23} \\ 
-3\delta a_{23} &-\delta a_{23} & 3a_{23}& a_{23} \\
3\delta a_{04} &\delta a_{04}& -3 a_{04} & - a_{04} \\
\delta a_{04}&- \delta a_{04} &a_{04} &-a_{04} \\
-3a_1&3a_1&-3a_1&3a_1 \\ 
-a_1&-3a_1&a_1&3a_1 \\ 
-\delta a_{23} & \delta a_{23} & -a_{23}& a_{23} \\ 
\delta a_{23}&- \delta a_{23} & -a_{23} & a_{23} \\ 
- \delta a_{04} & \delta a_{04} & a_{04}& -a_{04} \\
\end{array}
\right] 
\left[ 
\begin{array}{c}
f_1 \\
f_3 \\
f_2 \\
f_4 \\
\end{array} 
\right], \label{dc8-0} 
\end{eqnarray}
\begin{eqnarray}
&&\hspace{-0.5cm} 
\beta_{11} \equiv 
-3{a_1}^3(f^{-1}_4+f^{-1}_1)f_{1}f_{2}\big(-f_3+f_4\big), 
\label{dc8-1}
\end{eqnarray}
where $a_{04}$, $\delta a_{04}$, $a_{23}$, $\delta a_{23}$ and $f_{1,2,3,4}$ are 
defined in terms of $a_{0}$, $a_1$, $\cdots$ and $a_4$;
\begin{eqnarray}
&&\hspace{-0.5cm} 
\delta a_{23} \equiv a_2 - a_3 , \ \delta a_{04} \equiv a_0 - a_4 , \label{dc9-0} \\  
&&\hspace{-0.5cm} 
a_{23} \equiv a_2 + a_3, \ a_{04} \equiv a_0 + a_4, \label{dc9-1}   \\
&&\hspace{-0.5cm} 
f_1\equiv  \frac{1}{a^2_1 +(\delta a_{04}+\delta a_{23})(-\delta a_{04}+\delta a_{23})}, 
\label{dc9-2} \\ 
&&\hspace{-0.5cm} 
f_2\equiv  \frac{1}{a^2_1 +(- a_{04}+  a_{23})(a_{04}+  a_{23})},  
\label{dc9-3} \\
&&\hspace{-0.5cm} 
f_3\equiv  \frac{1}{9a^2_1 +(\delta a_{04}+\delta a_{23})(-\delta a_{04}+\delta a_{23})}, 
\label{dc9-4} \\ 
&&\hspace{-0.5cm} 
f_4\equiv  \frac{1}{9a^2_1 +(- a_{04}+  a_{23})(a_{04}+  a_{23})}.  \label{dc9-5} 
\end{eqnarray}

\subsection{Identification of the diffusion pole} 
Using eqs.~(\ref{r}-\ref{a}), 
we have summed up the ladder-type 
diagram in the particle-hole  
channel, only to obtain eq.~(\ref{dc7}). 
The coefficients $\beta_j$ appearing in eq.~(\ref{dc7}) 
are expressed
in terms of $F_0$ and $F_5$,  
by way of eqs.~(\ref{dc8-0}-\ref{dc9-5}) 
and eq.~(\ref{dc6-1}). 
When the self-consistent Born (scB) 
solution is used for 
$F_0$ and $F_5$, at least one of these $\beta_j$ is 
expected to have a diffusion pole structure.  
On the one hand, none of $a_j$ defined 
in eq.~(\ref{dc6-1}) does not diverge at $\omega=0$.  
As such, some of $f^{-1}_{j}$ 
should be zero at $\omega=0$.   
In this subsubsection, we will identify which 
$f_j$ diverges at small $\omega$.  This also 
determines the asymptotic 
{\it tensor}-form of the diffuson 
in the small $\omega$ limit.

To do this, let us first start from the self-consistent 
Born equations of $F_0$ and $F_5$, i.e. eqs.~(\ref{e26}-\ref{e28}).  
Or equivalently, begin with the following two; 
\begin{eqnarray}
&&\hspace{-1.0cm} 
(F_0-F_5) - \alpha \sum_{k}\frac{(F_0+F_5)}{k^2-(F^2_0-F^2_5)} =\mu - m + i\delta, 
\label{dc10-1} \\
&&\hspace{-1.0cm} 
(F_0+F_5) - \alpha \sum_{k}\frac{(F_0-F_5)}{k^2-(F^2_0-F^2_5)} =
\mu + m + i\delta. \label{dc10-2}  
\end{eqnarray} 
Then, subtracting eqs.~(\ref{dc10-1},\ref{dc10-2}) by their 
complex conjugates respectively, 
we can readily obtain    
\begin{eqnarray}
\left[\begin{array}{cc}
a_{04}-a_{23} &  3a_1 \\
3a_1 &a_{04}+a_{23}  \\  
\end{array} 
\right]_{|\omega= 0}  
\left[\begin{array}{c}
F^{\prime\prime}_0-F^{\prime\prime}_5 \\ 
F^{\prime\prime}_0+F^{\prime\prime}_5 \\ 
\end{array}
\right] = \left[\begin{array}{c}
0 \\ 
0 \\ 
\end{array}\right],  
\end{eqnarray} 
where eq.~(\ref{dc9-1}) and eq.~(\ref{dc6-1}) were used. 
This equation indicates that the determinant of the $2\times 2$ matrix 
in the left hand side should be zero, 
provided that either $F^{\prime\prime}_0$ or $F^{\prime\prime}_5$  
is non-zero. Any compressible phase having a finite density 
of state supports ${F^{\prime\prime}_0}^2+{F^{\prime\prime}_5}^2\ne 0$. 
As such, any scB solution in the compressible phase 
always guarantees the following identity; 
\begin{eqnarray}
\big\{9a^2_1 - a^2_{04} + a^2_{23}\big\}_{|\omega=0} 
\equiv 0. \label{dc10-5}
\end{eqnarray} 
Since $a_1$ defined in eq.~(\ref{dc6-1})  
is negative definite at  $\omega=0$, more accurately, 
eq.~(\ref{dc10-5}) should be replaced by;
\begin{eqnarray}
3{a_1}_{|\omega=0} = - 
\sqrt{a^2_{04}-a^2_{32}}_{|\omega=0}. \label{dc10-6}
\end{eqnarray} 
Observing eq.~(\ref{dc9-5}), notice that 
this is actually identical to the following,  
\begin{eqnarray}
\big\{f^{-1}_4\big\}_{|\omega=0} \equiv  0. \label{dc10-5-1}
\end{eqnarray}
Namely, $f_4$ carries the diffusion pole. 
 
$f_1$, $f_2$ and $f_3$ 
generally cannot have 
any pole structure for the small $\omega$ region. 
To see this explicitly, note 
first that, when generalized into 
the finite $q$ case, $f_4$ takes 
the following asymptotic 
form;  
\begin{eqnarray}
&&\hspace{-0.3cm} 
f_4(\omega) \simeq \frac{1}{a_0\tau}  
\frac{1}{i\omega} \nn \\
&&\hspace{-0.1cm} 
\rightarrow f_{4}(q,\omega) \simeq 
 \frac{1}{a_0\tau}  
\frac{1}{i\omega-D_0 q^2}.  \label{f4}
\end{eqnarray} 
In the right hand side, 
we have replaced $i\omega$ by $i\omega-D_0q^2$ with 
the bare diffusion constant $D_0$. 
By retaining the subleading 
contribution in small $q$ appearing in eq.~(\ref{dc4}),  
one can explicitly calculate its leading order 
expression in the large $\bar{\mu}\tau$ limit; 
\begin{eqnarray}
D_0 \equiv \frac{1}{6}\frac{\alpha_c-\alpha}{\alpha_c+\alpha} 
\tau, \label{baredif} 
\end{eqnarray} 
which is positive definite for $\alpha<\alpha_c$. 
Similarly, 
we can obtain the asymptotic form of 
$f_1$, $f_2$ and $f_3$ at $\omega, q \simeq 0$;
\begin{eqnarray} 
f_{1} &\simeq& \frac{1}{a_0\tau}  
 \frac{1}{i\omega-D_{1} q^2-\tau^{-1}_1}, \label{f1} \\ 
f_{2} &\simeq& \frac{1}{a_0\tau}   
\frac{1}{i\omega-D_{2} q^2-\tau^{-1}_2}, \label{f2} \\
f_{3} &\simeq& \frac{1}{a_0\tau}  
\frac{1}{i\omega-D q^2-\tau^{-1}_{\rm topo}}. \label{f3} 
\end{eqnarray}    
$\tau^{-1}_{1}$ and $\tau^{-1}_2$ above 
are positive definite;
\begin{eqnarray}
\tau^{-1}_{1} &\equiv& \tau^{-1}_{2} + \tau^{-1}_{\rm topo} > 0, \nn \\ 
\tau^{-1}_{2} &\equiv& 
\tau^{-1} \times 
\bigg\{ \frac{8 a^2_1}{a_0}
\bigg\}_{|\omega=0}
  > 0, \label{ntopo}  
\end{eqnarray}  
while $\tau^{-1}_{\rm topo}$ being positive semi-definite;
\begin{eqnarray}
\tau^{-1}_{\rm topo} \equiv 
\tau^{-1}\times \bigg\{ 
\frac{4(-a_0a_4+a_2a_3)}{a_0} 
\bigg\}_{|\omega=0} 
\ge 0. \label{topo}
\end{eqnarray}
The two inequalities in eq.~(\ref{ntopo}) 
and eq.~(\ref{topo}) are indeed 
supported by  eq.~(\ref{rv}) and 
eqs.~(\ref{rv}-\ref{p100}) respectively.  
These expressions indicate that $f_{1}$, $f_2$ 
and $f_3$ always experience the infrared 
cutoff for the low-energy and 
long wavelength region.

Comparing eq.~(\ref{topo}) with 
eq.~(\ref{p101}), notice also  
that $\tau^{-1}_{\rm topo}$ reduces to  
zero in the massless case, 
$\tau^{-1}_{\rm topo}\propto m^2$,   
since $a_2$, $a_3$ and $a_4$ being zero. 
As such, $f_3$ acquires a same diffusion 
pole as $f_4$ does in the absence of the topological mass.   
Meanwhile $f_1$ and $f_2$ always suffer from  
the (relatively large) finite infrared cutoff $\tau^{-1}_2$,  
irrespectively of the topological mass term.  
Thus, we will retain 
in eq.~(\ref{dc8-0}) only those terms 
proportional to 
$f_3$ and $f_4$.
Based on the same spirit,  we will also 
replace eq.~(\ref{dc8-1}) by 
its leading order contribution   
in small $\omega$ and $q$;
\begin{eqnarray}
\beta_{11} \simeq  \frac{3a_1}{8} \big(-f_3+f_4\big), \label{dc8-1-1}
\end{eqnarray} 
and use the following asymptotic expressions 
for $\{\beta_{0},\cdots,\beta_{11}\}$;
\begin{eqnarray}
&&\hspace{-0.3cm}
\big\{\beta_{0},\beta_{4},\beta_{5},\beta_{10}\big\} \nn \\
&&\hspace{0.2cm} \simeq \frac{{\delta a_{04}}_{|\omega=0}}{8}f_3 
\big\{-1,1,-1,1\big\} 
- \frac{{a_{04}}_{|\omega=0}}{8}f_4 
 \big\{1,1,1,1\big\}, \nn \\  
 &&\hspace{-0.3cm}
\big\{\beta_{1},\beta_{6},\beta_{7},\beta_{11}\big\} \nn \\
&&\hspace{0.2cm} \simeq \frac{{3a_{1}}_{|\omega=0}}{8}f_3 
\big\{1,1,-1,-1\big\} 
+ \frac{{3a_{1}}_{|\omega=0}}{8}f_4 
 \big\{1,1,1,1\big\}, \nn \\ 
 &&\hspace{-0.3cm}
\big\{\beta_{2},\beta_{3},\beta_{8},\beta_{9}\big\} \nn \\
&&\hspace{0.2cm} 
\simeq \frac{{\delta a_{23}}_{|\omega=0}}{8}f_3 
\big\{1,-1,1,-1\big\} 
+ \frac{{a_{23}}_{|\omega=0}}{8}f_4 
 \big\{1,1,1,1\big\}. \nn    
\end{eqnarray}

With these equations, the asymptotic form  
of the diffuson in small 
$\omega$ and $q$ will be derived out of 
eq.~(\ref{dc7}). It consists of the  
two {\it quasi}-degenerate dominant 
contributions; 
\begin{eqnarray}
\hat{\Gamma}^d(q,\omega) &\simeq& 
\frac{f_4}{8}\hat{\Gamma}^d_1  + \frac{f_3}{8} \hat{\Gamma}^d_2,    
\label{dc12-11} 
\end{eqnarray} 
where the two $\omega$,$q$-free  
tensors are given as follows;     
\begin{eqnarray}  
&&\hspace{-0.3cm} 
\hat{\Gamma}^d_1 \equiv \nn \\  
&&\hspace{-0.6cm} 
\left[\begin{array}{ccc}
-{a_{04}} & {3a_1} & {a_{23}} \\ 
\end{array}\right]_{|\omega=0} \cdot \left[\begin{array}{c}
(\hat{1}+\hat{T}_4)\cdot (\hat{1}+\hat{S}_1) \\ 
(\hat{1}+\hat{T}_4)\cdot (\hat{T}_1+\hat{S}_2) \\ 
(\hat{T}_2+\hat{T}_3)\cdot (\hat{1}+\hat{S}_1) \\
\end{array}\right], \label{dc12-13-a} \\  
&&\hspace{-0.3cm}  
\hat{\Gamma}^d_2 \equiv \nn \\ 
&&\hspace{-0.6cm}  
\left[\begin{array}{ccc}
-\delta {a_{04}} & {3a_1} & \delta {a_{23}} \\  
\end{array}\right]_{|\omega=0} \cdot \left[\begin{array}{c}
(\hat{1}-\hat{T}_4)\cdot (\hat{1}+\hat{S}_1) \\ 
(\hat{1}-\hat{T}_4)\cdot (\hat{T}_1+\hat{S}_2) \\ 
(\hat{T}_2-\hat{T}_3)\cdot (\hat{1}+\hat{S}_1) \\
\end{array}\right]. \label{dc12-13-b}  
\end{eqnarray}

\subsection{parity diffusion mode}  
To capture the physical meanings of the two members 
in eq.~(\ref{dc12-11}),  
notice first that, in the absence of the topological 
mass $m$, our hamiltonian,   
i.e. eq.~(\ref{e5-t}) with the chemical-potential type 
disorder, becomes invariant under the 
following $U(1)$ transformation; 
\begin{eqnarray}
 e^{i  \theta \int \psi^{\dagger}(r)\hat{\gamma}_{45}\psi(r)}
 \cdot \hat{\cal H} \cdot e^{-i \theta \int  
\psi^{\dagger}(r)\hat{\gamma}_{45}\psi(r) } 
 = \hat{\cal H},  \label{u1sym}
\end{eqnarray}
irrespectively of the strength of 
the disorder. As a result, each 
ensemble at $m=0$ acquires another conserved charge, 
$\psi^{\dagger}(r)\hat{\gamma}_{45}\psi(r)$, 
which is the parity density  
degree of freedom (see Table.~II). Observing  
this $U(1)$ symmetry, we can then  
expect that the diffuson $\hat{\Gamma}^d(q,\omega)$ 
calculated above 
should consist of two diffusive modes at $m=0$: One 
describes the usual charge diffusion and the other is 
for the diffusion of this parity density.  
These two physical modes actually correspond to the first term 
and the second term in eq.~(\ref{dc12-11}) 
respectively. In fact, the parity density 
becomes non-conserved in the presence of 
finite $m$, which is consistent with the finite infrared 
cut-off $\tau^{-1}_{\rm topo}\propto m^2$ appearing only 
in $f_3$ (see eqs.~(\ref{f3},\ref{f4})).  
    
To uphold this consideration more directly, 
one can also calculate the 
density correlation function and 
parity density correlation function  
at $m=0$;
\begin{eqnarray} 
\phi_{0}(q,\omega)&\equiv& \sum_{k,k',\alpha,\beta} 
\Phi_{\alpha\beta,\beta\alpha}(k,k';q,\omega), \nn \\
\phi^{\prime}_{45}(q,\omega)&\equiv& \sum_{\cdots} 
\big[\hat{\gamma}_{45}\big]_{\beta\alpha}
\Phi_{\alpha\delta,\gamma\beta}(k,k';q,\omega)
\big[\hat{\gamma}_{45}\big]_{\delta\gamma}, \nn   
\end{eqnarray}    
where $\hat{\Phi}(k,k';q,\omega)$ stands for 
the response function (see eq.~(\ref{e60}) for its definition). 
By noting that this response function for 
small $q$ and $\omega$ is dominated by the diffuson;  
\begin{eqnarray}
\Phi_{\alpha\delta,\gamma\beta}(k,k';q,\omega) 
&\simeq& -\frac{\alpha}{2\pi i} \hat{G}^{R}_{\alpha\alpha_1}(k_{+},\mu_{+})
\hat{G}^{A}_{\beta_1\beta}(k_{-},\mu_{-}) \nn \\
&&\hspace{-3.2cm}\times  
\big[\hat{\Gamma}^d(q,\omega)\big]_{\alpha_1\delta_1,\gamma_1\beta_1} 
\hat{G}^{R}_{\delta_1\delta}(k'_{+},\mu_{+})
\hat{G}^{A}_{\gamma\gamma_1}(k'_{-},\mu_{-}), \label{b1d} 
\end{eqnarray}
one can explicitly see that the two terms appearing 
in eq.~(\ref{dc12-11}) actually contribute the density 
correlation 
and parity density correlation 
separately;
\begin{eqnarray}  
\big(\phi_{0}(q,\omega),\phi^{\prime}_{45}(q,\omega)\big) \simeq 
- \frac{8 i a_0}{\alpha} \times 
\big(f_4,f_3\big).\ \label{pp} 
\end{eqnarray} 

\subsection{Cooperon and the quantum conductivity correction}
When the hole lines being time-reversed, eq.~(\ref{dc12-11}) 
will be transcribed into the two quasi-degenerate dominant contributions   
to the series sum of the ``fan''-type 
diagrams (see Fig.~\ref{t12-1-0}(b));
\begin{eqnarray}
&&\hspace{-1.1cm}\hat{U}^{\rm coop}(k+k',\omega) = 
\frac{\alpha}{8} 
\big\{f_4 \hat{U}^{\rm c}_1 + f_3  
\hat{U}^{\rm c}_2\big\}_{|q\rightarrow k+k'}. \label{162d}
\end{eqnarray}
where the $\omega$, $q$-free tensors $\hat{U}^{\rm c}_{1,2}$ 
are derived out of eqs.~(\ref{dc12-13-a}-\ref{dc12-13-b}) 
respectively;    
\begin{eqnarray}  
&&\hspace{-0.3cm} 
\hat{U}^{\rm c}_1 \equiv \nn \\  
&&\hspace{-0.6cm} 
\left[\begin{array}{ccc}
-{a_{04}} & {3a_1} & {a_{23}} \\ 
\end{array}\right]_{|\omega=0} \cdot \left[\begin{array}{c}
(\hat{1}+\hat{T}_4)\cdot (\hat{1} - \hat{S}_1) \\ 
(\hat{1} - \hat{T}_4)\cdot ( - \hat{T}_1 + \hat{S}_2) \\ 
(\hat{T}_2 + \hat{T}_3)\cdot ( \hat{1} - \hat{S}_1) \\
\end{array}\right], \label{dc12-13-c} \\  
&&\hspace{-0.3cm}  
\hat{U}^{\rm c}_2 \equiv \nn \\    
&&\hspace{-0.6cm}  
\left[\begin{array}{ccc}
-\delta {a_{04}} & {3a_1} & \delta {a_{23}} \\  
\end{array}\right]_{|\omega=0} \cdot \left[\begin{array}{c}
(\hat{1}-\hat{T}_4)\cdot (\hat{1} - \hat{S}_1) \\ 
(\hat{1} + \hat{T}_4)\cdot ( - \hat{T}_1 + \hat{S}_2) \\ 
(\hat{T}_2 - \hat{T}_3)\cdot (\hat{1} - \hat{S}_1) \\
\end{array}\right]. \label{dc12-13-d}  
\end{eqnarray}
Substituting these two Cooperon terms into the current-current 
correlation function, we can explicitly show that the 
two members in eq.~(\ref{162d}) lead 
the same magnitude of the anti-weak-localization (AWL) behaviour 
at the critical point ($m=0$);
\begin{widetext}
\begin{eqnarray}
&&\hspace{-1.2cm} \sum_{L^{-1}<|k+k'|<l^{-1}}
\frac{\alpha f_4}{8} \sum_k \{\hat{G}^{A}(k,\mu)\cdot 
\hat{\gamma}_{1}\cdot \hat{G}^{R}(k,\mu)\}_{\delta\alpha}
\hat{U}^{\rm c}_{1,\alpha\beta,\gamma\delta}
\{\hat{G}^{R}(-k,\mu)\cdot 
\hat{\gamma}_{1}\cdot \hat{G}^{A}(-k,\mu)\}_{\beta\gamma} \nn \\  
&&\hspace{-0.2cm} 
= \sum_{\cdots}\frac{\alpha f_4}{8} 
\sum_k \{\hat{G}^{A}(k,\mu)\cdot 
\hat{\gamma}_{1}\cdot \hat{G}^{R}(k,\mu)\}_{\delta\alpha}
\hat{U}^{\rm c}_{2,\alpha\beta,\gamma\delta}
\{\hat{G}^{R}(-k,\mu)\cdot 
\hat{\gamma}_{1}\cdot \hat{G}^{A}(-k,\mu)\}_{\beta\gamma} \nn \\  
&& \hspace{-0.2cm} = \frac{\alpha}{8} 
\frac{1}{D_0 \tau} \sum_{\cdots}  
\frac{1}{(k+k')^2}\sum_k 
\{\hat{G}^{A}(k,\mu)\cdot 
\hat{\gamma}_{1}\cdot \hat{G}^{R}(k,\mu)\}_{\delta\alpha}
\{\hat{1}-\hat{S}_1 - \hat{T}_1 + \hat{S}_2\}_{\alpha\beta,\gamma\delta}
\{\hat{G}^{R}(-k,\mu)\cdot 
\hat{\gamma}_{1}\cdot \hat{G}^{A}(-k,\mu)\}_{\beta\gamma} \nn \\ 
&& \hspace{-0.2cm} = c \times (l^{-1}-L^{-1})
\end{eqnarray}      
\end{widetext}
with $c=16\pi\cdot (1+\frac{\alpha}{\alpha_c})$ being positive definite.  
(we used $a_{2}=a_3=a_4=0$ and $-3a_1 = a_0$ in 
eqs.~(\ref{dc12-13-c}-\ref{dc12-13-d})). 
When the finite topological mass $m$ is introduced, however,  
the second member of eq.~(\ref{162d}) becomes suppressed, 
since the infrared divergence of $f_3$ becomes 
truncated by finite $\tau^{-1}_{\rm topo}$. As a result, 
one half of the AWL correction becomes ineffective in 
the presence of finite $m$ 
(``quantum correction doubling''). 

\section{Discussion}
\subsection{Summary of our findings}

In this paper, we have studied the effects of the 
time-reversal invariant disorder on the 
quantum spin Hall 
system~\cite{OAN, OFRM, Essin,Ryu07, Ostrovsky}. 
We have especially focused  
on the quantum critical point (QCP) 
which intervenes the $3$-d topological 
insulator (TI) and an $3$-d ordinary 
insulator. The topological 
insulator supports a single $2+1$ 
massless surface Dirac fermion 
for each boundary, while an ordinary insulator 
does not have any. 
As such, the bulk wavefunction in those parameter 
regions (or point) which intervene these two 
insulating phase should be extended, so as to 
mediate two opposite surfaces.  
Such extended bulk states
are stable against ${\cal T}$-invariant disorders, 
as far as each surface state in 
the TI phase is stable.  
In fact, Nomura {\it et al.}~\cite{NKR} and  
Bardarson {\it et al.}~\cite{BTBB} have recently 
calculated the $\beta$ function 
numerically, and  demonstrated that 
the single-copy of the $2+1$ massless 
Dirac fermion is topologically 
stable against the ${\cal T}$-invariant  
disorders.  
This observation strongly indicates that 
that  
there {\it always} exists delocalized 
(bulk-critical) region 
between the $3$-d topological insulator phase and 
an $3$-d ordinary insulator phase.
 
To uncover the nature of this peculiar 
quantum critical 
point (or region), we have studied 
the disorder effect on its minimal model, 
i.e.  the 1-copy of the $3+1$ 
Dirac fermion.  
As a basis for this, 
we first studied in the section~III 
how the chemical potential type disorder 
brings about a finite life time of the zero-energy wavefunction.
We then observed that there exists a certain critical 
disorder strength above which the DOS 
at the zero-energy becomes finite 
(see eq.~(\ref{e30})). 

When the finite topological 
mass is introduced, a system eventually enters  
either the TI or an ordinary insulator,  
depending on the sign of the topological mass. In 
Sec.~III, we studied
how this topological mass are renormalized by 
the chemical-potential-type disorder within the 
self-consistent Born approximation. 
By doing this, we have determined the phase 
boundary between the compressible 
phase and the gapped phase (see Figs.~\ref{t8},\ref{t9}). 
 
To further infer the low-energy structure 
in this compressible phase, we have derived in 
the section~IV the diffuson, Cooperon and 
the weak localization (WL) correction to 
the electric conductivity. 
We then observed that the charge diffusion mode 
and parity diffusion mode  
dominant the diffuson (see eq.~(\ref{dc12-11})); 
the charge channel always carries the diffusion pole 
structure, while the parity density channel becomes 
massless only in the absence of the topological mass. 
In the presence of the finite topological mass $m$, 
it generally suffers from the 
infrared cut-off $\tau^{-1}_{\rm topo}\propto m^2$. 

Corresponding to this feature in the diffuson, 
the Cooperon is also composed of 
two quasi-degenerate dominant contributions 
(see eqs.~(\ref{162d}-\ref{dc12-13-d})).  
In the zero topological mass limit, these 
two contributions bring about 
the same magnitude of the 
anti-weak-localization (AWL) correction 
with each other. 
When the finite topological mass $m$ 
is introduced, however, 
that from the parity density channel becomes 
truncated by the finite infrared cutoff  
$\tau^{-1}_{\rm topo}$.  As such, 
one half of the AWL correction becomes  
ineffective. 
As a result, on increasing $m$, 
the AWL correction exhibits 
a crossover into {\it one half} of 
its original value 
(``quantum correction doubling''). 
  
\subsection{``Levitation and pair annihilation'' phenomena}
\begin{figure}
\begin{center}
\includegraphics[width=0.4\textwidth]{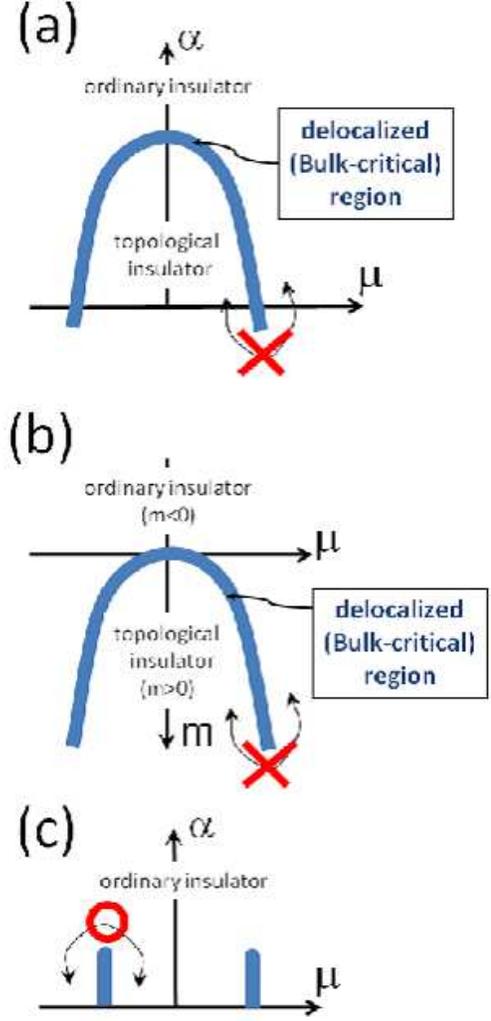}
\end{center}
\caption{(a); Onoda's phase diagram~\cite{OAN}  
in the $\alpha$-$\mu$ plane 
with $m>0$ (topological insulator side) 
(b); A schematic 
phase diagram in the $m$-$\mu$ plane, which 
is expected from the surface (edge) state's argument 
described in the section.~I. (c); A schematic 
phase diagram in the $\alpha$-$\mu$ plane 
of the ordinary insulator side ( $m<0$ ).   
In (a-c), we have 
two delocalized energy regions 
(blue filled regions), which locate 
at the center of the upper band and the lower band. 
In (a), these two delocalized regions eventually 
merge and annihilate in pair, when $\alpha$ increases. 
As a result, the topological insulator and ordinary insulator 
are always disconnected by the bulk-critical (delocalized) region.   
In (c), however, two delocalized regions registered at 
the upper band and the lower band annihilate without merging each other, 
when $\alpha$ increases. Thus, all the insulating regions appearing 
in (c) are adiabatically connected from one point to others.      
In (b), two delocalized regions merge and annihilate with 
each other, when the topological mass $m$ changes its sign from 
positive to negative. As a result, the topological insulator 
and ordinary insulator are again disconnected from each other by 
the bulk-critical (delocalized) region, as in (a).} 
\label{tlevi-1}
\end{figure}

Let us discuss open issues 
in the 3-d $Z_2$ QSH system in the view point of 
our findings. 
As a tightly related topic to
the stability of the QCP, the levitation and pair 
annihilations phenomena of the extended 
states\cite{AA}  
were recently observed 
in the 2-d $Z_2$ quantum spin Hall 
systems by Onoda {\it et al}.~\cite{OAN}  
They numerically studied the disorder effect on the
Kane-Mele model \cite{KM} on the honeycomb lattice.
In the clean case the system is set to be in the QSHI
phase; namely, the spectrum consists of two bands,
and there is a gap between them.
When the system is disordered, some states far from 
the band centers become localized, while
there are energy regions of delocalized states,
located at the centers of the upper (empty) band and 
lower (filled) band. What Onoda {\it et al.}~\cite{OAN} 
have found is that each of these two does not disappear 
{\it by itself}, when the disorder strength is increased. Instead, 
when the disorder becomes much stronger than 
the disorder strength for the localization in an ordinary insulator,
these two merge into one bundle of extended states 
energetically, and annihilate in pair (see Fig.~\ref{tlevi-1}(a)).     

\begin{figure}
\begin{center}
\includegraphics[width=0.45\textwidth]{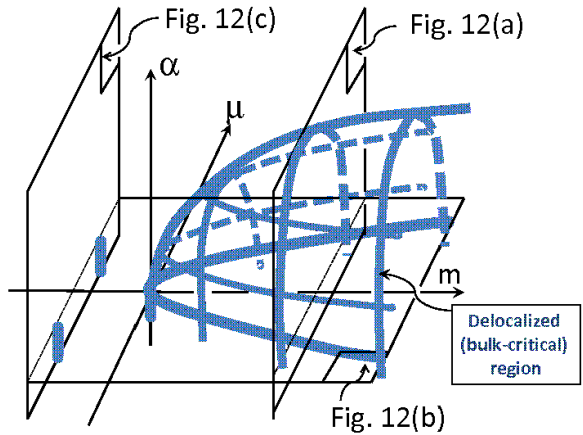}
\end{center}
\caption{A Schematic phase diagram in the 
$\alpha$-$\mu$-$m$ space.    
The vertical axis is the disorder strength $\alpha$, while 
the horizontal plane is spanned by the chemical potential 
$\mu$ and the topological mass $m$. The delocalized region 
(blue filled region) constitutes a {\it surface} 
in this 3-dimensional parameter space, so that 
an ordinary insulator phase and the topological insulator 
phase are adiabatically {\it disconnected} from each other.  
Namely, one cannot move from one phase to the other, without 
crossing the delocalized region, i.e. bulk-critical region.   
The phase diagram for the constant positive $m$ 
(topological insulator side) and that for the constant negative $m$ 
(ordinary insulator side) are separately 
described in Fig.~\ref{tlevi-1}(a) and (c) respectively. 
The phase diagram for the constant $\alpha$ corresponds to 
Fig.~\ref{tlevi-1}(b).} 
\label{tlevi-0}
\end{figure}   

To argue this phenomena  more generally, 
consider the 3-dimensional parameter space 
spanned by the topological mass term $m$, 
chemical potential $\mu$ and disorder 
strength $\alpha$. From the surface-state 
arguments described in the introduction,  
two insulating phases having 
different types of edge (surface) states, 
i.e. the topological insulator and an ordinary 
insulator, should be disconnected by the 
delocalized (bulk-critical) region. 
Then, when a finite topological mass term $m$ 
changes its sign from positive (topological 
insulator side) to negative (ordinary insulator side), 
we should also expect that 
a similar levitation and pair annihilation phenomena occurs.  
Namely, when a system transits from the topological 
insulator side to the ordinary insulator side, the region 
of extended states in the upper band and that in the lower band 
always merge and annihilate with each other    
(see Fig.~\ref{tlevi-1}(b)). Combining this picture 
with the Onoda's numerical observation~\cite{note1}, one 
can then expect that the delocalized (bulk-critical) 
region constitute a {\it surface} in the 3-d parameter space 
spanned by $\mu$, $m$ and $\alpha$, only to isolate  
the topological insulator phase from an ordinary insulator 
phase (see Fig.~\ref{tlevi-0}). 

\subsection{Possible microscopic scenario} 
Generally speaking, one has to go beyond 
our mean-field treatment of disorder 
in order to study the behaviors of mobility edges. 
However, we can still speculate the microscopic 
picture of the ``levitation and pair annihilation'' 
phenomena discussed above, in terms of 
the ``quantum correction doubling''  
found in this paper.  

We expect that the intervening 
bulk-critical region (blue filled region 
in Figs.~\ref{tlevi-1}(a-b) and  
Fig.~\ref{tlevi-0}) corresponds to the  
$\tau^{-1}_{\rm topo}\equiv 0 $ region.  
Namely, when a system transits from the topological insulator to 
an ordinary insulator, we surmise that {\it one of 
the high-energy modes, i.e. parity diffusion mode 
appearing in eq.~(\ref{dc12-11}), becomes massless once}, 
only to guarantee the existence of the 
bulk-critical region between these two insulating phases.  
This conjecture naturally leads to the following
microscopic scenario of the ``levitation and 
pair annihilation'' phenomena. 

\begin{figure}
\begin{center}
\includegraphics[width=0.45\textwidth]{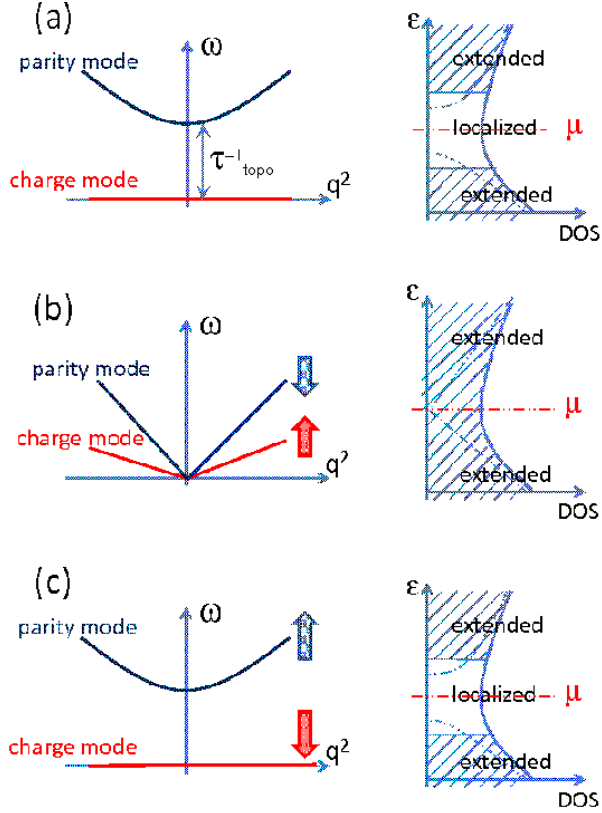}
\end{center}
\caption{(a); The low-energy spectrum in the topological insulator side 
contains two relevant diffusion modes. One is the usual charge diffusion 
mode, which disappears in the presence of relatively strong disorders. 
The other is the parity diffusion mode, which is protected by the infrared 
cut-off $\tau^{-1}_{\rm topo}$ from the disorders. (b); When 
a system transits from the topological insulator phase to the ordinary 
insulator phase, the parity diffusion mode becomes massless. Namely, when 
one further increases the disorder, starting from (a),  
$\tau^{-1}_{\rm topo}$ becomes renormalized by the disorder, only 
to reach zero. As a result, the Cooperon term 
corresponding to this parity diffusion mode becomes 
effective and induces the positive quantum interference effect on 
the backward scattering process of the charge-degrees of freedom. 
Because of this, the charge diffusion mode recovers at 
$\tau^{-1}_{\rm topo}=0$. (c); The low-energy spectrum  
in the ordinary insulator side. }
\label{le}
\end{figure}

Suppose that the  
${\cal T}$-symmetric 
disorder is introduced in 
the topological insulator. We assume that such disorder  
potential is strong enough to make the system localized. 
But it is not strong 
enough to make the upper (empty) band and 
low (occupied) band mixed with each other 
Namely, a system locates in the topological insulator 
side of Fig.~\ref{tlevi-1}(a), so that it 
can be adiabatically connected into the  
topological insulator phase in the clean limit.  
In such localized phase, we expect that  
a parity diffusion mode always exists in the 
high energy region and is {\it protected} by the 
infrared cut-off $\tau^{-1}_{\rm topo}$, while 
the charge diffusion mode disappear because of 
the relatively strong disorders    
(see Fig.~\ref{le}(a)).  
Starting from such localized phase, decrease the topological 
mass term (or further increase the disorder strength). 
Then, this infrared cutoff 
$\tau^{-1}_{\rm topo}$ associated with the parity diffusion mode 
decreases gradually, only to be renormalized to be zero at 
the transition point (see Fig.~\ref{le}(b)).  
Namely, at this transition point, 
the parity diffusion mode becomes massless. 
As a result, the Cooperon term corresponding to this 
parity diffusion mode, i.e. eq.~(\ref{dc12-13-d}),     
becomes effective and brings about the positive quantum 
interference effect on the back-scattering processes, 
in a same way as in the section.~IVD.  
Because of this positive quantum interference effect, 
which emerges only when the parity diffusion mode 
becomes massless, the charge diffusion constant recovers 
at around $\tau^{-1}_{\rm topo}\simeq 0$, 
even in the presence of the relatively strong disorder 
(see the red line in Fig.~\ref{le}(b)). 

However, when one further decreases the topological 
mass (or increases the disorder strength), 
the infrared cutoff $\tau^{-1}_{\rm topo}$
becomes finite again.  
As a result, this positive quantum interference effect 
due to the massless parity diffusion mode becomes 
ineffective again, only to lead a system into  
an insulating phase (see Fig.~\ref{le}(c)). 
This insulating phase is now adiabatically connected to 
an ordinary insulator in the clean limit.     
 
To uphold this microscopic picture, we need to consider
several ingredients missing in our approach~\cite{further}.  We will 
enumerate them in the following.  
As indicated in Figs.~\ref{tlevi-1}-\ref{tlevi-0},  
the pair-annihilation   
occurs only in the topological insulator side. 
Namely, the phase diagram is {\it asymmetric} 
with respect to the sign change of the 
topological mass term.  
On the other hand, all the findings 
in this paper are symmetric with respect to 
the sign change of this mass term. 
This is obviously because our starting model  
is the effective continuum model, 
describing only the {\it local} structure around 
a certain $k$-point.
On the other hand,
the $Z_2$ topological number 
is determined from the {\it global} information  
of the Bloch wavefunctions' phase 
in the $k$ space~\cite{FKM}. 
Therefore, in such an effective continuum model 
one cannot determine whether 
the topological insulator {\it by itself} 
corresponds to the $m>0$ phase or the $m<0$ phase. 
Instead, it simply dictates that   
one of these two should be the topological 
insulator, and the other  
is an ordinary insulator. 
As such, to describe the asymmetric 
behavior of the mobility edge as in 
Fig.~\ref{tlevi-0}, we clearly have to deal with 
lattice models.  

In the present work we treated disorder in the 
mean-field level, considering 
only the Cooperon 
correction. 
To verify the aforementioned scenario,  
we thus also need to 
deal with interactions among 
the various low-energy modes, 
beyond the mean-field treatment.   
In such situations, 
the inter-mode interaction
between the quasi-degenerate Goldstone modes  
found in the section~IV certainly plays an 
important role in  the ``levitation and 
pair annihilation'' phenomena. 
\begin{acknowledgments}
We are grateful to Leon Balents, A. P. Schnyder
, K-i Imura,  
Kentaro Nomura, Shinsei Ryu, Hideaki Obuse, Akira Furusaki 
and Hiroshi Kohno for helpful discussions. 
This research is supported in part 
by Grant-in-Aids  
from the Ministry of Education,
Culture, Sports, Science and Technology of Japan.  
RS was financially supported 
previously by the Osaka University and 
currently by the Institute of Physical and 
Chemical Research (RIKEN).   
Part of this work is done during the 
ISSP-YITP joint-workshop 
entitled as ``Topological Aspects of 
Solid State Physics (TASSP)''.    
\end{acknowledgments}

\appendix

\section{Effects of generic time-reversal invariant disorders}
\label{sec:Delta}
In this paper, we have restricted ourselves to the chemical potential type 
disorder for simplicity. However,  
there exist in general several other types of 
${\cal T}$-invariant disorder 
potentials, 
as described in section II (see eq.~(\ref{imp})). 
We basically expect 
that these additional time-reversal invariant disorders will not change our  
results drastically. To uphold this expectation, we study 
in this appendix how our self-consistent Born solution is modified 
in the presence of generic time-reversal invariant disorders, 
focusing on the zero-energy wavefunction at the critical point.  
As a result, we will obtain the following facts, which support 
this expectation. One is that, when only the diagonal 
correlations, $\Delta_{jj}$, are present, 
 our solutions derived in section ~III 
do not change at all (see eqs.(~\ref{type1},\ref{e18-1},\ref{g})).
When the off-diagonal correlation such as  
$\Delta_{05}$ is introduced, $F_5$ acquires a finite 
imaginary part, i.e. $F^{\prime\prime}_5\ne 0$, even at the 
zero-energy state at the critical point (see 
eqs.~(\ref{type1},\ref{e18-1},\ref{g})).
However, provided that $\Delta_{05}$ is not so large in comparison 
with the diagonal correlation such as $\Delta_{00}$, 
$\Delta_{55}$ and etc. the effect of the non-zero 
$F^{\prime\prime}_5$ is 
expected to be negligible.   
 
The generic ${\cal T}$-reversal invariant disorders 
bring about the coupling between $F_5$ and $F_0$ 
more explicitly in the self-consistent Born (scB) 
equation. Namely, instead of eqs.~(\ref{e26}-\ref{e27}), 
our scB equation reads;   
\begin{eqnarray}
&&\int_{0<|k|<1} d^3k\ \frac{A_{+} F_0 - B F_5}
{F^2_0 - \sum_{\mu=1}^5 F^2_{\mu}} 
= f_0 - F_0, \label{eb15} \\
&& \int_{0<|k|<1} d^3k\ 
\frac{ B F_0 - {A}_{-} {F}_5 
}{F^2_0 - \sum_{\mu=1}^5 F^2_{\mu}}
=f_5 - F_5, \label{eb16} 
\end{eqnarray}
where only the following three parameters are the 
relevant model parameters;    
\begin{eqnarray}
A_{\pm} &\equiv& \big\{ \Delta_{00}  + 
\Delta_{55} \pm \sum_{j\in \{15,\cdots,45\}} \Delta_{jj} \big\} 
\Lambda, \label{e18-0} \\
B &\equiv&  2\Delta_{05}\Lambda. \label{e18-1} 
\end{eqnarray}
The coefficients of $\hat{\gamma}_{1,2,3,4}$ 
in the 1-point Green function, on the other hand, are again  
free from renormalization;  
\begin{eqnarray} 
&& F_{1,2,3} \equiv f_{1,2,3}= -k_{1,2,3}, \ \ 
F_4 \equiv f_4 \equiv 0. \label{e17-2}
\end{eqnarray}  
In terms of $G$ defined in eqs.~(\ref{e22-1},\ref{e28}), 
We can rewrite eqs.~(\ref{eb15}-\ref{eb16}) more transparently;  
\begin{eqnarray}
\left\{\begin{array}{c}
2\pi(A_{+}F_0 - BF_5)\cdot G = f_0 - F_0, \\
2\pi(BF_0 - A_{-}F_5)\cdot G = f_5 - F_5. \\ 
\end{array}\right. \label{e21}
\end{eqnarray}

When it comes to the zero-energy wavefunction at the critical point, i.e. 
$f_0=f_5=0$, this coupled equation could be  
``diagonalized''; 
\begin{eqnarray}
\big(1-\eta_{\sigma}G\big)\big(\lambda_{\sigma}F_0-F_5\big)  = 0, \label{gap1}
\end{eqnarray}
with $\sigma=\pm$. $\eta_{\sigma}$ and $\lambda_{\sigma}$ are 
defined as follows 
\begin{eqnarray}
\lambda_{\pm} &\equiv& \frac{1}{B}
\big(\Delta_s \pm \sqrt{\Delta^2_s - B^2}\big),\label{g} \\ 
\eta_{\pm} &\equiv& -\big(\Delta_a \pm \sqrt{\Delta^2_s-B^2}\big),\label{tg}
\end{eqnarray}
with positive definite $\Delta_s$ and $\Delta_a$; 
\begin{eqnarray}
\Delta_s &\equiv& \frac{1}{2}\big(A_{+}+A_{-}\big) 
= \Delta_{00}\Lambda + \Delta_{55}\Lambda, \nonumber \\
\Delta_a &\equiv& \frac{1}{2} \big(A_{+} - A_{-}\big) 
= \sum_{j\in\{15,\cdots,45\}}\Delta_{jj}\Lambda 
. \nonumber  
\end{eqnarray}
Observing eqs.~(\ref{e9-2}), note also 
that $\Delta_s$ defined above is always 
greater than $|B|$ defined in eq.~(\ref{e18-1});
\begin{eqnarray}
\Delta^2_{s} - B^2 > 0.  \label{region}
\end{eqnarray} 

Eq.~(\ref{gap1}) with $\sigma = \pm$ can be trivially satisfied 
by $F_0=F_5\equiv 0$. In what follows, we will enumerate 
all possible non-trivial solutions of this coupled equation. Let 
us first write down the real part and imaginary part of Eq.~(\ref{gap1}) 
for both $\sigma=\pm$, separately. Noting 
that $\lambda_{\pm}$ and $\eta_{\pm}$ are real-valued,  
we have the following for $\sigma=+$,  
\begin{eqnarray}
\hspace{-0.4cm} 
\left[\begin{array}{cc}
1- \eta_{+}{\rm Re}G & \eta_{+}{\rm Im}G \\
-\eta_{+}{\rm Im}G & 1-\eta_{+}{\rm Re}G \\ 
\end{array}\right]\left[\begin{array}{c} 
\lambda_{+} F^{\prime}_0 -F^{\prime}_5 \\
\lambda_{+} F^{\prime\prime}_0 -F^{\prime\prime}_5 \\
\end{array}\right] = 0 \label{2by2+}. 
\end{eqnarray}
For $\sigma=-$, we have 
\begin{eqnarray}
\hspace{-0.4cm}
\left[\begin{array}{cc}
1- \eta_{-}{\rm Re}G & \eta_{-}{\rm Im}G \\
-\eta_{-}{\rm Im}G & 1-\eta_{-}{\rm Re}G \\ 
\end{array}\right]\left[\begin{array}{c} 
\lambda_{-} F^{\prime}_0 -F^{\prime}_5 \\
\lambda_{-} F^{\prime\prime}_0 -F^{\prime\prime}_5 \\
\end{array}\right] = 0 \label{2by2-}.  
\end{eqnarray}
Observing eq.~(\ref{region}), notice that $\lambda_{+}\ne \lambda_{-}$ 
in general. As such, $(F_0,F_5)$ 
cannot satisfy 
$F_5 = \lambda_{-} F_0$ and $F_5 = \lambda_{+} F_0$ {\it simultaneously}. 
Thus, when $F_5=\lambda_{-} F_0$ is adopted, the 
determinant of the $2\times 2$ matrix in 
eq.~(\ref{2by2+}) should be zero;
\begin{eqnarray}
\left|\begin{array}{cc}
1- \eta_{+}{\rm Re}G & \eta_{+}{\rm Im}G \\
-\eta_{+}{\rm Im}G & 1-\eta_{+}{\rm Re}G \\ 
\end{array}\right| = 0, 
\end{eqnarray} 
or equivalently
\begin{eqnarray}
1=\eta_{+} {\rm Re}G, \ \ {\rm Im}G=0. \nn 
\end{eqnarray}  
On the other hand, when 
$F_5=\lambda_{+} F_0$ holds true, we have the following 
in turn,  
\begin{eqnarray}
1=\eta_{-} {\rm Re}G, \ \ {\rm Im}G=0. \nn 
\end{eqnarray}   
We thus have the only two possible non-trivial solutions;
\begin{eqnarray} 
\left\{ 
\begin{array}{l}
{\rm (Bi)}: \ F_{5}=\lambda_{-} F_0, \ 1=\eta_{+} {\rm Re}G \ \ {\rm  and} \ \ {\rm Im}G=0, \\
{\rm (Bii)}: \ F_{5}=\lambda_{+} F_0, \ 1=\eta_{-} {\rm Re}G \ \ {\rm and} \ \ {\rm Im}G=0. \\
\end{array}\right. \nn
\end{eqnarray}

In either cases, ${\rm Im}G=0$ readily leads us to $a=0$ first. 
The reasoning of this was already  
described in section ~IIIA1. 
When $a=0$, 
the real part of the function $G$ becomes  
simplified; ${\rm Re}G_{|a=0} 
\equiv -2 + 2b{\rm ArcTan}\big[b^{-1}\big]$ (see eq.~(\ref{e24})).  
Thus, above two solutions will be transcribed into the following two;
\begin{eqnarray} 
b {\rm ArcTan}[b^{-1}]= 
\frac{1+2 \eta_{\mp}}{2 \eta_{\mp}}, \label{gap3} 
\end{eqnarray} 
with $F_{5}=\lambda_{\pm} F_0$ respectively.
   
\begin{figure}
\begin{center}
\includegraphics[width=0.45\textwidth]{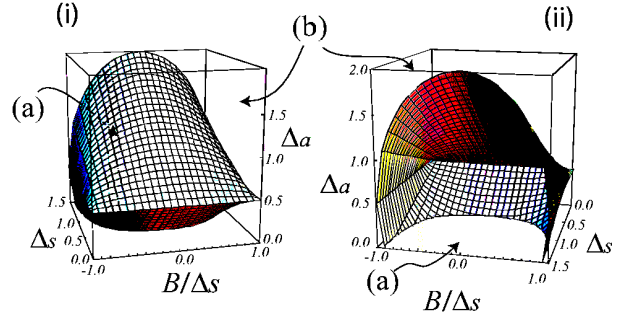}
\end{center}
\caption{The phase diagram of the scB solution in the presence of generic time-reversal 
invariant disorders. The region-$(a)$ includes the ``compressible phase'' argued in 
the section~III and IV, i.e. $\alpha>\alpha_c$ and $B=0$ and $\Delta_a$. 
The region-$(b)$ appears only when 
$\Delta_a\equiv \sum_{j=\{15,25,\cdots,45\}}\Delta_{jj}>0.5$. Note also that 
$\Delta_a,\Delta_s>0$ and $B<\Delta_s$ because of eq.~(\ref{e9-2})
 (see also eq.~(\ref{region})).}
\label{t100}
\end{figure}

Since the left hand side of eq.~(\ref{gap3}) is positive semi-definite, 
we have the following two parameter region supporting non-trivial solutions;
\begin{eqnarray} 
\left\{ 
\begin{array}{l}
(a): \ \eta_{+}<-\frac{1}{2}<\eta_{-}, \\ 
(b): \ \eta_{+}<\eta_{-}<-\frac{1}{2}. \\
\end{array}\right. \nn
\end{eqnarray} 
We also used $\eta_{-}>\eta_{+}$, which is trivially 
supported by eq.~(\ref{region}). These two parameter regions 
are depicted in Fig.~\ref{t100}, where the region-$(a)$ actually 
includes the ``compressible phase'' argued in the sections III and IV, 
i.e. $\alpha>\alpha_c$ and $B=\Delta_a=0$.    

In this region-$(a)$, only the type-(Bi) solution becomes possible;
\begin{eqnarray}
F_{5}=\lambda_{-}F_0, \ (a,b) \simeq \big(0,\frac{\pi + 2 \pi \eta_{+}}{4 \eta_{+}}\big). 
\end{eqnarray} 
Under $F^2_0-F^2_5\equiv (a+ib)^2$, this is identical to the following 
solution;   
\begin{eqnarray}
(F_{0},F_{5})=\pm i\frac{|b|}{\sqrt{1-\lambda^2_{-}}} (1,\lambda_{-}). \label{type1}
\end{eqnarray}  
This solution comprises continuously with the physical scB solution described 
in the section~III. 
Namely, when $\Delta_{a}$ taken 
to be zero, eq.~(\ref{type1}) 
precisely reduces to eq.~(\ref{e30}). 

When it comes to the region-$(b)$,   
type-(Bii) also becomes a possible solution;
\begin{eqnarray}
F_{5}=\lambda_{+}F_0, \ (a,b) \simeq \big(0,\frac{\pi+2\pi \eta_{-}}{4\eta_{-}}\big),
\end{eqnarray}
namely, 
\begin{eqnarray}
(F_{0},F_5)=\mp \frac{|b|}{\sqrt{\lambda^{-2}_{+}-1}}(\lambda^{-1}_{+},1). \label{type2}
\end{eqnarray}
In the absence of finite $B$ and $\Delta_{a}$, however, this solution is 
continued into eq.~(\ref{ty3}). Thus,  
this can never hold true therein. 
Because of this, we judge the type-(Bii) solution 
to be unphysical.  

\section{mode-mode coupling theory}
The weak-localization calculation (and also the self-consistent 
Born calculation) described in the section.~IV has the small 
coupling constant $1/(\mu\tau)$ only for the weakly disordered 
region, i.e. $\alpha <\alpha_c$, while it becomes an 
uncontrolled analysis for $\alpha > \alpha_c$. Bearing in mind 
this strong disorder region, we will employ in this appendix 
more phenomenological calculations, 
based on the mode-mode coupling theory~\cite{VW}. 
Without resorting to the Kubo formula, 
this theoretical framework gives us a 
mean-field equation for the diffusion constant $D$, 
where the quantum correction due to the Cooperon term  
are taken into account as in the standard 
weak-localization calculation~\cite{VW}.  The final results  
of this appendix.~B such as eqs.~(\ref{masslesssol},\ref{r5})  
indicate that this quantum correction becomes 
doubled, when the topological mass term $m$ are 
fined tuned to be zero.

The calculation consists of two steps. The first step 
begins with the Bethe-Salpeter (BS) equation for the 
response function 
$\Phi_{\alpha\delta,\gamma\beta}(k,k';q,\omega)$;  
\begin{widetext}
\begin{eqnarray}
&&\Phi_{\alpha\delta,\gamma\beta}(k,k';q,\omega) = \nn \\ 
&&\hspace{1cm} 
G^{R}_{\alpha\alpha_1}(k_+,\mu_{+})
G^{A}_{\beta_1\beta}(k_{-},\mu_{-}) \Big\{ - \frac{1}{2\pi i} 
\delta_{\alpha_1\delta}\delta_{\gamma\beta_1}\delta_{k,k'} 
+ \sum_{k_1} 
U^{\rm 2PIR}_{\alpha_1\delta_1,\gamma_1\beta_1}(k,k_1;q,\omega) \Phi_{\delta_1\delta,\gamma\gamma_1}
(k_1,k';q,\omega)\Big\}, \label{bs}  \\
&&\Phi_{\alpha\delta,\gamma\beta}(k,k';q,\omega)  
\equiv - \frac{1}{2\pi i}
\big\langle 
G^{R}_{\alpha\delta}(k_{+},k'_{+},\mu_{+})
G^{A}_{\gamma\beta}(k'_{-},k_{-},\mu_{-}) 
\big\rangle_{\rm imp}, \label{e60} 
\end{eqnarray} 
\end{widetext}  
with 
$k_{\pm} \equiv k \pm \frac{q}{2}$ and $\mu_{\pm} \equiv \mu \pm \frac{\omega}{2}$.    
Out of this equation, we first derive the linearized 
equations of motion (EOM's) for the density  
relaxation function $\phi_0(q,\omega)$, 
current relaxation function $\phi_j(q,\omega)$ 
and relaxation functions associated with 
other internal degrees of freedom;   
\begin{eqnarray}
\phi_0(q,\omega) &\equiv& \sum_{k,k'}\sum_{\alpha,\beta,\gamma}
\big[\hat{\gamma}_0\big]_{\beta\alpha}
\Phi_{\alpha\gamma,\gamma\beta}(k,k';q,\omega), \label{e61} \\
\phi_j (q,\omega) &\equiv& \sum_{k,k'}\sum_{\alpha,\beta,\gamma} 
\hat{q}_{\mu}\big[\hat{\gamma}_{\mu}\big]_{\beta\alpha}
\Phi_{\alpha\gamma,\gamma\beta}(k,k';q,\omega), \label{e62}  \\ 
&\cdots&  \nn 
\end{eqnarray}
with $\hat{q}$ normalized to be a unit vector. 
Since the EOM's thus derived 
are linearized, 
one can solve  
them for these relaxation functions, 
only to 
obtain their asymptotic expressions  
for the small $q$, $\omega$;     
\begin{eqnarray}
\phi_{0}(q,\omega) \sim \frac{1}{\omega + iDq^2}, \ \ 
\phi_{j}(q,\omega) \sim \frac{q}{\omega + iDq^2}, \cdots \label{e64-1} 
\end{eqnarray}
To be more specific, the diffusion constant $D$ appearing in 
the denominators will be expressed in 
terms of the {\it relaxation kernels}  
${\cal M}_{a,b}(q,\omega)$, latter of which are defined by 
the {\it two-particle irreducible} (2PIR) vertex function 
$\hat{U}^{\rm 2PIR}$ ({\bf step-(i)});  
\begin{eqnarray}
&&\hspace{-0.9cm} D \equiv i \bigg\{\frac{{\cal M}_{5j,5j}}
{{\cal M}_{j,j}{\cal M}_{5j,5j}-{\cal M}_{5j,j}
{\cal M}_{j,5j}}\bigg\}_{|q,\omega\equiv 0}.  \label{e65}  \\
&&\hspace{-0.9cm} 
{\cal M}_{a,b}(q,\omega) \equiv 2i \tau^{-1} \delta_{ab} + \frac{1}{2^4\pi}\times  
\nn \\
&&\hspace{-0.7cm}  
\sum_{k,k'}\big[\gamma^{\rm L}_{a}(k;q,\omega)\big]_{\beta\alpha} 
U^{\rm 2PIR}_{\alpha\delta,\gamma\beta}(k,k';q,\omega) 
\big[\gamma^{\rm R}_{b}(k';q,\omega)\big]_{\delta\gamma}. \label{e66} 
\end{eqnarray}
(see also eqs.~(\ref{leftvertex}-\ref{rightvertex}) 
for the definitions of 
$\hat{\gamma}^{\rm L,R}_{a}$). 

The 2PIR vertex function $\hat{U}^{\rm 2PIR}(k,k';q,\omega)$ in 
disordered media is usually dominated by the Cooperon 
at small $\omega$ and $k+k'$. The Cooperon 
is the series-sum of the ladder-type 
diagrams in the particle-particle channel, 
which is therefore obtained from the diffuson  
with the hole-line time-reversed.   
The diffuson is in turn responsible for 
the diffusion pole  
in the relaxation functions, i.e. the denominators 
in eq.~(\ref{e64-1}) . 
As such, in the presence of the 
${\cal T}$-symmetry,  
the asymptotic form of the Cooperon 
at small $\omega$ and 
$k+k'$ {\it should} be characterized 
by the {\it same} diffusion constant 
as that in eq.~(\ref{e64-1}). 
Based on this spirit,  
we will replace the 2PIR vertex 
function in eq.~(\ref{e66}) by this asymptotic 
form of the Cooperon. 
Through this approximation, 
eqs.~(\ref{e65}-\ref{e66})  
constitute a {\it self-consistent equation} for the 
diffusion constant $D$ 
({\bf step-(ii)}).

As we have seen explicitly in section.~IV,  
the diffuson consists of the charge diffusion mode 
and parity diffusion mode; 
\begin{eqnarray}
&&\hspace{-0.5cm} 
\hat{\Gamma}^d(q,\omega) \propto 
\frac{1}{\omega+iDq^2}\hat{\Gamma}^d_1 
+ \frac{1}{\omega+iDq^2+i\tau^{-1}_{\rm topo}}\hat{\Gamma}^d_2,  
\label{dif} 
\end{eqnarray}
with the positive semi-definite $\tau^{-1}_{\rm topo}$ 
proportional to $m^2$ (see 
eqs.~(\ref{topo},\ref{dc12-13-a}-\ref{dc12-13-b})). 
Namely, the second term, i.e. the parity diffusion mode, 
generally suffers from the {\it finite} infrared cut-off 
in the presence of the topological mass.   
while 
both of these two equally dominates the low-energy region 
for $m\simeq 0$;     
\begin{eqnarray}
\hat{\Gamma}^d(q,\omega) \propto 
\left\{\begin{array}{cc} 
\frac{1}{\omega + iDq^2} \big(\hat{\Gamma}^d_1 + \hat{\Gamma}^d_2\big) 
& \ \ {\rm for} \ \ \ Dl^{-2} \leq \tau^{-1}_{\rm topo}, \\ 
\frac{1}{\omega + iDq^2} \hat{\Gamma}^d_1  
& \ \ {\rm for} \ \ \ Dl^{-2} \geq \tau^{-1}_{\rm topo}, \\ 
\end{array} 
\right. \nn \\  
\label{sumdif} 
\end{eqnarray}  
(see eqs.~(\ref{f3},\ref{f4}) and 
eq.~(\ref{dc12-11})). In the presence of 
the ${\cal T}$-symmetry, this crossover behaviour will be 
transcribed onto the Cooperon term; the backward scattering 
process originated from the parity diffusion 
mode becomes ineffective, in the 
presence of the relatively large topological mass $m$; 
\begin{eqnarray}
&&\hspace{-0.5cm} 
\hat{U}^{\rm coop}(k+k',\omega) \propto  \nn \\ 
&&\hspace{-0.1cm} \left\{\begin{array}{cc} 
\frac{1}{\omega + iD(k+k')^2} \big(\hat{U}^{\rm c}_1 + 
\hat{U}^{\rm c}_2\big) 
& \ \ {\rm for} \ \ \ Dl^{-2} \leq \tau^{-1}_{\rm topo}, \\ 
\frac{1}{\omega + iD(k+k')^2} \hat{U}^{\rm c}_1  
& \ \ {\rm for} \ \ \ Dl^{-2} \geq \tau^{-1}_{\rm topo}, \\ 
\end{array} 
\right. \nn 
\end{eqnarray}  
(see eq.~(\ref{162d})).  

Corresponding to these two-mode features,  
we will derive in this appendix 
the two limiting gap equations; one is valid 
for $Dl^{-2} \geq \tau^{-1}_{\rm topo}$, while 
the other is for $Dl^{-2} \leq \tau^{-1}_{\rm topo}$.  
This appendix is organized as follows. 
The appendix.~B1 is devoted  
for the {\bf step-(i)}, in which the linearized coupled EOM's for the 
relaxation functions and eqs.~(\ref{e65}-\ref{e66}) will be derived. 
Using eqs~(\ref{e65}-\ref{e66}), we will derive 
in the appendix.~B2 the gap equations for the 
two-limiting cases ({\bf step-(ii)}). By solving 
these gap equations, we will finally see   
how the diffusion constant 
for $\alpha>\alpha_c$ behaves as a 
function of $\bar{\mu}$ 
and $m$ (see eqs.~(\ref{masslesssol},\ref{r5}) 
and Fig.~\ref{t23}).  
  
\subsection{Coupled EOM's for relaxation functions}
The EOM's derived henceforth are  
{\it linearized}  
with respect to the relaxation functions (unknown 
quantities). Namely, the mode-mode interactions among  
various bosonic degrees of freedom (density, current and so on) 
will be represented by the `mean-field' 
induced by the corresponding relaxation functions. 
This mean-field for the relaxation function is 
analogous to the self-energy for 
a 1-point Green function, so that 
it is often dubbed as the relaxation kernel~\cite{VW}.      
Being linearized,  such EOM's can be easily  
solved, only to let us express relaxation 
functions in terms of the relaxation kernel.  

These linearized EOM's  
also have to be {\it closed} with respect to  
a set of unknown relaxation functions.
Consider, for example, the EOM of the 
density relaxation function, which is 
nothing but the continuity equation. 
This equation contains 
the current relaxation function. Accordingly, to make  
coupled EOM's to be closed, we
further need the EOM for this current 
relaxation function, i.e. constitutive equation.   
The constitutive equation usually 
involves interactions between the current 
and other degrees of freedom (DOF's) 
such as the spin density, sublattice density and  
so forth. As such, 
we further need to derive the EOM's 
of the relaxation functions associated with 
these internal DOF's. In this way,  
we need to make our entire coupled EOM's 
to be {\it closed} with respect to 
a set of unknown relaxation functions.  
     
Let us begin with the continuity equation. Apply the following 
differential operator from the left hand side of the Bethe-Salpeter 
equation eq.~(\ref{bs}); 
\begin{eqnarray}
\delta_0 \hat{G}^{-1}(k;q,\omega) 
&\equiv& \hat{G}^{R,-1}(k_+,\mu_{+}) -\hat{G}^{A,-1}(k_-,\mu_{-})   \nn  \\ 
&=& \omega \hat{1} - q_{\lambda}\hat{\gamma}_{\lambda}
 - \hat{\Sigma}^{R}(\mu_{+}) + 
 \hat{\Sigma}^{A}(\mu_{-}). \nn 
\end{eqnarray}   
Taking the summation over repeated band indices, we then 
have;  
\begin{eqnarray}
&&\hspace{-0.5cm} 
\big[\omega\hat{1} - q_{\mu}\hat{\gamma}_{\mu} 
- \hat{\Sigma}^{R} + 
\hat{\Sigma}^{A}\big]_{\beta\alpha}
\Phi_{\alpha\gamma,\gamma\beta}(k,k';q,\omega) \nn \\
&&\hspace{-0.5cm}= \ - \big[\hat{G}^{R}(k_{+},\mu_{+}) - 
\hat{G}^{A}(k_{-},\mu_{-})\big]_{\beta_1\alpha_1} 
\Big\{-\frac{1}{2\pi i}\delta_{\alpha_1\beta_1} 
\delta_{k,k'}  \nn \\
&&\hspace{0.1cm} 
+ \sum_{k_1} U^{\rm 2PIR}_{\alpha_1\delta_1,\gamma_1\beta_1} 
(k,k_1;q,\omega)\Phi_{\delta_1\gamma,\gamma\gamma_1}(k_1,k';q,\omega)
\Big\}. \label{bs2}  
\end{eqnarray}
Under the integrals over $k$ and $k'$, the vertex function 
and the self-energy in eq.~(\ref{bs2}) set off each other;    
\bea
&&\hspace{-0.5cm} 
\omega\phi_{0}(q,\omega) - q\phi_{j}(q,\omega)
\nn \\   
&& = \frac{1}{2\pi i}\sum_{k',\gamma} 
\big\{\hat{G}^{R}_{\gamma\gamma}(k'_{+},\mu_{+}) 
- \hat{G}^{A}_{\gamma\gamma}(k'_{-},\mu_{-})\big\}.   
\label{bs3}
\eea
Namely, we used the following Ward identity;
\bea 
&& \big[\Sigma^{R}(k_{+},\mu_{+})- 
\Sigma^{A}(k_{-},\mu_{-})\big]_{\beta\alpha} \nn \\ 
&&\ \ \equiv  \sum_{k'} \delta 
\hat{G}_{\beta'\alpha'}(k';q,\omega)  
U^{\rm 2PIR}_{\alpha'\alpha,\beta\beta'}
(k',k;q,\omega), \nn 
\eea
with $\delta\hat{G}(k;q,\omega) \equiv 
\hat{G}^{R}(k_{+},\mu_{+}) 
- \hat{G}^{A}(k_{-},\mu_{-})$.

Recall that we are interested in the relaxation functions for 
sufficiently low-energy and long wave-length region; only 
to derive their diffusion pole structure. Thus, 
regarding $\omega$ and $q$ as sufficiently small 
quantities, we can replace the right hand side 
of eq.~(\ref{bs3}) by the spectral function;     
\begin{eqnarray}
\omega\phi_{0}(q,\omega) - q\phi_{j}(q,\omega) =  
A_0 + {\cal O}(q,\omega), \label{cont1}   
\end{eqnarray}
where $|A_0|$ stands for the density 
of state at $\epsilon=\mu$;
\begin{eqnarray}
A_0 \equiv \frac{1}{2\pi i}\sum_k {\rm Tr}
\big[\delta \hat{G}(k;0,0)\big] \simeq 
-16F^{\prime\prime}_0\Lambda.  \label{dos}  
\end{eqnarray}  
Eq.~(\ref{cont1})  
is nothing but the continuity equation.
%
%

The continuity equation derived above contains the current 
relaxation function. Thus, we need to next derive an equation 
of motion for this. The derivation goes along 
in a quite analogous way 
as that of the continuity equation. Specifically, 
to end up with an 
equation having $\omega\phi_{j}(q,\omega)$, 
we will apply the following 
onto the Bethe-Salpeter equation, 
instead of $\delta_0 \hat{G}^{-1}$;    
\begin{eqnarray}
\delta_j \hat{G}^{-1} (k;q,\omega)  
\equiv  \frac{1}{2} \big[
\delta_0 \hat{G}^{-1}(k;q,\omega), \ 
\hat{q}_{\mu}\hat{\gamma}_{\mu}\big]_{+}. \nn  
\end{eqnarray}
Since $\omega$ and $q$ being sufficiently small,  
we will keep only its leading-order contributions;  
\begin{eqnarray}
\delta_j \hat{G}^{-1} (k;q,\omega)  
\simeq  \Big(\omega 
\frac{\partial F'_0}{\partial \mu} + 2i F^{''}_0 \Big)
\hat{q}_{\mu}\hat{\gamma}_{\mu} - q \hat{1}. \label{bs33} 
\end{eqnarray}  
Apply this onto eq.~(\ref{bs}) and 
take the sum over $k$, $k'$ and 
the band indices. By way of this,  
we obtain the following constitutive equation; 
\begin{eqnarray}
&&\hspace{-0.5cm}
(\omega\frac{\partial F'_0}{\partial \mu} + 
2i F^{''}_0)\phi_{j}(q,\omega) - q\phi_{0}(q,\omega) \nn \\ 
&&\hspace{-0.1cm} = A_j -  
\sum_{k,k_1}\big[\hat{\gamma}^{\rm L}_{j}(k;q,\omega)\big]_{\beta_1\alpha_1} \nn \\
&&\hspace{-0.5cm} \times 
U^{\rm 2PIR}_{\alpha_1\delta_1,\gamma_1\beta_1}(k,k_1;q,\omega) 
\sum_{k'}\Phi_{\delta_1\gamma,\gamma\gamma_1}(k_1,k';q,\omega).  
\label{bs6} 
\end{eqnarray} 
$A_j$ and $\hat{\gamma}^{\rm L}_j(k;q,\omega)$ 
are defined as follows;
\begin{eqnarray}
&&\hspace{1.15cm} A_j \equiv \frac{1}{2\pi i}  
\sum_{k} {\rm Tr}\big[
\hat{\gamma}^{\rm L}_{j}(k;q,\omega)\big], \label{bs7} \\ 
&& \hspace{0.cm} \hat{\gamma}^{\rm L}_{j}(k;q,\omega) \equiv \frac{1}{2} \times \nn \\
&&\hspace{-0.8cm}  
\big\{\delta \hat{G}(k;q,\omega)
\cdot \hat{G}^{R,-1}(k_{+},\mu_{+})\cdot \hat{q}_{\mu}
\hat{\gamma}_{\mu} \cdot \hat{G}^{R}(k_{+},\mu_{+}) +  \nn \\  
&&\hspace{-0.6cm} \ \hat{G}^{A}(k_{-},\mu_{-})\cdot \hat{q}_{\mu}
\hat{\gamma}_{\mu}\cdot \hat{G}^{A,-1}(k_{-},\mu_{-})
\cdot \delta \hat{G}(k;q,\omega) \big\}. \label{bs7-1}
\end{eqnarray}

Contrary to the continuity equation, this equation 
of motion contains 
the convolution between the 2PIR vertex function 
and the response function explicitly.  
This convolution part describes  
the interactions  
between the current relaxation function and 
the other types of relaxation 
functions. We will linearize this convolution part 
with respect to relaxation functions in the 
following three paragraphs. 

\begin{figure}
\begin{center}
\includegraphics[width=0.5\textwidth]{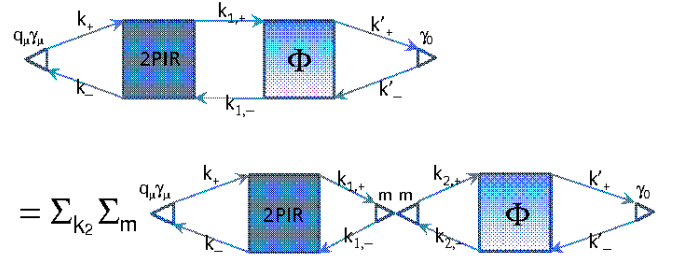}
\end{center}
\caption{The convolution between the 2PIR vertex function and 
the response function is replaced by the direct product 
between the relaxation kernels 
and relaxation functions, where we 
used the complete set for a function of $k_1$, i.e. 
$f(k_1) \equiv \sum_{k_2} \delta (k_1-k_2) f(k_2) \equiv  
\sum_{k_2} \sum_{m} u_m(k_1) \times u^{*}_m(k_2)f(k_2)$. Namely, 
$u_{m}$ constitutes the 
bare vertex part described by 
``$\triangleright$''-mark in the figure, 
while $u^{\ast}_{m}$ constitutes that 
described by ``$\triangleleft$''-mark}
\label{t12-1}
\end{figure}

To do this, note first the completeness of the 
$\gamma$-matrices, 
\begin{eqnarray}
\delta_{\gamma\gamma'}\delta_{\delta\delta'} \equiv 
\frac{1}{4}\sum_{\mu=\big\{0,1,\cdot,5,15,\cdots,42\big\}}\big[\hat{\gamma}_{\mu}\big]_{\gamma\delta}
\big[\hat{\gamma}_{\mu}\big]_{\delta'\gamma'} \label{bs9} 
\end{eqnarray}
and that of the spherical harmonic function $Y_{lm}(\hat{\Omega})$, 
\begin{eqnarray}
f(x\hat{\Omega}) \equiv \sum_{l=0}^{\infty}\sum_{m=-l}^{l} Y_{lm}(\hat{\Omega})
\sum_{\hat{\Omega}^{\prime}} Y^{*}_{lm}(\hat{\Omega}^{\prime}) 
f(x\hat{\Omega}^{\prime}),  
\label{bs10} 
\end{eqnarray}
where $\hat{\Omega}$ denotes 
the normalized vector and 
$\sum_{\hat{\Omega}}\cdots$  
stands for the 2-dimensional integral over 
the angle-direction; 
$\sum_{\hat{\Omega}}\equiv 4\pi$. 
Using these two completeness relations, 
we can decouple the convolution part in eq.~(\ref{bs6}) 
into the sum over the countable numbers of modes 
(see also Fig.~\ref{t12-1}); 
\begin{eqnarray}
&&\hspace{-0.7cm}
(\omega\frac{\partial F'_0}{\partial \mu} + 
2i F^{''}_0)\phi_{j}(q,\omega) - q\phi_{0}(q,\omega) \nn \\ 
&&\hspace{-0.5cm} = A_j -\frac{1}{4}  
\sum_{k,k_1}\big[
\hat{\gamma}^{\rm L}_{j}(k;q,\omega)
\big]_{\beta_1\alpha_1}
U^{\rm 2PIR}_{\alpha_1\delta_1,\gamma_1\beta_1}(k,k_1;q,\omega)  
\nn \\
&&\hspace{-0.5cm} \times \sum_{\mu=0}^{15} \sum_{l=0}^{\infty}\sum_{m=-l}^{l}
Y_{lm}(\hat{k}_1)  
\big[\hat{\gamma}_{\mu}\big]_{\delta_1\gamma_1} 
\overline{\phi}_{lm,\mu}(|k_1|;q,\omega). \label{bs6-1}
\end{eqnarray}
Namely, the $k_1$-dependence of the response function is 
decomposed 
into the dependence on its radial coordinate $|k_1|$ and 
that on the angle coordinate $\hat{k}_1$. At a price for this, 
the right hand side contains the summation over the azimuthal 
and magnetic quantum numbers, $l$ and $m$. 
For each $l$, $m$ and $\mu$, $\overline{\phi}_{lm,\mu}(x;q,\omega)$ 
is defined as follows;  
\begin{eqnarray}
&&\hspace{-0.6cm}
\overline{\phi}_{lm,\mu}(x;q,\omega) \equiv  \nn \\
&&\hspace{-0.34cm} 
\sum_{\hat{k}}\sum_{k^{\prime}} 
\sum_{\alpha,\beta,\gamma} 
\big[\hat{\gamma}_{\mu}\big]_{\beta\alpha} 
Y^{*}_{lm} (\hat{k}) \Phi_{\alpha\gamma,\gamma\beta}
(x\hat{k},k';q,\omega). \label{bs6-2}
\end{eqnarray}

Observing this definition, notice that   
the $x$-dependence of 
$\overline{\phi}_{lm,\mu}(x;q,\omega)$  and its $\omega,q$-dependence 
can be further {\it factorized} 
for {\it small} $\omega$ and $q$; 
\begin{eqnarray}
\overline{\phi}_{lm,\mu}(x;q,\omega) 
= g_{lm,\mu}(x) \phi_{lm,\mu}(q,\omega).  \label{bs13}
\end{eqnarray}   
This is because, for such small $\omega$ and $q$, the response 
function appearing in eq.~(\ref{bs6-2}) 
is dominated by the diffuson, which depends 
only on $\omega$ and $q$; 
\begin{eqnarray}
\Phi_{\alpha\delta,\gamma\beta}(k,k';q,\omega) 
&\simeq& -\frac{\alpha}{2\pi i} \hat{G}^{R}_{\alpha\alpha_1}(k_{+},\mu_{+})
\hat{G}^{A}_{\beta_1\beta}(k_{-},\mu_{-}) \nn \\
&&\hspace{-3.cm}\times \  
\big[\hat{\Gamma}^d(q,\omega)\big]_{\alpha_1\delta_1,\gamma_1\beta_1} 
\hat{G}^{R}_{\delta_1\delta}(k'_{+},\mu_{+})
\hat{G}^{A}_{\gamma\gamma_1}(k'_{-},\mu_{-}). \label{bs12} 
\end{eqnarray}  
By taking the integrals over 
$\hat{k}$ and $k'$ in eq.~(\ref{bs6-2}) 
and keeping only the leading order 
in small $\omega$ and $q$, 
one can actually verify this factorization   
for any $l$, $m$ and $\mu$     
(consult also the appendix.~D  
for several examples.) 

Without loss of generality, we 
can assume that $g_{lm,\mu}(x)$ thus obtained  
is normalized with respect to the integral  
over the radial direction;
\begin{eqnarray}
\int_{0}^{\Lambda} x^2 dx g_{lm,\mu}(x) \equiv 1. \nn 
\end{eqnarray}
Then, corresponding $\phi_{lm,\mu}(q,\omega)$ given 
in eq.~(\ref{bs13})
becomes %
the non-zero azimuthal number $(l\ne 0)$
generalizations of the relaxation functions defined 
in eqs.~(\ref{e61},\ref{e62});
\begin{eqnarray}
\phi_{lm,\mu}(q,\omega) \equiv 
\sum_{k,k^{\prime}} 
\sum_{\alpha,\beta,\gamma} 
\big[\hat{\gamma}_{\mu}\big]_{\beta\alpha} 
Y^{*}_{lm} (\hat{k}) \Phi_{\alpha\gamma,\gamma\beta}
(k,k';q,\omega). \label{bs666} 
\end{eqnarray}   
Thus, substitute 
eq.~(\ref{bs13}) back into eq.~(\ref{bs6-1}). Then,  
we finally obtain the constitutive equation,  
which is fully linearized 
with respect to these relaxation functions; 
\begin{eqnarray}
&&\hspace{-0.3cm}
(\omega\frac{\partial F'_0}{\partial \mu} + 
2i F^{''}_0)\phi_{j}(q,\omega) - q\phi_{0}(q,\omega) =  A_j - 
\frac{1}{4}  \nonumber \\
&&\hspace{-0.5cm}\   
\times \sum_{l=0}^{\infty}\sum_{m=-l}^{l}\sum_{\mu=0}^{15}
\Big\{\sum_{k,k'}
\big[\hat{\gamma}^{\rm L}_{j}(k;q,\omega) 
\big]_{\beta\alpha} 
U^{\rm 2PIR}_{\alpha\delta,\gamma\beta}
(k,k';q,\omega) \nn \\ 
&&\hspace{0.5cm} \times 
\big[\hat{\gamma}_{\mu}\big]_{\delta\gamma} 
g_{lm,\mu}(|k^{\prime}|) 
Y_{lm}(\hat{k}^{\prime})\Big\} \ \phi_{lm,\mu}(q,\omega). \label{bs15} 
\end{eqnarray}  
The second member of its right hand side 
described the `mean-field' 
induced by other relaxation functions, 
into which the 2PIR vertex function is encoded. 
This situation is quite analogous to how  
the 1-particle irreducible function (i.e. self-energy) 
describes the interaction among 
1-point Green functions.

Due to this interaction, however, 
the constitutive equation 
above also contains relaxation 
functions $\phi_{lm,\mu}(q,\omega)$ 
assigned to the higher order  
harmonic ($l \ge 1$) sector. 
Thus, to make the final coupled EOM's 
to be closed, we must also derive 
the EOM for all of 
these functions. This is, however, limitless.   
To make it tractable, we thus need to truncate 
interactions among these too many modes. 
In this paper, we will  
consider the interactions only 
within the ``$s$-wave'' sector. 
Namely, we will restrict the 
summation over $l$, $m$ and $\mu$ 
in eq.~(\ref{bs15}) 
to the $l=0$ sector;
\begin{eqnarray} 
\phi_{00,\eta}(q,\omega) &\equiv& \sum_{k,k'} 
\sum_{\alpha,\beta,\gamma} \big[\hat{\gamma}_{\eta}\big]_{\beta\alpha}
\Phi_{\alpha\gamma,\gamma\beta}(k,k';q,\omega). \nn 
\end{eqnarray} 

These 16 modes in the $s$-wave sector 
further reduce into the 8 modes, 
when the rotational symmetry is taken into account.  
Namely, by noting that the response 
function is invariant under  
the simultaneous rotation 
in the pseudo-spin space and in the 
momentum space; 
\begin{eqnarray} 
&&\hspace{-0.4cm} 
\big[\hat{U}_{n,\phi}\big]_{\alpha'\alpha}
\big[\hat{U}_{n,\phi}\big]_{\gamma'\gamma} 
\Phi_{\alpha\delta,\gamma\beta}(k,k;q,\omega)
\big[\hat{U}^{\dagger}_{n,\phi}\big]_{\delta\delta'}
\big[\hat{U}^{\dagger}_{n,\phi}\big]_{\beta\beta'}\nn \\
&&\hspace{1.4cm} 
\equiv \Phi_{\alpha'\delta',\gamma'\beta'}
(R_{n,\phi} k,R_{n,\phi} k';R_{n,\phi} q,\omega), \nn \\ 
&&\hspace{0.65cm} 
\hat{U}_{n,\phi} \equiv e^{\frac{\phi}{4} 
\epsilon_{\mu\nu\rho}n_{\mu}\hat{\gamma}_{\nu}\hat{\gamma}_{\rho}},  
\label{rot}
\end{eqnarray} 
we can derive the following identity; 
\begin{eqnarray}
&&\hspace{-1.1cm} 
\sum_{\mu=0}^{15} \big[\hat{\gamma}_{\mu}\big]_{\alpha\beta}
\phi_{00,\mu}(q,\omega) 
\equiv \frac{1}{\sqrt{4\pi}} 
\sum_{a=0}^{45j} 
\big[\hat{v}_{a}(q)\big]_{\alpha\beta}\phi_{a}(q,\omega), \label{red}
\end{eqnarray} 
with:
\begin{eqnarray} 
&&\hspace{-0.5cm} 
\phi_{a}(q,\omega) \equiv \sum_{k,k'} 
\sum_{\alpha,\beta,\gamma} \big[\hat{v}_{a}(q)\big]_{\beta\alpha}
\Phi_{\alpha\gamma,\gamma\beta}(k,k';q,\omega), \nn \\ 
&&\hspace{-0.5cm} (\hat{v}_{0},\hat{v}_{5},\hat{v}_{4},\hat{v}_{45}) 
\equiv \big(\hat{\gamma}_0,\hat{\gamma}_{5},\hat{\gamma}_4,\hat{\gamma}_{45}
\big) \nn \\ 
&&\hspace{-0.5cm} 
(\hat{v}_{j},\hat{v}_{5j},\hat{v}_{4j},\hat{v}_{45j})  
\equiv \big(\hat{q}_{\mu}\hat{\gamma}_{\mu},
\hat{q}_{\mu}\hat{\gamma}_{5\mu},
\hat{q}_{\mu}\hat{\gamma}_{4\mu},
\frac{1}{2}\hat{q}_{\mu}\epsilon_{\mu\nu\lambda}
\hat{\gamma}_{\nu\lambda}\big). \nn \\
\label{vde}
\end{eqnarray}
\begin{table}[htbp]
\begin{center} 
\begin{tabular}{|c||c|c|c||c||c|c|c|} \hline
label & density & ${\cal T}$ & ${\cal I}$ & label & ``current''  & ${\cal T}$ & ${\cal I}$  \\ \hline \hline 
``$0$'' & $\hat{\gamma}_{0}$ & $+$ & $+$ & ``$j$'' & $\hat{q}_{\mu}\hat{\gamma}_{\mu}$ & $-$ & $-$  \\ 
``$5$'' & $\hat{\gamma}_{5}$ & $+$ & $+$ & ``$5j$'' & $\hat{q}_{\mu}\hat{\gamma}_{\mu5}$ & $+$ & $-$  \\ \hline   
``$4$'' & $\hat{\gamma}_{4}$ & $-$ & $-$ & ``$4j$'' & $\hat{q}_{\mu}\hat{\gamma}_{\mu4}$ & $-$ & $+$  \\ 
``$45$''& $\hat{\gamma}_{45}$ & $+$ & $-$ &``$45j$'' & $\hat{q}_{\mu}\frac{1}{2}\epsilon_{\mu\nu\rho} 
\hat{\gamma}_{\nu\rho}$ & $-$ & $+$ \\ 
\hline 
\end{tabular}
\end{center}
\caption{Symmetry of eight modes in the $s$-wave sector and their symmetries  
under the spatial inversion ${\cal I}$ and the time-reversal ${\cal T}$. 
Since $\hat{\gamma}_0$, $\hat{\gamma}_4$, $\hat{\gamma}_5$ and 
$\hat{\gamma}_{45}$  
behave as a scalar quantity under the rotation 
defined in eq.~(\ref{a80}),  
we regard them as the ``density'' 
associated with the sublattice and spin 
degrees of freedom. Corresponding 
to these four types of density, 
we have 4 types of ``current'', 
which in turn behave as a vector 
quantity under the rotation. } 
\end{table}  
By use of this equality, 
eq.~(\ref{bs15}) turns out to 
consist only of those 8 functions 
defined in eq.~(\ref{vde});  
\begin{eqnarray}
&&\hspace{-0.6cm} 
\omega\frac{\partial F'_0}{\partial \mu} 
\phi_{j} - q\phi_0 
+ \sum_{a=0,5,\cdots,45j} {\cal M}_{j,a}(q,\omega) \phi_{a} =  A_j, \label{bs18} \\ 
&&\hspace{-0.6cm} 
{\cal M}_{a,b}(q,\omega) \equiv 
2iF^{\prime\prime}_0 \delta_{ab} + \ \frac{1}{2^4 \pi} \sum_{k,k'} \nn \\ 
&&\hspace{-0.7cm} 
\times 
\big[\hat{\gamma}^{\rm L}_{a}(k;q,\omega) 
\big]_{\beta\alpha} 
U^{\rm 2PIR}_{\alpha\delta,\gamma\beta}
(k,k';q,\omega) \big[\hat{\gamma}^{\rm R}_{b}(k';q,\omega)
\big]_{\delta\gamma}. \label{bs19}    
\end{eqnarray} 
$\hat{\gamma}^{\rm L,R}_{a}(k;q,\omega)$ above  
are given as follows; 
\begin{eqnarray}
&& \hspace{-0.8cm} \hat{\gamma}^{\rm L}_{a}(k;q,\omega) \equiv \frac{1}{2} \times \nn \\
&&\hspace{-0.6cm}  \big\{\delta 
\hat{G}(k;q,\omega)
\cdot \hat{G}^{R,-1}(k_{+},\mu_{+})\cdot \hat{v}_a(q) 
\cdot \hat{G}^{R}(k_{+},\mu_{+}) + \nn \\  
&&\hspace{-0.4cm}  \hat{G}^{A}(k_{-},\mu_{-})\cdot \hat{v}_a(q) 
\cdot \hat{G}^{A,-1}(k_{-},\mu_{-}) 
\cdot \delta \hat{G}(k;q,\omega)\big\} \label{leftvertex} \\
&&\hspace{-0.7cm} 
\hat{\gamma}^{\rm R}_{a}(k;q,\omega) \equiv \hat{v}_{a}(q) 
g_{00,a}(|k|). \label{rightvertex} 
\end{eqnarray}

${\cal M}_{a,b}(q,\omega)$ defined in eq.~(\ref{bs19}) 
generally appears in the EOM for the $v_a$-type relaxation 
function and plays role of the `mean-field' induced by 
the $v_{b}$-type relaxation functions. 
Namely, this $8\times 8$ matrix is nothing but 
the `self-energy' in the matrix-formed  
EOM's for the $s$-wave sector (see eq.~(\ref{eom1})).   
Thus, we will refer to $\hat{\cal M}(q,\omega)$ 
as the {\it relaxation kernel} henceforth. 
Before deriving the remaining 6 constitutive equations,  
let us remark on the general property of this kernel.    
Observing eq.~(\ref{bs19}), notice that 
each element of this matrix becomes 
pure imaginary, when its two arguments 
taken to be zero;  
\begin{eqnarray}
{{\cal M}_{a,b}(q,\omega)}^{*} = 
- {\cal M}_{a,b}(-q,-\omega).   \label{bs20} 
\end{eqnarray} 
This can be directly seen from;  
\begin{eqnarray}
\big\{\delta\hat{G}(k;q,\omega)\big\}^{*}  
\equiv - \big\{\delta\hat{G}(k;-q,-\omega)\big\}^t.  \nn \\
\big\{U^{\rm 2PIR}_{\alpha\delta,\gamma\beta}(k,k';q,\omega)\big\}^{*} 
= U^{\rm 2PIR}_{\beta\gamma,\delta\alpha}(k,k';-q,-\omega). 
\end{eqnarray}
\\

The EOM's for the other 6 $s$-wave modes 
can be derived in parallel with that for the current 
relaxation function. Specifically, 
we will 
begin with the Bethe-Salpeter equation applied 
by the following, instead of eq.~(\ref{bs33}); 
\begin{eqnarray}
\delta_{a} \hat{G}^{-1} (k;q,\omega)  
&\equiv&  \frac{1}{2} \big[
\delta_0 \hat{G}^{-1}(k;q,\omega), \ 
\hat{v}_{a}(q)\big]_{+}, \nn   
\end{eqnarray}
with $\hat{v}_a (q)$ taken to be 
$\hat{\gamma}_5,\hat{\gamma}_4, 
\cdots, \frac{1}{2}\hat{q}_{\mu}\epsilon_{\mu\nu\rho}
\hat{\gamma}_{\nu}\hat{\gamma}_{\rho}$ respectively. 
Going through the same procedure as described so far,  
we will reach the constitutive equations for these 
remaining 
6 modes. 
Combined with eq.~(\ref{cont1}) and eq.~(\ref{bs18}),  
such equations consist of the following $8$ by $8$ 
matrix-formed EOM's; 
\begin{eqnarray} 
\big[\hat{\cal K}(q,\omega) + \hat{\cal M}(q,\omega)\big] 
\cdot \hat{\phi}(q,\omega) \equiv  
\hat{A}(q,\omega). \label{eom1}  
\end{eqnarray}
$\hat{\phi}(q,\omega)$ and
$\hat{A}(q,\omega)$ have the eight components; 
\begin{eqnarray}
\hat{\phi}^{t} 
&\equiv& 
\big[ \phi_0,\ 
\phi_j,\  
\phi_5, \ 
\phi_{5j}, \ 
\phi_4, \  
\phi_{4j}, \ 
\phi_{45}, \  
\phi_{45j} 
\big],  \nn \\
\hat{A}^t 
&\equiv& \big[
A_{0}, \ 
A_{j}, \ 
A_{5}, \ 
A_{5j}, \ 
A_{4}, \ 
A_{4j}, \ 
A_{45}, \ 
A_{45j}\big],  \nn    
\end{eqnarray}
latter of which is defined as follows;
\begin{eqnarray}
A_{a}(q,\omega) \equiv \frac{1}{2\pi i}\sum_{k}
{\rm Tr}\big[\hat{\gamma}^{\rm L}_{a}(k;q,\omega)\big]. 
\label{othera}
\end{eqnarray} 
$\hat{\cal K}(q,\omega)$ and 
$\hat{\cal M}(q,\omega)$ are defined as follows; 
\begin{eqnarray} 
\hat{\cal K}&\equiv& \left[\begin{array}{cc}
\hat{\cal K}_1 & \hat{0} \\ 
\hat{0} & \hat{\cal K}_2 \\ 
\end{array}\right], \ \ \hat{\cal M} \equiv 
\left[\begin{array}{cc} 
\hat{\cal M}_1 & \hat{0} \\ 
\hat{0} & \hat{\cal M}_2 \\ 
\end{array}\right], \nn \\ 
\hat{\cal K}_1  
&\equiv& \left[ 
\begin{array}{cccc}
\omega & -q & &  \\
-q & \omega c^{-1} & & \\
\omega d^{-1}
 + 2i e^{-1} & &\omega c^{-1} &  \\
 & & & \omega c^{-1} \\ 
\end{array}\right], \label{eom2} \\  
\hat{\cal K}_2 &\equiv& \left[\begin{array}{cccc}
\omega c^{-1} & & & \\
& \omega c^{-1} & & \omega d^{-1}
 + 2i e^{-1} \\
& & \omega c^{-1} & -q \\
 &\omega d^{-1}
 + 2ie^{-1} & -q & \omega c^{-1} \\ 
\end{array} 
\right], \nn \\ 
\hat{\cal M}_1 &\equiv&  
\left[ 
\begin{array}{cccc}
0 &  & 0 &  \\
 & {\cal M}_{j,j} & & {\cal M}_{j,5j} \\
{\cal M}_{5,0}& &{\cal M}_{5,5} &  \\
 & {\cal M}_{5j,j} & & {\cal M}_{5j,5j}\\ 
\end{array}\right], \label{eom3} \\ 
\hat{\cal M}_2 &\equiv&  
\left[\begin{array}{cccc}
{\cal M}_{4,4} & & {\cal M}_{4,45} & \\
& {\cal M}_{4j,4j} & & {\cal M}_{4j,45j} \\
{\cal M}_{45,4} & & {\cal M}_{45,45} &  \\
& {\cal M}_{45j, 4j} &  & {\cal M}_{45j,45j} \\ 
\end{array}\right],  \nn
\end{eqnarray} 
with 
\begin{eqnarray}
c^{-1} \equiv \frac{\partial F^{\prime}_0}{\partial \mu},  
\ \ \ d^{-1} \equiv 
\frac{\partial F^{\prime}_5 }{\partial \mu},  \ \ \ 
e^{-1} \equiv F^{\prime\prime}_5. 
\end{eqnarray}  

By solving eq.~(\ref{eom1}),  
one can obtain the asymptotic 
expressions for the relaxation functions for 
small $\omega$ and $q$.; 
\begin{eqnarray}
&&\hspace{-1.5cm} 
\phi_{0}(q,\omega) \simeq \frac{A_0}{\omega + i D q^2}, \  
\phi_{5}(q,\omega) \simeq \frac{A_0 B_0 }{\omega + iDq^2}, \cdots \label{bs24} 
\end{eqnarray} 
where $|A_0|$ stands for the density of state 
(see eq.~(\ref{dos})).  The (renormalized) 
diffusion constant $D$ used above 
and other coupling constants are 
expressed only in terms of the 
relaxation kernels estimated 
at $\omega, q=0$;  
\begin{eqnarray}
&&\hspace{-0.6cm} 
D \equiv  i \bigg\{\frac{{\cal M}_{5j,5j}}{{\cal M}_{jj}{\cal M}_{5j,5j}
-{\cal M}_{5j,j}{\cal M}_{j,5j}}\bigg\}_{|q,\omega=0}. \label{bs25} \\ 
&&\hspace{-0.7cm}  
B_0 \equiv - \bigg\{
\frac{2iF^{\prime\prime}_5+{\cal M}_{5,0}}
{{\cal M}_{5,5}}\bigg\}_{q,\omega=0}, \label{bs26} 
\end{eqnarray} 
Eq.~(\ref{bs25}) and eq.~(\ref{bs19}) become 
the essential bulding blocks of our 
gap equation (see below).  

\subsection{gap equation and its solution} 
When the non-crossing approximation is employed for the 
1-point Green function, the diffusion pole  
in eq.~(\ref{bs24}) should be attributed to the  
ladder-type diagram whose long-wavelength expressions   
were already obtained in sec.~IV.  Especially, 
we have observed that in the section.~IVB  
that the charge diffusion  
mode and parity diffusion  
mode equally dominates the diffuson 
in the massless case ($Dl^{-2}\geq \tau^{-1}_{\rm topo}$),  
while the parity mode becomes ineffective in the presence of  
the relatively large topological mass 
($Dl^{-2}\leq \tau^{-1}_{\rm topo}$) 
(see eqs.~(\ref{dc12-11},\ref{f4},\ref{f3})).   
Corresponding to these two limiting cases, 
we will derive two types of gap equations and their 
solutions in this section.   
\subsubsection{for $m=0$ case} 
Let us begin with the zero topological mass case first.  
In this case,  we will sum up eq.~(\ref{dc12-13-a}) and 
eq.~(\ref{dc12-13-b}), since 
$f_3 \equiv f_4$. 
With use of eq.~(\ref{dc10-6}) and $a_{2,3,4}\equiv 0$, 
such a summand takes on a following form; 
\begin{eqnarray} 
\hat{\Gamma}^d(q,\omega)_{|F_5\equiv 0} 
= -\frac{a_0 f_4}{4} 
\Big(\hat{1}+\hat{T}_1 + \hat{S}_1 + \hat{S}_2 \Big).     
\label{masslessd}
\end{eqnarray}  
In section.~IVB, we have observed that 
the overall factor, $a_0 f_4$,    
has the diffusion pole as in eq.~(\ref{f4}), 
where its {\it bare} 
expression were calculated explicitly. 
Namely, by  keeping track of the small $q$ effect 
in eq.~(\ref{dc6-1}), 
we obtained the bare  
diffusion constant as in eq.~(\ref{baredif}).  
Instead of such bare expressions,  
however, we will describe henceforth this 
$a_0f_4$ in terms of the 
{\it renormalized} 
diffusion constant 
defined by eq.~(\ref{bs25}). 
Namely, we want $a_0f_4$ 
to be given by the 
relaxation kernels, only to 
obtain the {\it self-consistent} 
equation for the diffusion constant.   

To do this, notice that relaxation functions 
for {\it small} $q$ and $\omega$ are dominated by the diffuson 
as in eq.~(\ref{bs12}). 
Thus, by substituting eq.~(\ref{masslessd}) 
into eqs.~(\ref{bs12},\ref{e61}), 
we will first express 
the density relaxation 
function 
in terms of $a_0f_4$; 
\begin{eqnarray}
\phi_0(q,\omega) \simeq - 64\pi i \alpha \Lambda^2 a_0 f_4. \label{dc13} 
\end{eqnarray}
The factor $\Lambda^2$ in the  
right hand side stems from the momentum 
integral over $k$ and $k'$ in eq.(\ref{e61}); $\Lambda$ is the ultraviolet cutoff 
of the momentum-integral. Then, 
we will equate this with $\phi_0(q,\omega)$ 
obtained in the {\bf step-(i)}, i.e. eq.~(\ref{bs24}). 
By doing this, $a_0f_4$ is expressed in 
terms of relaxation kernels;  
\begin{eqnarray}
-64\pi i \alpha a_0 f_4 \equiv  \frac{1}{\Lambda^2} 
\frac{A_0}{\omega + iDq^2}. \label{dc14}
\end{eqnarray} 
Namely, the diffusion constant $D$ in the right hand 
side was already given by the relaxation kernels 
as in eq.~(\ref{bs25}). 

Substituting this back into eq.~(\ref{masslessd}), 
we obtain the asymptotic form of the diffusion;
\begin{eqnarray}
&& \hspace{-0.2cm} 
\hat{\Gamma}^d(q,\omega) \nn \\  
&& \hspace{0.0cm} 
= - \frac{ a_0 f_4}{4}
\Big(\hat{1}+\hat{T}_1+\hat{S}_1+\hat{S}_2\Big) \nn \\ 
&& \hspace{0.0cm} 
= \frac{1}{2^8 \pi \alpha i} \frac{1}{\Lambda^2} 
\frac{A_0}{\omega + iDq^2}
\Big(\hat{1}+\hat{T}_1+\hat{S}_1+\hat{S}_2\Big). \label{dc14-1}
\end{eqnarray}
When its hole line time-reversed, the corresponding  
Cooperon at small $\omega$ and $k+k'$ is also 
derived;   
\begin{eqnarray}
&&\hspace{-0.7cm}\hat{U}^{\rm coop}(k+k',\omega) \nn \\
 \nn \\  
&& \hspace{-0.5cm} 
= - \frac{ \alpha a_0 f_4}{4}
\Big(\hat{1}-\hat{T}_1-\hat{S}_1+\hat{S}_2\Big) \nn \\ 
&&\hspace{-0.5cm} 
= \frac{1}{2^8\pi i} \frac{1}{\Lambda^2} 
\frac{A_0}{\omega + iD(k+k')^2} 
\Big(\hat{1}-\hat{T}_1 -\hat{S}_1+\hat{S}_2\Big),   \label{c7}  
\end{eqnarray}
where we used  the following identities;
\begin{eqnarray}
\big[\hat{1}\otimes \hat{s}_y\big]_{\gamma^{\prime}\gamma}
\hat{T}_{1,\alpha\delta,\beta\gamma}
\big[\hat{1}\otimes \hat{s}_y\big]_{\beta\beta^{\prime}} 
&\equiv& - \hat{T}_{1,\alpha\delta,\gamma^{\prime}\beta^{\prime}}, \nn \\  
\big[\hat{1}\otimes \hat{s}_y\big]_{\gamma^{\prime}\gamma}
\hat{S}_{1,\alpha\delta,\beta\gamma}
\big[\hat{1}\otimes \hat{s}_y\big]_{\beta\beta^{\prime}} 
&\equiv& - \hat{S}_{1,\alpha\delta,\gamma^{\prime}\beta^{\prime}},  \nn \\ 
\big[\hat{1}\otimes \hat{s}_y\big]_{\gamma^{\prime}\gamma}
\hat{S}_{2,\alpha\delta,\beta\gamma}
\big[\hat{1}\otimes \hat{s}_y\big]_{\beta\beta^{\prime}} 
&\equiv& \hat{S}_{2,\alpha\delta,\gamma^{\prime}\beta^{\prime}}.  \nn  
\end{eqnarray}

The diffusion constant $D$ in eq.~(\ref{c7}) 
is now given by the relaxation kernels, 
via eq.~(\ref{bs25}). These relaxation kernels
are in turn defined by the 2PIR vertex function, via eq.~(\ref{bs19}). 
The 2PIR vertex function is usually dominated by 
the Cooperon given by eq.~(\ref{c7}), 
at around $k+k'\simeq 0$.   
As such, we will replace (approximate) 
the 2PIR vertex function in eq.~(\ref{bs19}) 
by this 
asymptotic form of the Cooperon, i.e. eq.~(\ref{c7}). 
By way of this, we obtain  
closed coupled equations for the (renormalized) 
diffusion constant $D$; 
\begin{eqnarray}
D &\equiv&  i \frac{{\cal M}_{5j,5j}}{{\cal M}_{jj}{\cal M}_{5j,5j}
-{\cal M}_{5j,j}{\cal M}_{j,5j}}, \label{sc1}  \\ 
{\cal M}_{a,b} &\equiv & 2iF^{\prime\prime}_0 \delta_{a,b}
+  \frac{|A_0|}{2^{12} \pi^2 D \Lambda^2} 
\int \int _{L^{-1}<|k+k'|<l^{-1}}  
d^3 k d^3 k' \ \nn \\
&& \hspace{-1.5cm}  
\frac{\big\{\hat{\gamma}^{\rm L}_{a}(k)\big\}_{\beta\alpha} 
\big\{\hat{1}-\hat{T}_1 -\hat{S}_1+\hat{S}_2\big\}_{\alpha\delta,\gamma\beta}
\big\{\hat{\gamma}^{\rm R}_{b}(k')
\big\}_{\delta\gamma}}
{|k+k'|^2} . \label{sc2}  
\end{eqnarray}
Since eq.~(\ref{c7}) is valid only for small $|k+k'|$, 
we have imposed the additional constraint $|k+k'|<l^{-1}$ 
into these integral variables. One might regard  
this upper limit as the mean-free path. 
We have already taken in eq.~(\ref{sc2}) both 
$\omega$ and $q$ to be zero.  
Thus, $\hat{\gamma}^{\rm L,R}(k)$ in the right 
hand side stands for  
$\hat{\gamma}^{\rm L,R}(k;q,\omega)$ estimated 
there;  
\begin{eqnarray}
\hat{\gamma}^{\rm L}_{a}(k)
&\equiv& \hat{\gamma}^{\rm L}_a(k;0,0), \nn \\
\hat{\gamma}^{\rm R}_{a}(k) & \equiv & 
\hat{v}_{a}\times g_{00,a}(|k|), \nn 
\end{eqnarray}
where $\hat{v}_a$ was already defined in eq.~(\ref{vde}). 
The normalized real-valued function $g_{00,a}(x)$ used above  
is given only in terms of $F_0$. 
For example, $g_{00,j}(x)$ is given as follows,
\begin{eqnarray}
g_{00,j}(x) \equiv \frac{4}{\pi}\frac{1}{{\cal N}_j}\frac{1}{|F^2_0-x^2|^2}
\Big\{1+\frac{8}{3}\frac{x^2 {F^{\prime}_0}^2}{|F^2_0-x^2|^2}\Big\}, \label{g00j}
\end{eqnarray}  
with its normalization factor $N_{j}$; 
\begin{eqnarray}
{\cal N}_j &=& \frac{1}{F^{\prime\prime}_0}
\Big\{1 + \frac{1}{3}\Big(\frac{F^{\prime}_0}{F^{\prime\prime}_0}\Big)^2\Big\}. \label{ng00j} 
\end{eqnarray}  
(see the appendix.~D for its derivation). Thus,  
eqs.~(\ref{sc1}-\ref{sc2}) constitute closed coupled equations 
for the diffusion constant. 

To solve this gap equation, notice first that the coupling  
between the current and the $\hat{\gamma}_5$-type current 
is disconnected in the massless case; 
${\cal M}_{j,5j}= 0$. 
This can be seen directly from  
\begin{eqnarray}
\big(\hat{1}-\hat{T}_1-\hat{S}_1 +\hat{S}_2\big)_{\alpha\delta,\gamma\beta}
\hat{q}_{\mu}\big[\hat{\gamma}_{\mu 5}\big]_{\delta\gamma} &=& 0. \label{use1}   
\end{eqnarray} 
which leads to $D \equiv i/{\cal M}_{j,j}$. 
As a result of this, eqs.~(\ref{sc1}-\ref{sc2}) 
becomes linear in $D$; 
\begin{eqnarray} 
&&\hspace{-0.7cm} 
\frac{1}{D} = 2F^{\prime\prime}_0 -   
\frac{A_0 F^{\prime\prime}_0}{2^7 \pi^2 D} \frac{1}{\Lambda^2} 
\int \int_{L^{-1}<|k+k'|<l^{-1}} 
d^3 k d^3 k' \nn \\ 
&&\hspace{-0.9cm} \times \  
 \frac{g_{00,j}(|k'|)}{|k+k'|^2} \frac{(-(F^{\prime}_0)^2
-(F^{\prime\prime}_0)^2+k^2)-2(k\cdot \hat{q})^2}
{((F^{\prime}_0)^2-(F^{\prime\prime}_0)^2-k^2)^2
+4(F^{\prime}_0)^2(F^{\prime\prime}_0)^2}. \label{ml1} 
\end{eqnarray}  

Using eqs.~(\ref{g00j}-\ref{ng00j}), 
we can readily evaluate the momentum integral in 
the right hand side of eq.~(\ref{ml1}). To do this, 
introduce a new integral variable $q'\equiv k+k'$, 
so that 
$dk dk' \equiv dk dq'$. Moreover, 
we approximate $g_{00,j}(|k-q'|)$ in 
the integrand by $g_{00,j}(|k|)$, since $g_{00,j}(x)$ 
is a slowly varying function in the scale of $l^{-1}$.  
These treatments give us the following 
expression for $2D\tau^{-1} $;
\begin{eqnarray}
2D \tau^{-1} \equiv  1 + \frac{1}{6} 
\frac{l^{-1}-L^{-1}}{\Lambda}  
\frac{\tau^{-2}+\frac{1}{2}\overline{\mu}^2}
{\tau^{-2}+\frac{1}{3}\overline{\mu}^2},   \label{masslesssol}
\end{eqnarray} 
where we used 
$A_0 \equiv -16F^{\prime\prime}_0\Lambda$ 
and $F_0 \equiv \bar{\mu} + i\tau^{-1}$.  
Observing this expression, notice  
that the second member of the r.h.s. 
is nothing but the quantum correction to 
the diffusion constant, which basically 
corresponds to the AWL correction to 
the conductivity. 
\begin{figure}
\begin{center}
\includegraphics[width=0.4\textwidth]{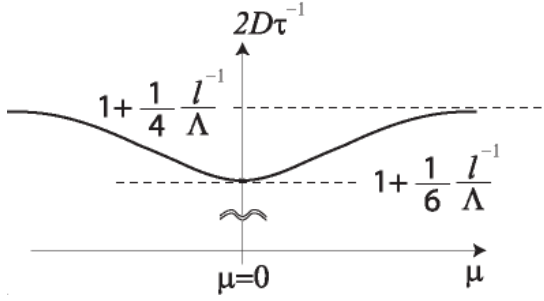}
\end{center}
\caption{A schematic plot of $2D\tau^{-1}$ as a function of the 
chemical potential in the massless case.} 
\label{t22}
\end{figure}
%
%
%
\subsubsection{for $m \ne 0$ case}
In the presence of relatively large topological mass, i.e. 
$Dl^{-2} \le \tau^{-1}_{\rm topo}$,  
the parity diffusion mode becomes 
the high-energy degree of freedom. As such, 
only the first term in eq.~(\ref{dc12-11})  
contributes the diffuson. 
With eq.~(\ref{dc10-5}), such diffuson is 
given as follows; 
\begin{eqnarray} 
&& \hspace{-0.5cm} 
\hat{\Gamma}^d(q,\omega)_{|F_5\ne 0}  
= - \frac{s f_4}{8} \Big\{ \frac{1+t^2}{2} 
\big(\hat{1}+\hat{T}_4\big) 
\big(\hat{1}+\hat{S}_1\big) \nn \\
&&\hspace{-1.1cm} + \frac{1-t^2}{2} \big(\hat{1}+\hat{T}_4\big)
\big(\hat{T}_1+\hat{S}_2\big)
+  t \big(\hat{T}_2+\hat{T}_3\big)
\big(\hat{1}+\hat{S}_1\big)\Big\}, \label{massived1}  
\end{eqnarray} 
where $s$ and $t$ are defined by $a_{04}$, $a_1$ and 
$a_{23}$;
\begin{eqnarray}
a_{04} \equiv s\frac{1+t^2}{2}, \ \ 
-3a_{1} \equiv s \frac{1-t^2}{2}, \ \ 
-a_{23} \equiv s t. \nn 
\end{eqnarray}
Contrary to the previous subsection, 
the tensor-part of the diffuson  
depends on the model-parameters 
through the ``tensor-form factor'' $t$. 
As such, we will 
employ in this case not only the 
diffuson constant $D$ but also this tensor-form factor  $t$
as the ``mean-field parameters'', which should be 
self-consistently determined. 
In other words, both of them should  
be given by the 
relaxation kernels, as in eq.~(\ref{bs25}).  

To do this, we will first calculate  
both the density 
relaxation function $\phi_0$ and 
the sublattice density relaxation 
function $\phi_5$,  
by the use of eq.~(\ref{massived1}).  
Namely, we will substitute eq.~(\ref{massived1}) 
into eqs.~(\ref{bs12},\ref{vde}), 
only to obtain these two functions in terms 
of $sf_4$ and $t$ first.  
The relaxation functions 
thus calculated read as follows;   
\begin{eqnarray} 
\phi_0(q,\omega) &\simeq& - 32\pi \alpha \Lambda^2  i s f_4 , \label{0IVE} \\ 
\phi_5(q,\omega) &\simeq& 32\pi \alpha \Lambda^2 i s t f_4.  \label{5IVE}  
\end{eqnarray}   
Then, we will equate eqs.~(\ref{0IVE},\ref{5IVE}) 
with the first two members of  
eq.~(\ref{bs24}) respectively.   
By way of this, $f_4$ and $t$ can be 
given in terms of the relaxation kernels;  
\begin{eqnarray}
 s f_4 &\equiv& 
\frac{1}{32\pi \alpha}\frac{1}{\Lambda^2} 
\frac{|A_0|}{i\omega  - Dq^2}, \label{dc12-2} \\ 
t &\equiv& - B_0 \equiv 
\bigg\{\frac{2iF^{\prime\prime}_5+{\cal M}_{5,0}}
{{\cal M}_{5,5}}\bigg\}_{|q,\omega=0}. \label{dc12-5}
\end{eqnarray}   

By substituting these two 
back into eq.~(\ref{massived1}), 
we can express the 
diffuson only in terms of the relaxation kernels.   
When its hole-line 
time reversed, the corresponding Cooperon 
is readily derived; 
\begin{eqnarray}
&&\hspace{-0.4cm} U^{\rm coop}(k+k',\omega) \simeq  
 \nn \\ 
&&\hspace{-0.9cm} \frac{1}{2^9 \pi  i} \frac{1}{\Lambda^2} 
\frac{A_0}{\omega + iD(k+k')^2} 
\Big\{(1+t^2)
\big(\hat{1}+\hat{T}_4\big) 
\big(\hat{1}-\hat{S}_1\big) \nn \\
&&\hspace{-1.2cm} 
- (1-t^2)
\big(\hat{1}-\hat{T}_4\big)\big(\hat{T}_1-\hat{S}_2\big) 
+ 2t \big(\hat{T}_2+\hat{T}_3\big)\big(\hat{1} 
- \hat{S}_1\big)\Big\}. \label{c6}
\end{eqnarray} 
The tensor-form factor $t$ and 
the diffusion constant $D$ appearing in 
the right hand side above are already given by 
the relaxation kernels, 
via eq.~(\ref{dc12-5}) 
and eq.~(\ref{bs25}).  
Such relaxation kernels are 
given by the 2PIR vertex 
function (see eq.~(\ref{bs19})).  
Thus, as in the previous subsection, 
we will approximate the 
2PIR vertex function by eq.~(\ref{c6}). 
In terms of this substitution, 
we arrive at a closed coupled equation 
for the diffusion 
constant $D$ and the tensor-form factor $t$, 
whose explicit expressions are given 
in the appendix.~C.   

\begin{figure}
\begin{center}
\includegraphics[width=0.4\textwidth]{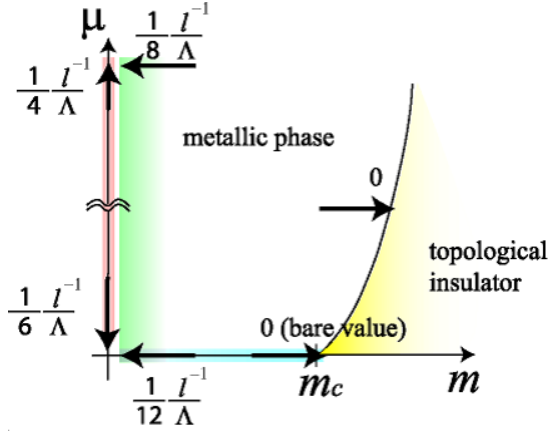}
\end{center}
\caption{A summary of several limiting values of the quantum 
correction of the diffusion constant 
in the presence of the topological mass.  
Eq.~(\ref{r5}) corresponds to the green shaded region.}
\label{t23}
\end{figure}

When solving this gap equation, we can 
see how the quantum correction to the diffusion 
constant behaves as a function of $m$ and 
$\mu$. Several limiting values are 
summarized in Fig.~\ref{t23}. 
Especially, in the zero mass limit 
i.e. $m=0+$, the solution of the gap equation 
reduces to a following simple function of  
$F_0\equiv \bar{\mu} + i\tau^{-1}$; 
\begin{eqnarray}
\lim_{\overline{m}\rightarrow 0+} 2D\tau^{-1} 
= 1 + \frac{1}{12}\frac{l^{-1}-L^{-1}}{\Lambda} 
\frac{\tau^{-2}+\frac{1}{2}{\overline{\mu}}^2}
{\tau^{-2}+\frac{1}{3}{\overline{\mu}}^2}, \label{r5}  
\end{eqnarray} 
Comparing this with eq.~(\ref{masslesssol}), one can easily 
see that the quantum correction to the diffusion constant 
is actually {\it half} of that 
for $m=0$ case. 

The discrepancy between eq.~(\ref{masslesssol})
and eq.~(\ref{r5}) is responsible for the Cooperon 
term associated with the parity diffusion mode, 
i.e. the second 
member of the right hand side of eq.~(\ref{162d}). 
To see this explicitly, note 
first that $a_{2}$, $a_3$ and $a_4$ appearing in 
eqs.~(\ref{dc12-13-c}-\ref{dc12-13-d})
reduce to zero in the limit of $m\rightarrow 0+$. 
Then, 
eqs.~(\ref{dc12-13-c},\ref{dc12-13-d})
in this limit read as follows;
\begin{eqnarray}
&& \lim_{m\rightarrow 0+} \hat{U}^{\rm c}_1   
= - \Big\{\big(\hat{1}-\hat{T}_1-\hat{S}_1 + \hat{S}_2\big) \nn \\ 
&&\hspace{2.0cm} + \ 
\hat{T_4}\cdot \big(\hat{1}+\hat{T}_1-\hat{S}_1-\hat{S}_2\big)\Big\}, \label{r25} \\ 
&&\lim_{m\rightarrow 0+} \hat{U}^{\rm c}_2   
= - \Big\{\big(\hat{1}-\hat{T}_1-\hat{S}_1 + \hat{S}_2\big) \nn \\ 
&&\hspace{2.0cm} - \ 
\hat{T_4}\cdot \big(\hat{1}+\hat{T}_1-\hat{S}_1-\hat{S}_2\big)\Big\}. \label{r26} 
\end{eqnarray}
Observing these expressions,  
notice that the second members 
of 
both eq.~(\ref{r25}) and eq.~(\ref{r26})  
are totally ineffective in the current-type relaxation kernels;  
\begin{eqnarray} 
&& \big\{\hat{T_4}\cdot\big(\hat{1}+\hat{T}_1-\hat{S}_1-\hat{S}_2\big)
\big\}_{\alpha\delta,\gamma\beta}\hat{q}_{\mu}
\big[\hat{\gamma}_{\mu}\big]_{\delta\gamma} = 0, \label{use2} \\ 
&& \big\{\hat{T_4}\cdot\big(\hat{1}+\hat{T}_1-\hat{S}_1-\hat{S}_2\big)
\big\}_{\alpha\delta,\gamma\beta}\hat{q}_{\mu}
\big[\hat{\gamma}_{\mu5}\big]_{\delta\gamma} = 0, \label{use3} 
\end{eqnarray} 
The consequence of these two equations are 
two-fold. The second equation 
in combination with 
eq.~(\ref{use1}) leads 
${\cal M}_{j,5j}\equiv 0$ first. Thus, we have 
$D\equiv \frac{i}{{\cal M}_{jj}}$ again, which 
indicates that   
${\cal M}_{j,j}$ originated from eq.~(\ref{r25}) 
and that from eq.~(\ref{r26}) contribute to 
the quantum correction in an {\it additive} way. 
Eq.~(\ref{use2}) moreover indicates that 
these two quantum corrections have 
the {\it same magnitude} and {\it sign}.
In other words, the quantum correction derived 
in Appendix.~B2a, i.e. eq.~(\ref{masslesssol}), 
can be divided into two parts;
\begin{eqnarray}
&&\hspace{-0.5cm} 2 {D}_{|\overline{m}=0}\tau^{-1} = \nn \\
&&\hspace{-0.2cm} 1 + \frac{1}{12}\frac{l^{-1}-L^{-1}}{\Lambda}
\frac{\tau^{-2}+\frac{1}{2}\overline{\mu}^2}{\tau^{-2}+\frac{1}{3}\overline{\mu}^2}
+ \frac{1}{12}\frac{l^{-1}-L^{-1}}{\Lambda}
\frac{\tau^{-2}+\frac{1}{2}\overline{\mu}^2}{\tau^{-2}+\frac{1}{3}\overline{\mu}^2}. \nn 
\end{eqnarray}  
Each of these two quantum corrections is 
originated from eq.~(\ref{r25}) 
and eq.~(\ref{r26}) respectively. Since we have
already ignored eq.~(\ref{r26}) for the $m \ne 0$ case, 
the resulting solution has only 
single $1/12$, as in eq.~(\ref{r5}).  

\section{mean-field equation for $m\ne 0$ case} 
The mean-field equation for the diffusion constant 
$D$ and $t$ in the presence of the 
finite topological mass $m$ is given as follows;
\begin{eqnarray}
&& \hspace{-0.5cm} 
D \equiv  i \frac{{\cal M}_{5j,5j}}{{\cal M}_{jj}{\cal M}_{5j,5j}
-{\cal M}_{5j,j}{\cal M}_{j,5j}}, \ \ t  \equiv  
\frac{2iF^{\prime\prime}_5+{\cal M}_{5,0}}
{{\cal M}_{5,5}}, \nn 
\end{eqnarray} 
with the relaxation kernels ${\cal M}_{a,b}$ being given by $D$ and $t$ 
self-consistently;   
\begin{widetext}
\begin{eqnarray} 
&& \hspace{-0.5cm} 
{\cal M}_{a,b} \equiv 2iF^{\prime\prime}_0 \delta_{a,b}
+ \frac{|A_0|}{2^{13} \pi^2 D}  \frac{1}{\Lambda^2}  
\int 
\int_{0<|k+k'|<l^{-1}}  
d^3 k d^3 k' \frac{1}{|k+k'|^2}   \nn \\
&& \hspace{-0.4cm}  
\times \big\{\hat{\gamma}^{\rm L}_{a}(k)\big\}_{\beta\alpha} 
\big\{(1+t^2)
\big(\hat{1}+\hat{T}_4\big) 
\big(\hat{1}-\hat{S}_1\big) 
- (1-t^2)\big(\hat{1}-\hat{T}_4\big)\big(\hat{T}_1-\hat{S}_2\big) 
+  2t \big(\hat{T}_2+\hat{T}_3\big)\big(\hat{1} 
- \hat{S}_1\big)\big\}_{\alpha\delta,\gamma\beta} 
 \big\{\hat{\gamma}^{\rm R}_{a}(k')
\big\}_{\delta\gamma}. \nn 
\end{eqnarray}
\end{widetext}  
Note that $\hat{\gamma}^{\rm L,R}(k)$ above are previously defined; 
\begin{eqnarray} 
&&\hspace{-0.5cm} 
\hat{\gamma}^{\rm L}_{a}(k)
\equiv \frac{1}{2}\big\{\delta \hat{G}(k;0,0) 
\cdot \hat{G}^{R,-1}(k,\mu) 
\cdot \hat{v}_a \cdot
\hat{G}^{R}(k,\mu)  \nn \\
&&\hspace{0.1cm} + \ \hat{G}^{A}(k,\mu) 
\cdot \hat{v}_a \cdot 
\hat{G}^{A,-1}(k,\mu) \cdot 
\delta \hat{G}(k;0,0)  \big\}, \nn \\  
&&\hspace{-0.5cm}  
\hat{\gamma}^{\rm R}_{a}(k)  \equiv  
\hat{v}_{a}\times g_{00,a}(|k|), \nn  
\end{eqnarray} 
with $\hat{v}_a$ for $a=0,5,j$ and $5j$ given in 
eq.~(\ref{vde}).  
$g_{00,a}(x)$ used in $\hat{\gamma}^{\rm R}_a(k)$ 
are given in terms of the 1-point Green functions ($F_0$ and $F_5$) and
the tensor-form factor $t$;    
\begin{eqnarray}  
&&\hspace{-0.8cm}
 g_{00,0}(x) \propto  
\frac{x^2+\big\{(|F_0|^2+|F_5|^2)-t(F^{*}_0F_5+F_0F^{*}_5)\big\}}{|(a+ib)^2-x^2|^2}, \nn \\   
&&\hspace{-0.8cm}
g_{00,5}(x)  \propto   
\frac{tx^2-\big\{t(|F_0|^2+|F_5|^2)-(F^{*}_0F_5+F_0F^{*}_5)\big\}}{|(a+ib)^2-x^2|^2} \label{msg0005} \nn \\   
&&\hspace{-0.8cm}  
g_{00,j}(x) \propto 
\bigg\{\frac{F^{\prime\prime}_0-tF^{\prime\prime}_5}{|(a+ib)^2-x^2|^2}
+ \frac{8}{3}\frac{ab(F^{\prime}_0-tF^{\prime}_5)x^2}{|(a+ib)^2-x^2|^4}\bigg\}, \nn \\ 
&&\hspace{-0.8cm}
g_{00,5j}(x)  \propto 
\bigg\{\frac{F^{\prime}_5-tF^{\prime}_0}{|(a+ib)^2-x^2|^2}
- \frac{8}{3}\frac{ab(F^{\prime\prime}_5-tF^{\prime\prime}_0)x^2}
{|(a+ib)^2-x^2|^4}\bigg\}, \nn  
\end{eqnarray}  
with $(a+ib)^2\equiv F^2_0-F^2_5$.   
\section{Derivation of $g_{00,a}(x)$}
\label{sec:g00}
Starting from the Bethe-Salpeter (BS) equation 
for the response function, we have derived in the 
section~B1 the EOM's for 
the various types of relaxation functions. Such coupled EOM's 
have two features; they are closed and {\it linearized} 
with respect to the relaxation functions. Because of these two 
features, we can solve them for the relaxation 
functions.  
Out of this solution, we can relate the renormalized 
diffusion constant with the 2PIR (two-particle irreducible) vertex 
function. This relation in turn becomes an essential building-block 
of the self-consistent loop of the diffusion constant (see appendix.~B). 

To obtain such linearized EOM's, 
we need to reduce the convolution part between the 2PIR 
vertex function and the response function 
into the simple product between relaxation kernels and 
relaxation functions. For this purpose, we have introduced  
the completeness in the space of 
the integral variable, say $y$ or $w$, associated with 
this convolution;  
\begin{eqnarray}
\sum_{a} u_{a}(y)\cdot u^{*}_{a}(w) \equiv \delta(y-w). \nn 
\end{eqnarray} 
Namely, by use of this, 
any convolution in principle 
can be decomposed into 
a simple product;  
\begin{eqnarray}
&&\hspace{-0.2cm} 
\int dy f(\cdots,y) g(y,\cdots) \equiv \nn \\
&&\hspace{0.5cm} \  \sum_{a} \int dy f(\cdots,y) u_{a}(y) 
\cdot \int dw u^{*}_{a}(w) g(w,\cdots). \nn 
\end{eqnarray} 
In the current context, $f(\cdot \cdot,y)$ corresponds to the 
2PIR vertex function, while $g(w,\cdot \cdot)$ to the response function. 
Therefore, $\int dy \cdots f(\cdot \cdot, y) u_{a}(y)$ corresponds 
to the relaxation kernels, while $\int dw u^{*}_{a}(w) g(w,\cdot \cdot) \cdots $ 
does to the relaxation functions (see also Appendix.~B1). 
The trade-off for this decomposition 
is therefore the sum over {\it infinite}  
(but countable) numbers of modes specified 
by $a$.   
   
To be more specific, we did this decomposition systematically, 
based on the completeness relation of the $\gamma$-matrices and 
the spherical harmonic function $Y_{lm}(\hat{\Omega})$;  
\begin{eqnarray}
&&\hspace{-0.4cm} 
\int dw u^{*}_{a}(w) g(w,\cdots) \rightarrow  \nn \\ 
&&\hspace{0.4cm} 
\sum_{\hat{k}} \sum_{\alpha,\beta} \big[\hat{\gamma}_{\mu}\big]_{\beta\alpha}
Y^{*}_{lm}(\hat{k}) \Phi_{\alpha\cdots,\cdots\beta}(|k|\hat{k},\cdots). \label{b0}
\end{eqnarray}  
As such, the momentum integral only over the 
angle-direction, $\hat{k}$, is taken, while that over its radial 
direction, $|k|$, is {\it not} taken. 
As for the convolution with respect to this radial direction, 
we simply replace the $|x|$-dependence of $\Phi_{\cdots}(|x|\hat{x},\cdots)$ by 
some real-valued function $g_{\cdots}(|x|)$. Namely, we rewrite 
the right hand side of eq.~(\ref{b0}) as follows;  
\begin{eqnarray}
&&\hspace{-0.8cm} 
\sum_{\hat{k}} \sum_{\alpha,\beta} \big[\hat{\gamma}_{\mu}\big]_{\beta\alpha}
Y^{*}_{lm}(\hat{k}) \Phi_{\alpha\cdots,\cdots\beta}(|k|\hat{k},\cdots) \nn \\
&&\hspace{-0.5cm}  
=g_{lm,\mu}(|k|)\sum_{k} 
\sum_{\alpha,\beta} \big[\hat{\gamma}_{\mu}\big]_{\beta\alpha}
Y^{*}_{lm}(\hat{k}) \Phi_{\alpha\cdots,\cdots\beta}(k,\cdots). \label{b1} 
\end{eqnarray}

Let us justify this treatment of the radial direction.  
In the response function,  
$\Phi_{\alpha\delta,\gamma\beta}(k,k';q,\omega)$, 
$q$ and $\omega$ are associated with the external momentum 
and frequency for the bosonic 
degrees of freedom. We can take these two to be 
small, as far as the relaxation functions 
for the long wave-length and 
low-energy region is concerned.  
Then, such a response function is 
usually dominated by the diffuson  
$\hat{\Gamma}^d(q,\omega)$ (see eq.~(\ref{b1d})). 
As a result of this, 
the $|k|$-dependence 
in the left hand side of eq.~(\ref{b1}) and 
its $q$, $\omega$-dependence can be factorized 
at the leading order in small $q$ and $\omega$; 
\begin{eqnarray}
&&\hspace{-0.7cm} 
\sum_{k'}\sum_{\hat{k}} \sum_{\alpha,\beta,\gamma} 
\big[\hat{\gamma}_{\mu}\big]_{\beta\alpha}
Y^{*}_{lm}(\hat{k}) \Phi_{\alpha\gamma,\gamma\beta}(x\hat{k},k';q,\omega) \nn \\&&\hspace{-0.4cm} 
=g_{lm,\mu}(x)\sum_{k,k'} \sum_{\alpha,\beta,\gamma} \big[\hat{\gamma}_{\mu}\big]_{\beta\alpha}
Y^{*}_{lm}(\hat{k}) \Phi_{\alpha\gamma,\gamma\beta}(k,k';,q,\omega) \nn \\ 
&&\hspace{-0.4cm} 
= g_{lm,\mu}(x) \times \phi_{lm,\mu}(q,\omega). \label{b3}
\end{eqnarray} 
To see this factorization more explicitly,  
one can take the following steps: (i) substitute the asymptotic 
tensor-form of the diffuson into eq.~(\ref{b1d}) and the left hand side 
of eq.~(\ref{b3}), (ii) take the integral and the 
sum over $\hat{k}$, $k^{\prime}$, 
$\alpha$, $\beta$ and $\gamma$ in eq~(\ref{b3}) 
, and (iii) retain the leading 
order in small $q$ and $\omega$. By way of this, one can reach 
the factorization given in the right hand side of 
eq.~(\ref{b3}) with a specific 
$g_{lm,\mu}(x)$. 

For example, let us follow these prescriptions 
in the case of zero topological mass case.  
Observing eq.~(\ref{masslessd}), 
notice first the following relation; 
\begin{eqnarray}
\hat{\Gamma}^d_{\alpha\delta,\gamma\beta}(q,\omega)_{|F_5\equiv 0}
\big[\hat{\gamma}_{\mu}\big]_{\delta,\gamma} \equiv 0,  \label{mlsim}
\end{eqnarray} 
for $\mu=1,2,3$. Using this, one can readily checked that 
the diffuson in this case turns out to be proportional to 
the unit matrix, when its right hand side is traced out;
\begin{eqnarray}
&&\hspace{-0.6cm} 
\sum_{\delta,\gamma}
\hat{\Gamma}^d_{\alpha\delta,\gamma\beta}(q,\omega)_{|F_5=0} 
\sum_{k',\epsilon} \hat{G}^{R}_{\delta\epsilon}(k'_{+},\mu_{+})
\hat{G}^{A}_{\epsilon\gamma}(k'_{-},\mu_{-}) \nn \\
&&\hspace{0.5cm} = \frac{1}{\omega+iDq^2} 
\sum_{k'}\frac{|F_0|^2+{k^{\prime}}^2}{|F^2_0-{k^{\prime}}^2|^2}  
\delta_{\alpha\beta}. \nn 
\end{eqnarray}
As such, to obtain the normalized function $g_{00,a}(x)$ 
in the massless case,  we have only to 
calculate the following quantity up 
to the leading order in small $\omega$ and $q$; 
\begin{eqnarray} 
&&\hspace{-0.6cm} 
\sum_{\hat{k},k',\delta,\alpha,\beta}
\big[\hat{v}_{a}(q)\big]_{\beta\alpha}\Phi_{\alpha\delta,\delta\beta}(k,k^{\prime};q,\omega) \nn \\
&&\hspace{-0.5cm} 
 \propto \frac{1}{i\omega-Dq^2}
\sum_{\hat{k}} \sum_{\alpha,\delta,\rho} 
\hat{G}^{R}_{\delta\alpha}(k_{+},\mu_{+})
\hat{G}^{A}_{\alpha\rho}(k_{-},\mu_{-}) \big[\hat{v}_{a}(q)\big]_{\rho\delta}.  \nn 
\end{eqnarray}
For example, taking the current 
component as $\hat{v}_{a}(q)$ above, we have;
\begin{eqnarray}
&&\hspace{-0.5cm} 
\sum_{\hat{k},k',\alpha,\beta,\delta}
\big[\hat{q}_{\mu}\hat{\gamma}_{\mu}\big]_{\beta\alpha}
\Phi_{\alpha\delta,\delta\beta}(k,k^{\prime};q,\omega) \nn \\
&&\hspace{-0.2cm} 
\propto - \frac{q}{i\omega-Dq^2} 
\Big\{\frac{1}{|F^2_0-k^2|^2} + \frac{8}{3} 
\frac{{F^{\prime}_0}^2 k^2}{|F^2_0-k^2|^4}\Big\} + {\cal O}(q^2,\omega). \nn
\end{eqnarray}  
Observing the right hand side, one can then convince oneself of 
eq.~(\ref{b3}). Moreover, the normalized real-valued 
function $g_{00,j}(x)$ will be obtained as in 
eqs.~(\ref{g00j},\ref{ng00j}).

\end{document}